\documentclass[twoside]{article}

\usepackage[accepted]{aistats2022}
\usepackage[round]{natbib}

\usepackage{graphicx}
\usepackage{bbold}
\usepackage{bm}
\usepackage{amsmath,amscd,amssymb}
\usepackage[linesnumbered,ruled,vlined]{algorithm2e}
\usepackage{xcolor,colortbl}
\usepackage{tabu}
\usepackage{hyperref}

\usepackage{lipsum}
\usepackage{mathtools}
\usepackage{cuted}

\usepackage{enumitem}

\usepackage{thm-restate,cleveref}
\usepackage{wrapfig,caption}

\usepackage[mathscr]{euscript}

\newcommand{\EE}{\mathbb{E}}
\newcommand{\thalf}{\tfrac{1}{2}}
\newcommand{\DD}{\mathcal{D}}
\newcommand{\II}{\mathcal{I}}

\newcommand{\KK}{\mathcal{K}}
\newcommand{\NN}{\mathcal{N}}
\newcommand{\XX}{\mathcal{X}}

\newcommand{\kk}{\mathscr{K}}
\newcommand{\ik}{{\II_\kk}}

\newcommand{\ZZt}{\tilde{\mathcal{Z}}}
\newcommand{\XXt}{\tilde{\mathcal{X}}}
\newcommand{\inv}{^{-1}}
\newcommand{\invT}{^{- {\rm T}}}
\newcommand{\TT}{^{\rm T}}
\newcommand{\ZZ}{\mathcal{Z}}

\newcommand{\OO}{\mathcal{O}}
\newcommand{\RR}{\mathbb{R}}
\newcommand{\ZZZ}{\mathbb{Z}}

\newcommand{\bhat}{\hat{\beta}}
\newcommand{\psihat}{\hat{\psi}}

\newcommand{\id}{\mathbb{1}}
\newcommand{\gmi}{\gamma_{-i}}
\newcommand{\gmj}{\gamma_{-j}}
\newcommand{\olya}{\`olya}

\newcommand{\omhat}{\hat{\omega}}
\newcommand{\Tburn}{T_{\rm burn}}
\newcommand{\TC}{C}
\newcommand{\gone}{{\gamma + 1}}
\newcommand{\gabsone}{{|\gamma| + 1}}
\newcommand{\gabs}{{|\gamma|}}

\newcommand{\AAA}{\mathscr{A}}
\newcommand{\PPP}{\mathscr{P}}
\newcommand{\SSS}{\mathscr{S}}

\begin{document}

%

%

\twocolumn[

\aistatstitle{Bayesian Variable Selection in a Million Dimensions}

\aistatsauthor{Martin Jankowiak }

\aistatsaddress{ Broad Institute  \;\;\;\; Basis Research Institute} ]

\begin{abstract}
Bayesian variable selection is a powerful tool for data analysis, as
it offers a principled method for variable selection that accounts for prior information and uncertainty. 
However, wider adoption of Bayesian variable selection has been hampered by computational challenges, 
especially in difficult regimes with a large number of covariates $P$ or non-conjugate likelihoods.
To scale to the large $P$ regime we introduce an efficient MCMC scheme whose cost per iteration is sublinear in $P$. 
In addition we show how this scheme can be extended to generalized linear models for count data, 
which are prevalent in biology, ecology, economics, and beyond.
In particular we design efficient algorithms for variable selection in binomial and negative binomial regression, which 
includes logistic regression as a special case.
In experiments we demonstrate the effectiveness of our methods, including on cancer and maize genomic data.
\end{abstract}

\section{Introduction}
\label{sec:intro}

Generalized linear models are ubiquitous throughout applied statistics and data analysis \citep{mccullagh2019generalized}.
One reason for their popularity is their interpretability:
they introduce explicit parameters that encode how the observed response depends
on each covariate. 
In the scientific setting this interpretability is of central importance.
Indeed model fit is often a secondary concern, and the primary goal is to identify
a \emph{parsimonious} explanation of the observed data. 
This is naturally viewed as a model selection problem, in particular one
in which the model space is defined as a nested set of models, with
distinct models including distinct sets of covariates.

The Bayesian formulation of this approach, known as Bayesian variable selection 
in the literature, 
offers a powerful set of techniques for realizing Occam's razor in this 
setting \citep{george1993variable, george1997approaches, chipman2001practical}. 
Despite the intuitive appeal of this approach, approximating the
resulting posterior distribution can be computationally challenging. 
A principal reason for this is the astronomical size of the model space that results
whenever there are more than a few dozen covariates.
Indeed for $P$ covariates the total number of distinct models, namely $2^P$, 
exceeds the estimated number of atoms in the known universe ($\sim\!10^{80}$) for $P \gtrsim 266$.
In addition for many models of interest non-conjugate likelihoods make it infeasible to integrate
out real-valued model parameters, resulting in a challenging high-dimensional 
inference problem defined on a transdimensional mixed discrete/continuous latent space. 
In this work we develop efficient MCMC methods for Bayesian variable selection. 
Our contributions include:
\begin{enumerate}[itemsep=0pt]
\item We introduce an efficient MCMC sampler for large $P$ whose cost per iteration is sublinear in $P$. 
\item We develop efficient MCMC samplers for two generalized linear models 
for count data: i) binomial regression and ii) negative binomial regression.
\item We show how our algorithmic strategies can be combined and how they can accommodate inference over the prior inclusion probability.
\end{enumerate}
\vspace{-1mm}

\section{Background}
\label{sec:bg}

\subsection{Problem setup}
\label{sec:setup}

Consider linear regression with $X \in \RR^{N\times P}$ and $Y \in \RR^{N}$
and define the following space of models: 
\begin{align}
\label{eqn:linmodeldefn}
&{\rm [inclusion \; variables]} \qquad &&\gamma_i \sim {\rm Bernoulli}(h) \\ 
&{\rm [noise \; variance]} \qquad  &&\sigma^2 \sim {\rm InvGamma}(\thalf \nu_0, \thalf \nu_0 \lambda_0) \nonumber \\ 
&{\rm [coefficients]} \qquad &&\beta_\gamma  \sim \NN(0, \sigma^2 \tau \inv \id_\gabs)  \nonumber\\
&{\rm [responses]} \qquad  &&Y_n \sim \NN(\beta_\gamma  \cdot X_{n\gamma}, \sigma^2) \nonumber
\end{align}
where $i=1,...,P$ and $n=1,...,N$.
Here each $\gamma_i \in \{0, 1\}$ controls whether the coefficient $\beta_i$
and the $i^{\rm th}$ covariate are included ($\gamma_i=1$) or excluded ($\gamma_i=0$) from the model.
In the following we use $\gamma$ to refer to the vector $(\gamma_1, ..., \gamma_P)$.
The hyperparameter $h \in (0, 1)$ controls the overall level of sparsity; in particular $hP$ is the expected number of 
covariates included a priori.
The $|\gamma|$ coefficients $\beta_\gamma \in \RR^{|\gamma|}$ are governed by a Normal prior 
with precision proportional to $\tau > 0$.\footnote{We usually drop the $\gamma$ subscript on $\beta_\gamma$ to simplify the notation.}
Here $|\gamma| \in \{0, 1, ..., P \}$ denotes the total number of included covariates. 
The response $Y_n$ is generated from a Normal distribution with variance governed by an Inverse Gamma prior.\footnote{Throughout
we take the limit $\nu_0 \rightarrow 0$ and $\lambda_0 \rightarrow 0$, which corresponds 
to an improper prior $p(\sigma^2) \propto \sigma^{-2}$.}
Note that we do not include a bias term in Eqn.~\ref{eqn:linmodeldefn}, but doing so may be desirable in practice.
An attractive feature of Eqn.~\ref{eqn:linmodeldefn} is that it explicitly reasons about variable inclusion
and allows us to define \emph{posterior inclusion probabilities} or PIPs:
\begin{align}
    \rm{PIP}(i) \equiv p(\gamma_i = 1 | \DD) \in [0, 1]
\end{align}
where $\DD = \{X, Y \}$ is the observed dataset.

\subsection{Inference}
\label{sec:bginf}

Conjugacy in Eqn.~\ref{eqn:linmodeldefn} implies that the coefficients $\beta$ and the variance $\sigma^2$ can be integrated out, 
resulting in a discrete inference problem over $\{0, 1\}^P$ \citep{chipman2001practical}.
Inference over $\{0, 1\}^P$ readily admits a Gibbs sampling scheme;
however, the resulting sampler is notoriously slow in high dimensions.
For example, consider the scenario in which the two covariates corresponding to $i=1$ and $i=2$ are highly correlated and
each on its own is sufficient for explaining the responses $Y$. In this scenario the posterior concentrates on models with
$\gamma = (1, 0, 0, ...)$ and $\gamma = (0, 1, 0, ...)$. Single-site Gibbs updates w.r.t.~$\gamma_i$ will move between the two modes very
infrequently, since they are separated by low probability models like $\gamma = (0, 0, 0, ...)$.

A recently developed inference algorithm---Tempered Gibbs Sampling (TGS) \citep{zanella2019scalable}---utilizes
\emph{coordinatewise tempering} to cope with this kind of problematic stickiness. 
In the following we describe a variant of TGS called wTGS that is particularly 
well-suited to Bayesian variable selection \citep{zanella2019scalable}. As we will see,
this algorithm will serve as a powerful subtrate for building MCMC samplers for Bayesian variable selection 
that can accommodate large $P$ and count-based likelihoods.

\subsection{wTGS}
\label{sec:wtgs}

Consider the (unnormalized) target distribution
\begin{align}
\label{eqn:fulllintarget}
f(\gamma, i)
&\equiv p(\gamma | \DD) \frac{ \eta(\gmi) U(\gamma_i)}{ p(\gamma_i | \gmi, \DD)} \\
& = U(\gamma_i) \eta(\gmi) p(\gmi | \DD)
\end{align}
where we have introduced an auxiliary variable $i \in \{ 1, ..., P \}$.
Here $U(\cdot)$ is the uniform distribution on $\{0, 1\}$ and $\gmi$ denotes all components of $\gamma$ apart
from $\gamma_i$. Finally $\eta(\gmi)$ is an additional weighting factor to be defined below.
The key property of Eqn.~\ref{eqn:fulllintarget} is that for any $i$ the distribution over $\gamma_i$ is 
uniform and factorizes across $\{ \gamma_i, \gmi \}$.
wTGS proceeds by defining a sampling scheme for the target Eqn.~\ref{eqn:fulllintarget} that utilizes Gibbs updates 
w.r.t.~$i$ and Metropolized-Gibbs updates w.r.t.~$\gamma_i$.

\paragraph{$i$-updates}

If we marginalize $i$ from Eqn.~\ref{eqn:fulllintarget} we obtain
\begin{align}
f(\gamma) = p(\gamma | \DD) \phi(\gamma)
\label{eqn:fphilin}
\end{align}
where we define
\begin{align}
 \phi(\gamma) \equiv \sum_{i=1}^P  \frac{\tfrac{1}{2} \eta(\gmi)}{ p(\gamma_i | \gmi, \DD)}
\label{eqn:phidefnlin}
\end{align}
As is clear from Eqn.~\ref{eqn:fphilin}, $\phi(\gamma)\inv$ is an \emph{importance weight} that can be used to obtain samples
from the non-tempered target of interest, i.e.~$p(\gamma | \DD)$.
Additionally Eqn.~\ref{eqn:fulllintarget} implies that we can do Gibbs updates w.r.t.~$i$ using the distribution\footnote{Note that Eqn.~\ref{eqn:phidefnlin}-\ref{eqn:iupdatelin}-\ref{eqn:etadefn} depends on conditional PIPs $p(\gamma_i =1| \gmi, \DD)$;
as discussed in Sec.~\ref{app:mll} these can be computed efficiently with careful linear algebra.}
\begin{align}
\label{eqn:iupdatelin}
f(i | \gamma)  =  \frac{1}{ \phi(\gamma) }  \frac{\tfrac{1}{2}  \eta(\gmi)}{ p(\gamma_i | \gmi, \DD)} 
\end{align}

\paragraph{$\gamma$-updates}

The auxilary variable $i$ is used to control which component of $\gamma$ we update. 
Since the posterior conditional w.r.t.~$\gamma_i$ is the uniform distribution $U(\gamma_i)$,
Metropolized-Gibbs \citep{liu1996peskun} updates w.r.t.~$\gamma_i$ result in deterministic flips that are accepted with probability one:
$\gamma_i \to 1 - \gamma_i$.

\paragraph{Weighting factor $\eta$}

To finish specifying wTGS we need to define the weighting factor $\eta(\gmi)$ in Eqn.~\ref{eqn:fulllintarget}:
\begin{align}
    \eta(\gmi)= p(\gamma_i = 1 | \gmi, \DD) + \tfrac{\epsilon}{P}
\label{eqn:etadefn}
\end{align}
Here $p(\gamma_i = 1 | \gmi, \DD) $ is a conditional PIP, and $\epsilon$ 
trades off between exploitation ($\epsilon \to 0$) and exploration ($\epsilon \to \infty$).
Indeed since the marginal $f(i)$ is given by
\begin{align}
\label{eqn:marglin}
    f(i) \propto \EE_{p(\gmi | \DD)} \left[ \eta(\gmi) \right] = \rm{PIP}(i) + \tfrac{\epsilon}{P}
\end{align}
this choice of $\eta$ ensures that the sampler focuses its computional effort on large PIP 
covariates.\footnote{See Sec.~\ref{app:wtgsdisc} for further discussion of the mix of exploration and exploitation enabled
by weighted tempering.}
For the full algorithm see Algorithm~\ref{linalgo} in the supplement.

\paragraph{Rao-Blackwellization}

A side benefit of computing conditional PIPs in Eqn.~\ref{eqn:iupdatelin} 
is that they can be repurposed to compute lower variance Rao-Blackwellized PIP estimates.
See Sec.~\ref{app:rao} for details.

\section{The large $P$ regime: Subset wTGS}
\label{sec:largep}

Running wTGS in the large $P$ regime can be prohibitively expensive,
since it involves computing $P$ conditional PIPs per MCMC iteration. 
We would like to devise an algorithm that, like wTGS, utilizes conditional PIPs to 
make informed moves in $\gamma$ space while avoiding this $\OO(P)$ computational cost.

\paragraph{Subset wTGS}
\label{sec:sswtgs}

\begin{algorithm*}[t!]  
\DontPrintSemicolon 
\KwIn{Dataset $\DD = \{X, Y\}$ with $P$ covariates;
     prior inclusion probability $h$;
     prior precision $\tau$;
     subset size $S$; anchor set size $A$;
     total number of MCMC iterations $T$; number of burn-in iterations $\Tburn$.
     }
    \KwOut{Approximate weighted posterior samples $\{\rho^{(t)}, \gamma^{(t)} \}^T_{t=\Tburn +1}$}
    Let $\gamma^{(0)} = (0, ..., 0)$ and choose $\AAA$ to be the $A$ covariate indices 
    exibiting the largest correlations with $Y$. \\
    Choose $i^{(0)}$ randomly from $\{1, ..., P\}$ and $\SSS^{(0)} \sim  U(\cdot | i^{(0)}, \AAA)$.\\
\For{$t =1, ..., T$} {
    Sample $i^{(t)} \sim f(\cdot | \gamma^{(t-1)}, \SSS^{(t-1)})$ using Eqn.~\ref{eqn:iupdateSS} \\
    Let $\gamma^{(t)} = {\rm flip}(\gamma^{(t-1)} | i^{(t)})$ where ${\rm flip}( \gamma | i)$ 
        flips the $i^{\rm th}$ coordinate of $\gamma$: $\gamma_i \to 1 - \gamma_i$. \\
    Sample $\SSS^{(t)} \sim U(\cdot | i^{(t)}, \AAA)$ and compute the unnormalized weight $\tilde{\rho}^{(t)} = \phi(\gamma^{(t)}, \SSS^{(t)})^{-1}$ using Eqn.~\ref{eqn:phidefnSS}. \\
    If $t \le \Tburn$ adapt $\AAA$ using the scheme described in Sec.~\ref{app:anchor}.
}
    Compute the normalized weights $\rho^{(t)} = \frac{ \tilde{\rho}^{(t)}  } { \sum_{s > \Tburn} \tilde{\rho}^{(s)} }$ for 
    $t=\Tburn + 1,...,T$. \\
    \Return{$\{\rho^{(t)}, \gamma^{(t)} \}_{t=\Tburn + 1}^T$}
    \caption{We outline the main steps in Subset wTGS. See Sec.~\ref{sec:largep} for details. Subset wTGS
    reduces to wTGS in the limit $S \rightarrow P$, in which case $\SSS$ becomes redundant.
    Superscripts indicate MCMC iterations.}
\label{algo}
\end{algorithm*}

To do so we leverage a simple augmentation strategy. In effect, we introduce
an auxiliary variable $\SSS \subset \{1, ..., P \}$ that controls which conditional
PIPs are computed in a given MCMC iteration. Since we can
choose the size $S$ of $\SSS$ to be much less than $P$, this can result in significant speed-ups. 

In more detail, consider the following (unnormalized) target distribution:
\begin{align}
\label{eqn:fulltargetSS}
f(\gamma, i, \SSS)
    &\equiv p(\gamma | \DD) \frac{ \eta(\gmi) U(\gamma_i)}{ p(\gamma_i | \gmi, \DD)} U(\SSS | i, \AAA)
\end{align}
Here $\SSS$ ranges over all the subsets of $\{1, ..., P \}$ of size $S$ that also contain a fixed
`anchor' set $\AAA \subset \{1, ..., P \}$ of size $A < S$.  
Moreover $ U(\SSS | i, \AAA)$ is the uniform distribution over all size $S$ subsets of $\{1, ..., P \}$ that contain
both $i$ and $\AAA$.\footnote{This is reminiscent of the Hamming ball construction in \citet{titsias2017hamming}.} 
We choose the same weighting function $\eta$ as in wTGS (see Eqn.~\ref{eqn:etadefn}).
In practice we adapt $\AAA$ during burn-in,
but for now the reader can suppose that $\AAA$ is chosen at random.
Subset wTGS proceeds by defining a sampling scheme for the target distribution Eqn.~\ref{eqn:fulltargetSS} that utilizes Gibbs updates 
w.r.t.~$i$ and $\SSS$ and Metropolized-Gibbs updates w.r.t.~$\gamma_i$.

\paragraph{$i$-updates}

Marginalizing $i$ from Eqn.~\ref{eqn:fulltargetSS} yields 
\begin{align}
f(\gamma, \SSS) = p(\gamma | \DD) \phi(\gamma, \SSS)
\label{eqn:fphiSS}
\end{align}
where we define
\begin{align}
 \phi(\gamma, \SSS) \equiv \sum_{i \in \SSS}  \frac{\tfrac{1}{2} \eta(\gmi)}{ p(\gamma_i | \gmi, \DD)}  U(\SSS | i, \AAA)
\label{eqn:phidefnSS}
\end{align}
and have leveraged that $ U(\SSS | i, \AAA) =0$ if $i \notin \SSS$.
Crucially, computing $\phi(\gamma, \SSS)$ is $\OO(S)$ instead of $\OO(P)$.
We can do Gibbs updates w.r.t.~$i$ using the distribution
\begin{align}
\label{eqn:iupdateSS}
f(i | \gamma, \SSS)  \propto  \frac{\eta(\gmi)}{ p(\gamma_i | \gmi, \DD)}  U(\SSS | i, \AAA)
\end{align}

\paragraph{$\gamma$-updates}

Just as for wTGS we utilize Metropolized-Gibbs updates w.r.t.~$\gamma_i$ that result in deterministic flips $\gamma_i \to 1 - \gamma_i$.
Likewise the marginal $f(i)$ is proportional to $\rm{PIP}(i) + \tfrac{\epsilon}{P}$ so that the sampler focuses on large PIP covariates. 

\paragraph{$\SSS$-updates}

$\SSS$ is updated with Gibbs moves, $\SSS \sim U(\cdot | i, \AAA)$.
For the full algorithm see Algorithm~\ref{algo}.

\paragraph{Importance weights}

The Markov chain in Algorithm~\ref{algo} targets the auxiliary distribution Eqn.~\ref{eqn:fulltargetSS}.
To obtain samples from the desired posterior $p(\gamma | \DD)$ we reweight each sample $(\gamma, \SSS)$
with an importance weight $\tilde{\rho} = \phi(\gamma, \SSS)^{-1}$ and discard $\SSS$; see Eqn.~\ref{eqn:fphiSS}.
Crucially, the importance weights are upper bounded and exhibit only moderate variance. 
Ultimately this moderate variance can be traced to the coordinatewise tempering, which keeps the tempering to a 
modest level; see Sec.~\ref{app:iw} for additional discussion. Moreover, we can show that samples obtained with Algorithm \ref{algo}  
can be used to estimate posterior quantities of interest:

\begin{restatable}{prop}{proplarge}
\label{prop:large}
The Subset wTGS estimator 
\begin{equation}
    \label{eqn:prop}
    {\textstyle \sum}_{t=1}^T  \rho^{(t)} h(\gamma^{(t)})
    \rightarrow \EE_{p(\gamma | \DD)} \left[ h(\gamma) \right] \;\; {\rm as} \;\; T \rightarrow \infty 
\end{equation}
almost surely for every test function $h(\gamma): \{0,1\}^P \rightarrow \RR$, 
    where $\rho^{(t)} \propto \phi^{-1}(\gamma^{(t)}, \SSS^{(t)})$ are normalized weights.
    Moreover, we can use a (partially) Rao-Blackwellized PIP estimator in Eqn.~\ref{eqn:prop}.
    See Sec.~\ref{app:proof} in the supplement for the proof and additional details.
\end{restatable}

\section{Binomial Regression: PG-wTGS}
\label{sec:pginf}

For simplicity we focus on the binomial regression case, leaving a discussion of the
negative binomial case to Sec.~\ref{app:nb} in the supplement.
Let $X \in \RR^{N\times P}$, $\TC \in \ZZZ^{N}_{> 0}$, and $Y \in \ZZZ^{N}_{\ge 0}$ with $Y \le \TC$ 
and consider the following space of generalized linear models: 
\begin{align}
\label{eqn:pgmodeldefn}
&{\rm [inclusion \; variables]} \qquad &&\gamma_i \sim {\rm Bernoulli}(h) \\ 
&{\rm [bias \; term]} \qquad &&\beta_0 \sim \NN(0, \tau_{\rm bias} \inv)  \nonumber\\
&{\rm [coefficients]} \qquad &&\beta_\gamma  \sim \NN(0, \tau \inv \id_\gabs)  \nonumber\\
&{\rm [success \; logits]} \qquad  &&\psi_n \equiv \beta_0 + \beta_\gamma  \cdot X_{n\gamma} \nonumber \\ 
&{\rm [responses]} \qquad  &&Y_n \sim {\rm Binomial}(\TC_n, \sigma(\psi_n)) \nonumber
\end{align}
where $i=1,...,P$ and $n=1,...,N$.
Note that we introduce a bias term $\beta_0$ governed by a Normal prior with precision $\tau_{\rm bias} > 0$; 
we assume that $\beta_0$ is always included in the model.\footnote{For simplicity we take $\tau = \tau_{\rm bias}$ throughout.}
The response $Y_n$ is generated from a Binomial distribution with total count $\TC_n$ and success probability $\sigma(\psi_n)$,
where $\sigma(\cdot)$ denotes the logistic function $\sigma(x) \equiv \{1 + \exp(-x) \}\inv$. 
This reduces to logistic regression with binary responses if $\TC_n\!=\!1 \forall n$.

\subsection{P\olya-Gamma augmentation}
\label{sec:pg}

wTGS relies on conditional PIPs to construct informed moves; unfortunately these cannot be computed in closed form
for non-conjugate likelihoods like that in Eqn.~\ref{eqn:pgmodeldefn}.
To get around this we introduce P\olya-Gamma auxiliary variables, which rely on the identity
\begin{align}
    \frac{(e^\psi)^a}{(1 + e^\psi)^b} = \tfrac{1}{2^b} e^{(a-\tfrac{1}{2}b)\psi} \EE_{{\rm PG}(\omega |  b, 0)} \left[ \exp(-\tfrac{1}{2} \omega \psi^2) \right] \nonumber
\end{align}
noted by \citet{polson2013bayesian}. Here $a, \psi \in \RR$, $b >0$, and ${\rm PG}(\omega |  b, 0)$ is the P\olya-Gamma distribution, which
has support on the positive real axis. Using this identity we can introduce a $N$-dimensional vector of P\olya-Gamma (PG) 
variates $\omega$ governed by the prior $\omega_n \sim {\rm PG}(\TC_n, 0)$
and rewrite the Binomial likelihood in Eqn.~\ref{eqn:pgmodeldefn} as follows
\begin{align}
    p(Y_n | \TC_n, \sigma(\psi_n)) &\propto \sigma(\psi_n)^{Y_n} (1 -  \sigma(\psi_n))^{\TC_n - Y_n}  \\
&=  \frac{(\exp(-\psi_n))^{\TC_n - Y_n}}{(1 + \exp(-\psi_n))^{\TC_n}}  = \frac{(\exp(\psi_n))^{Y_n}}{(1 + \exp(\psi_n))^{\TC_n}}  \nonumber 
\end{align}
so that each likelihood term in Eqn.~\ref{eqn:pgmodeldefn} is replaced with
\begin{equation}
\label{eqn:pgfactor}
\exp(\kappa_n \psi_n -\tfrac{1}{2} \omega_n \psi_n^2)  \;\;\; {\rm with} \;\;\;  \kappa_n \equiv Y_n - \tfrac{1}{2} \TC_n
\end{equation}
This augmentation leaves the marginal distribution w.r.t.~$(\gamma, \beta)$ unchanged.
Crucially each factor in Eqn.~\ref{eqn:pgfactor} is Gaussian w.r.t.~$\beta$, with
the consequence that P\olya-Gamma augmentation establishes conjugacy.

\subsection{PG-wTGS}
\label{sec:pgwtgs}

\begin{algorithm*}[t!]
\DontPrintSemicolon 
\KwIn{Dataset $\DD = \{X, Y, \TC\}$ with $P$ covariates;
     prior inclusion probability $h$;
     prior precision $\tau$;
     total number of MCMC iterations $T$; 
     number of burn-in iterations $\Tburn$;
     hyperparameter $\xi > 0$ (optional)
     }
    \KwOut{Approximate weighted posterior samples $\{\rho^{(t)}, \gamma^{(t)}, \omega^{(t)} \}^T_{t=\Tburn +1}$}
    Let $\gamma^{(0)} = (0, ..., 0)$ and $\omega^{(0)} \sim {\rm PG}(\TC, 0)$. \\
\For{$t =1, ..., T$} {
    Sample $i^{(t)} \sim f(\cdot | \gamma^{(t-1)}, \omega^{(t-1)})$ using Eqn.~\ref{eqn:iupdatepg}. \\
    If $i^{(t)} >0$ let $\omega^{(t)}= \omega^{(t-1)}$ and $\gamma^{(t)} = {\rm flip}(\gamma^{(t-1)} | i^{(t)})$.\\
    Otherwise if $i^{(t)} =0$ let $\gamma^{(t)}= \gamma^{(t-1)}$ and 
               sample $\omega^{\prime (t)} \sim p(\cdot | \gamma^{(t-1)}, \bhat(\gamma^{(t-1)}, \omega^{(t-1)}), \DD)$.
    Set $\omega^{(t)} = \omega^{\prime(t)}$ with probability $\alpha(\omega^{(t)}  \! \to \! \omega^{\prime(t)} | \gamma^{(t)})$ given in Eqn.~\ref{eqn:alphaomega} (accept); otherwise set $\omega^{(t)} = \omega^{(t-1)}$ (reject). \\
    Compute the unnormalized weight $\tilde{\rho}^{(t)} = \phi(\gamma^{(t)}, \omega^{(t)})^{-1}$ using Eqn.~\ref{eqn:phidefnpg}. \\
    If $\xi$ is not provided and $t \le \Tburn$ adapt $\xi$ using the scheme described in Sec.~\ref{app:xi}.
}
    Compute the normalized weights $\rho^{(t)} = \frac{ \tilde{\rho}^{(t)}  } { \sum_{s > \Tburn} \tilde{\rho}^{(s)} }$ for 
    $t=\Tburn + 1,...,T$. \\
    \Return{$\{\rho^{(t)}, \gamma^{(t)}, \omega^{(t)} \}_{t=\Tburn + 1}^T$}
    \caption{We outline the main steps in PG-wTGS. 
    See Sec.~\ref{sec:pginf} for details.
    }
\label{pgalgo}
\end{algorithm*}

We can now adapt wTGS to our setting.
The augmented target distribution in Sec.~\ref{sec:pg} is given by
\begin{align}
\!\!\!\!\! p(Y | \beta, \gamma, \omega, X, \TC)  p(\beta)  p(\gamma)p(\omega|\TC)  \propto p(\beta,\gamma, \omega | \DD) 
\label{eqn:augtargetbeta}
\end{align}
where we define $\DD \equiv \{X, Y, \TC\}$. 
We marginalize out $\beta$ to obtain
\begin{align}
p(Y | \gamma, \omega, X, \TC) p(\gamma) p(\omega|\TC) \propto p(\gamma, \omega | \DD) 
\label{eqn:augtargetpg}
\end{align}
Thanks to PG augmentation we can compute $p(Y | \gamma, \omega,  X, \TC)$ in closed form.
Next we introduce an auxiliary variable $i \in \{0, 1, 2, ..., P\}$ that controls which variables, if any, are
tempered (note the additional state $i=0$).
We define the (unnormalized) target distribution $f(\gamma, \omega, i)$ as follows:
\begin{align}
\label{eqn:fulltargetpg}
p(\gamma, \omega | \DD)  \left \{ \delta_{i0} \xi
                   + \tfrac{1}{P} \Sigma_{j=1}^P \frac{\delta_{ij} \eta(\gmj, \omega) U(\gamma_j)}{ p(\gamma_j | \gmj, \omega, \DD)} \right\} 
\end{align}
Here $\xi > 0$ is a hyperparameter whose choice we discuss below. 
We note two important features of Eqn.~\ref{eqn:fulltargetpg}.
First, by construction when $i>0$ the posterior conditional w.r.t.~$\gamma_i$ is the uniform distribution $U(\gamma_i)$.
Second, as we discuss in more detail in Sec.~\ref{app:mll}, the posterior conditional $p(\gamma_i | \gmi, \omega, \DD)$ in Eqn.~\ref{eqn:fulltargetpg}
can be computed in closed form thanks to PG augmentation. This is important because computing $p(\gamma_i | \gmi, \omega, \DD)$ 
is necessary for importance weighting and Rao-Blackwellization.
We proceed to construct a sampler for the target distribution Eqn.~\ref{eqn:fulltargetpg} that utilizes Gibbs updates w.r.t.~$i$,
Metropolized-Gibbs updates w.r.t.~$\gamma_i$, and Metropolis-Hastings updates w.r.t.~$\omega$.

\paragraph{$i$-updates}

If we marginalize $i$ from Eqn.~\ref{eqn:fulltargetpg} we obtain
$f(\gamma, \omega) = p(\gamma, \omega | \DD) \phi(\gamma, \omega)$
where we define 
\begin{align}
    \phi(\gamma, \omega) \equiv \xi + \frac{1}{P} \sum_{i=1}^P  \frac{\tfrac{1}{2} \eta(\gmi, \omega)}{ p(\gamma_i | \gmi, \omega, \DD)}
\label{eqn:phidefnpg}
\end{align}
Evidently $\phi(\gamma, \omega)\inv$ is the importance weight that is used to obtain samples 
from the non-tempered target Eqn.~\ref{eqn:augtargetpg}. 
Moreover we can do Gibbs updates w.r.t.~$i$ using the distribution
\begin{align}
\label{eqn:iupdatepg}
f(i | \gamma, \omega)  \propto \delta_{i0} \xi + \frac{1}{P} \sum_{j=1}^P \delta_{ij} \frac{\tfrac{1}{2}  \eta(\gmj, \omega)}{ p(\gamma_j | \gmj, \omega, \DD)} 
\end{align}

To better understand the behavior of the auxiliary variable $i$, we compute the marginal 
distribution w.r.t.~$i$ for the special case $\eta(\cdot)=1$,
\begin{align}
    f(i) \propto \delta_{i0} \xi + \tfrac{1}{P} {\textstyle \sum}_{j=1}^P \delta_{ij} 
\label{eqn:fmarg}
\end{align}
which clarifies that $\xi$ controls how often we visit $i=0$.

\paragraph{$\gamma$-updates}

Whenever $i > 0$ we do a Metropolized-Gibbs update of $\gamma_i$, resulting in a flip $\gamma_i \to 1 - \gamma_i$.

\paragraph{$\omega$-updates}

Whenever $i = 0$ we update $\omega$.  To do so we use a simple proposal that can be computed in closed form.
Importantly $f(\gamma, \omega, i=0)$ is not tempered by construction
so we can rely on the conjugate structure that is made manifest when we condition
on a value of $\beta$.
In more detail, we first compute the \emph{mean} of the conditional posterior $p(\beta | \gamma, \omega, \DD)$ of Eqn.~\ref{eqn:augtargetbeta}:
\begin{align}
    \bhat(\gamma, \omega) \equiv \EE_{p(\beta | \gamma, \omega, \DD)} \left[ \beta \right]
\end{align}
Using this (deterministic) value we then form the conditional posterior distribution $p(\omega^\prime | \gamma, \bhat, \DD)$, which is a P\olya-Gamma distribution
whose parameters are readily computed.
We then sample a proposal $\omega^\prime \sim p(\cdot | \gamma, \bhat, \DD)$ and 
compute the corresponding MH acceptance probability $\alpha(\omega \! \to \! \omega^\prime | \gamma)$. The proposal is then accepted
with probability $\alpha(\omega  \! \to \! \omega^\prime | \gamma)$; otherwise it is rejected. The acceptance probability 
can be computed in closed form and is given by
\begin{align}
\label{eqn:alphaomega}
    \!\!\!    &\alpha(\omega  \! \to \! \omega^\prime | \gamma) = \min \bigg(1, 
\frac{p(Y | \gamma, \omega^\prime, X, \TC) }{p(Y | \gamma, \omega, X, \TC)} \times \\
    &\frac{p(Y | \gamma, \omega, \bhat(\gamma, \omega^\prime), X, \TC) }{p(Y | \gamma, \omega^\prime, \bhat(\gamma, \omega), X, \TC)}
    \frac{p(Y | \gamma,  \bhat(\gamma, \omega), X, \TC) }{p(Y | \gamma, \bhat(\gamma, \omega^\prime), X, \TC)} \bigg) \nonumber
\end{align}
Eqn.~\ref{eqn:alphaomega} is readily computed; conveniently there is no need to compute
the PG density, which can be challenging in some regimes. See Sec.~\ref{app:omega} for details.

We note that the proposal distribution $p(\omega^\prime | \gamma, \bhat(\gamma, \omega), \DD)$ can be thought of as an approximation to the posterior conditional
$p(\omega^\prime | \gamma, \DD) = \int \! d\beta \, p(\omega^\prime | \gamma, \beta, \DD) p(\beta | \gamma, \DD)$ that would be used in a Gibbs update. 
Since this latter density is intractable, we instead opt for this tractable option.
One might worry that the resulting acceptance probability
could be low, since $\omega$ is $N$-dimensional and $N$ can be large. 
However, $p(\omega^\prime | \gamma, \beta, \DD)$ only depends
on $\beta$ through $\psi_n = \beta_\gamma \cdot X_{n\gamma}$; the induced posterior over $\psi_n$ is typically somewhat narrow, since 
the $\psi_n$ are pinned down by the observed data, and consequently  $p(\omega^\prime | \gamma, \bhat, \DD)$ is a reasonably good approximation to the
exact posterior conditional. 
In practice we observe high mean acceptance probabilities $\sim50\% - 95\%$ for all the experiments in this work,\footnote{For the experiment in Sec.~\ref{app:runtime} $N$ varies between $100$ and $4000$ and $P$ varies between $134$ and $69092$ and the
average acceptance prob.~ranges between $49\%$ and $89\%$.}
even for $N \gg 10^2$. 
 
\paragraph{Weighting factor $\eta$}

We choose
$\eta(\gmj, \omega) = p(\gamma_j = 1 | \gmj, \omega) + \tfrac{\epsilon}{P}$.
For the full algorithm see Algorithm~\ref{pgalgo}.

\section{Further extensions}
\label{sec:extensions}

We briefly discuss how to accommodate inference over the prior inclusion probability $h$
in Eqn.~\ref{eqn:linmodeldefn} and Eqn.~\ref{eqn:pgmodeldefn}. It is natural to place
a Beta prior on $h$ \citep{steel2007effect}, since in the non-tempered setting this allows
for conjugate Gibbs updates of $h$. Unfortunately this is spoiled by the tempering in wTGS.
We adopt a simple workaround. For example in the case of linear regression, Eqn.~\ref{eqn:linmodeldefn},
we add an additional state $i=0$ to wTGS analogous to the $i=0$ state in PG-wTGS in Sec.~\ref{sec:pginf}.
By construction when $i=0$ the target distribution is not tempered, thus allowing for conjugate updates
of $h$.  See Algorithm \ref{inferhalgo} in the supplement for details. Subset wTGS and PG-wTGS are also readily combined;
see Algorithm~\ref{pglargealgo} in the supplement for details.

\section{Related work}
\label{sec:related}

Some of the earliest approaches to Bayesian variable selection (BVS) were introduced by \citet{george1993variable, george1997approaches}.
\citet{chipman2001practical} provide an in-depth discussion of BVS for linear regression and CART models.
\citet{zanella2019scalable} introduce Tempered Gibbs Sampling (TGS) and apply it to BVS for linear regresion.
\citet{griffin2021search} introduce an efficient adaptive MCMC method for BVS in linear regression. \citet{wan2021adaptive} extend this approach to logistic regression and accelerated failure time models. We include this approach (ASI) in our empirical validation in Sec.~\ref{sec:exp}.
\citet{dellaportas2002bayesian} and \citet{o2009review} review various methods for BVS. 
\citet{polson2013bayesian} introduce P\olya-Gamma augmentation and use it to develop efficient Gibbs samplers.

\section{Experiments}
\label{sec:exp}

\begin{figure}[ht]
    \centering
    \includegraphics[width=0.76\linewidth]{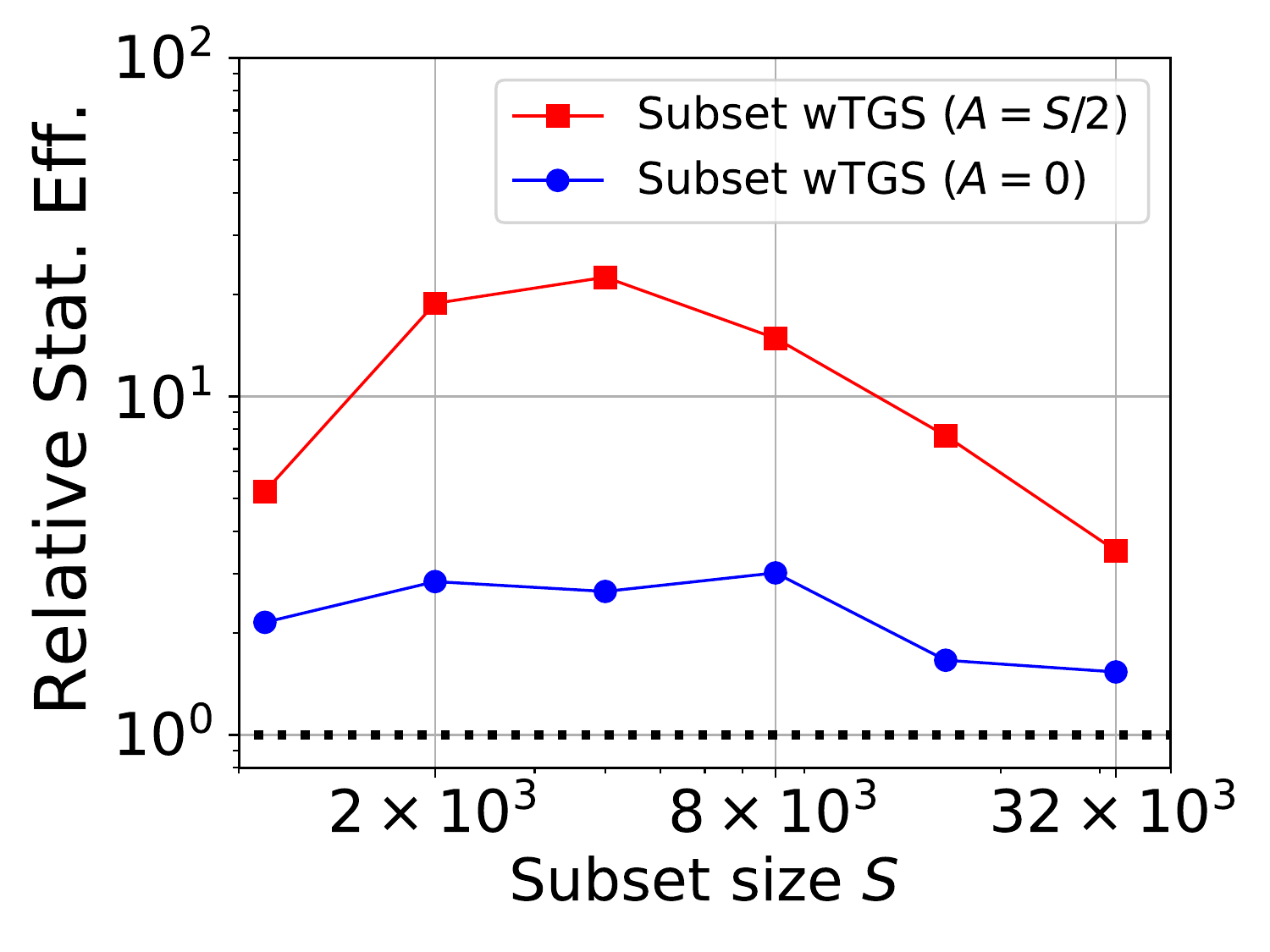}
    \includegraphics[width=0.76\linewidth]{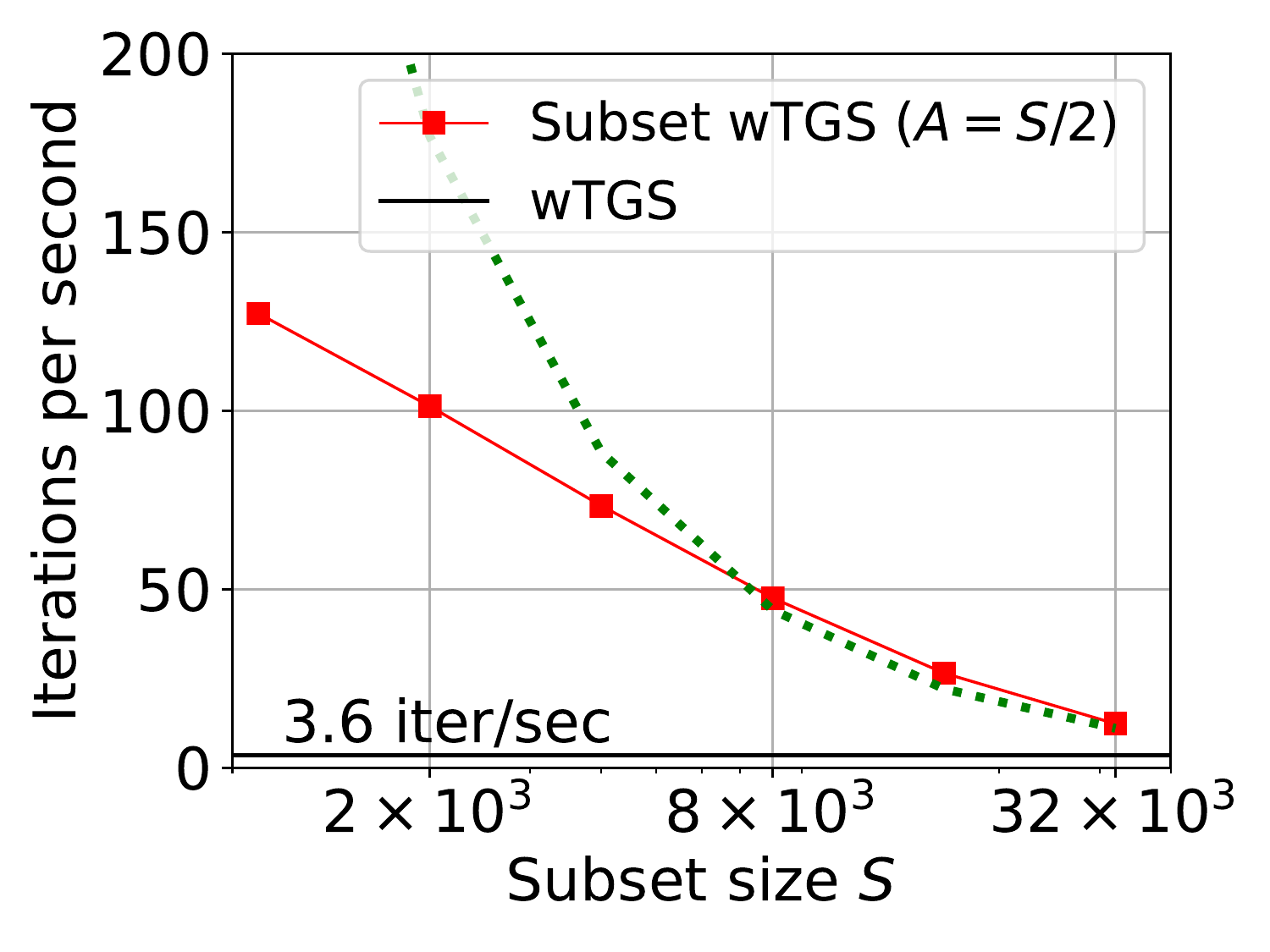}
    \caption{We report results for the experiment in Sec.~\ref{sec:largepexp} with $P=98385$.
    {\bf (Top)}  We depict the relative statistical efficiency of Subset wTGS with subset size $S$ compared to wTGS.
    {\bf (Bottom)} We depict the number of iterations per second (IPS) for Subset wTGS as a function of $S$. The green dotted line
    depicts the IPS that would be expected if the latter scaled like $S^{-1}$.
    }
\label{fig:largep}
\end{figure}

\begin{figure*}[ht]
    \centering
    \includegraphics[width=0.325\linewidth]{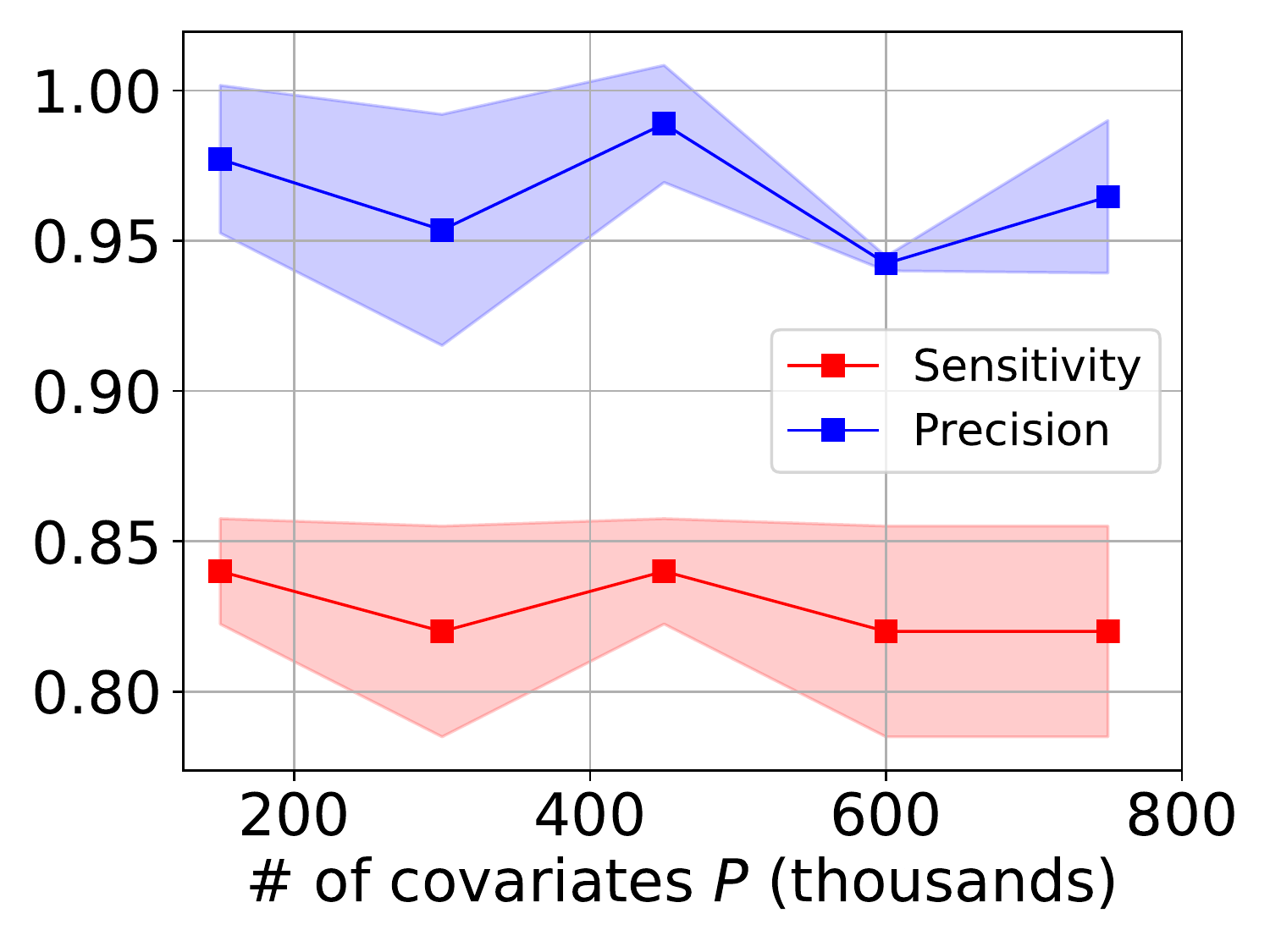}
    \includegraphics[width=0.325\linewidth]{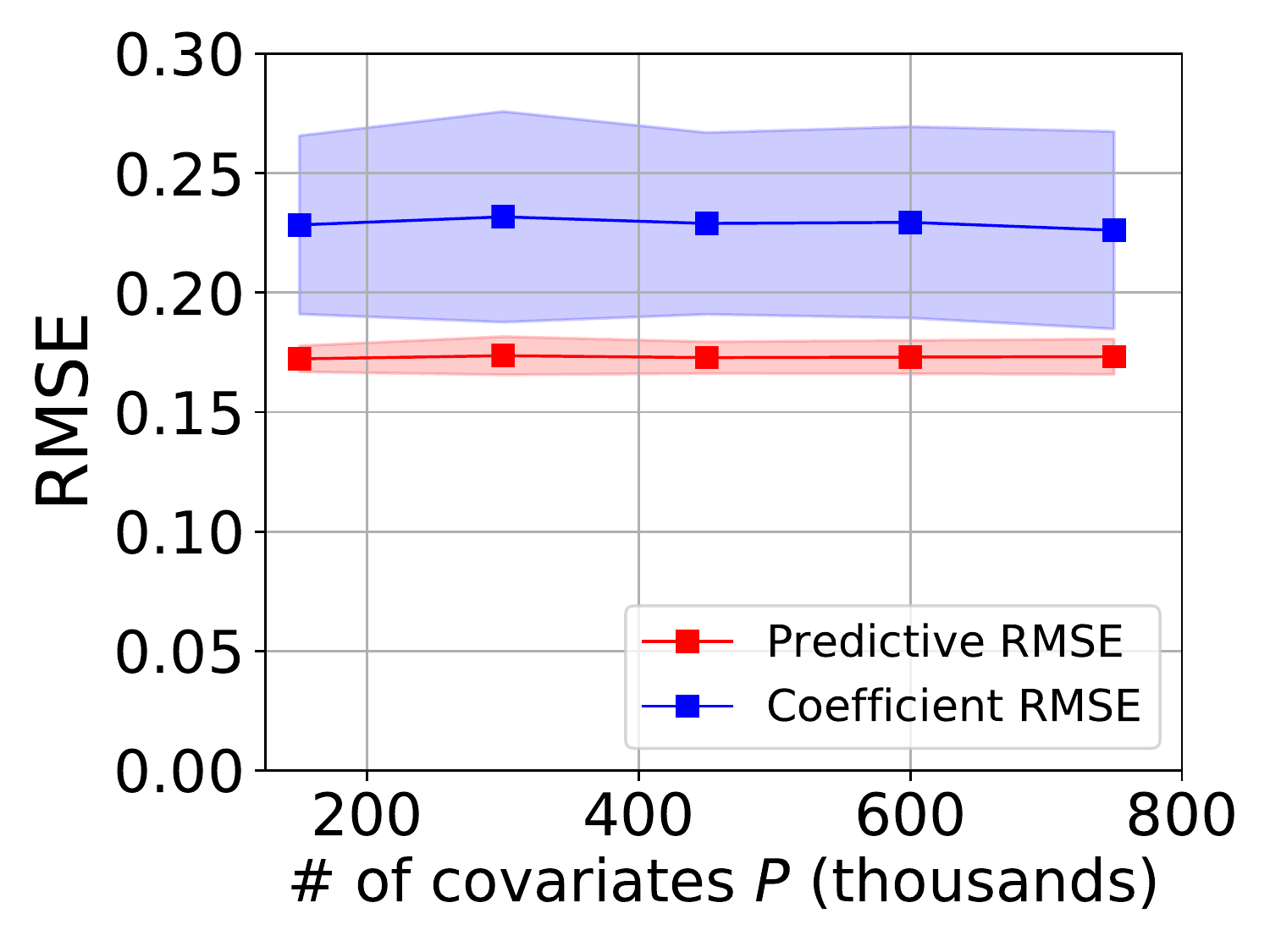}
    \includegraphics[width=0.325\linewidth]{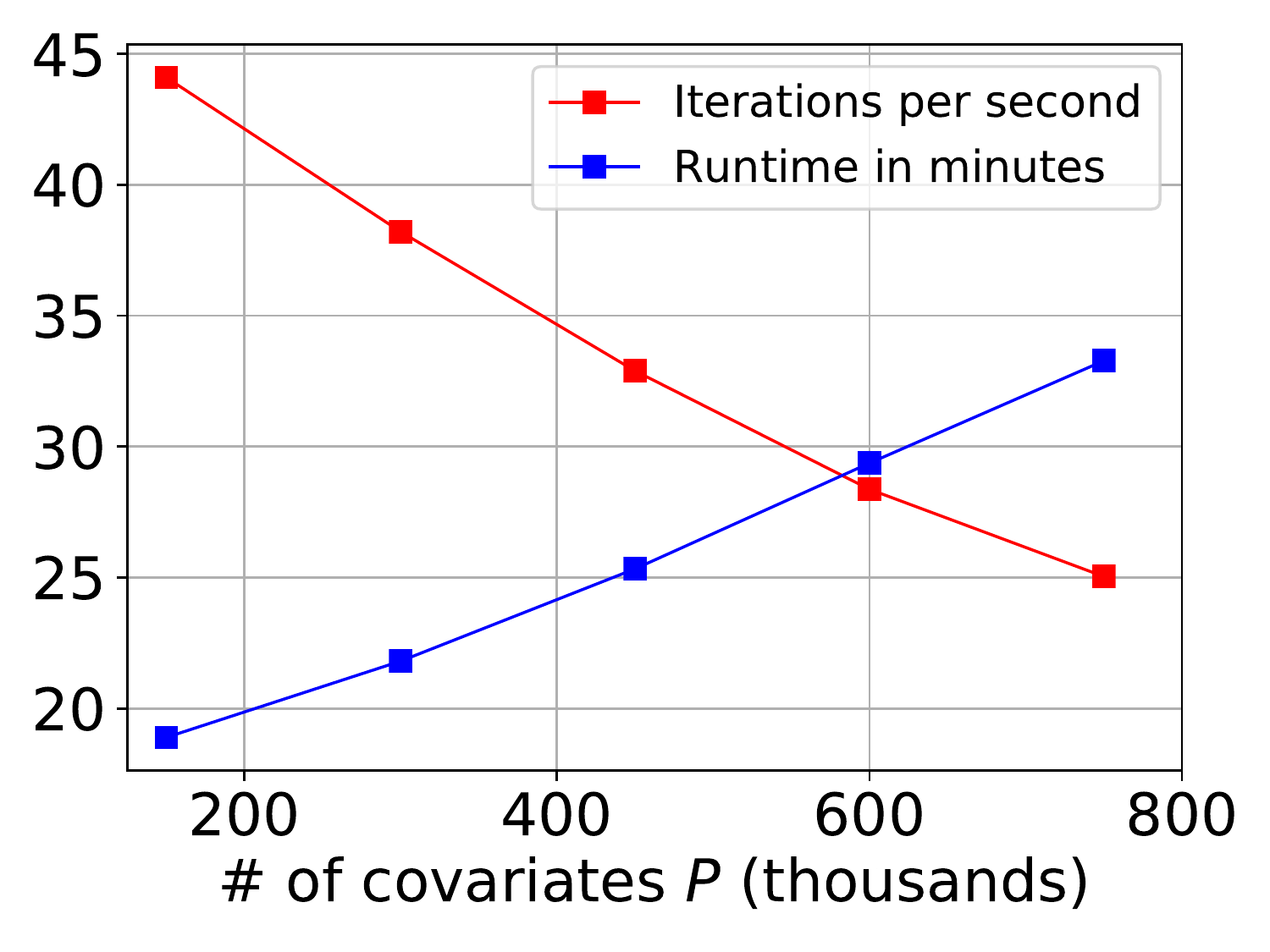}
    \caption{We report results for the Subset wTGS experiment in Sec.~\ref{sec:largepexp} with $P$ ranging
    from $1.5 \times 10^5$ to $7.5 \times 10^5$ and $S=16384$. We report 95\% confidence intervals across $5$ train/test splits.
    {\bf (Left)}  We depict the sensitivity and precision obtained if covariates with PIPs above a threshold of $0.5$ are declared hits. 
    {\bf (Middle)} We depict the root mean squared error (RMSE) for held out responses $Y^*$ and the inferred
    posterior mean over coefficients $\beta$.
    {\bf (Right)} We depict runtimes obtained on a NVIDIA Tesla T4 GPU for collecting $50$k MCMC samples. 
    }
\label{fig:big}
\end{figure*}

We validate the performance of Algorithms \ref{algo}, \ref{pgalgo}, \& \ref{inferhalgo} on synthetic and real world data.
We implement all algorithms using PyTorch \citep{paszkepytorch} and
the \texttt{polyagamma} package for sampling from PG distributions \citep{bleki_2021}.
We provide a unit-tested, easy-to-use, and open source implementation of our methods at \texttt{https://github.com/BasisResearch/millipede}.
See Sec.~\ref{app:addexp}-\ref{app:exp} for additional experimental details and experiments (e.g.~negative
binomial results in Sec.~\ref{sec:hospital}-\ref{sec:health}).

\subsection{Subset wTGS performance for large $P$}
\label{sec:largepexp}

We conduct two semi-synthetic experiments using maize genomic data from \citet{romay2013comprehensive} that have also
been analyzed by \citet{zeng2017non,pmlr-v162-biswas22a}. This dataset serves as a good benchmark for our method,
since it has large $P$ ($P=98385$), moderately large $N$ ($N=2267$), 
and complex correlation structure in the covariates $X$. We use synthetic
responses $Y$ so we have access to ground truth.\footnote{Note that we do not include comparisons to ASI \citep{griffin2021search}, since we were unable to obtain results using ASI that were remotely competitive with wTGS.}

In the first experiment we examine statistical efficiency and runtime, see Fig.~\ref{fig:largep}.
We find that Subset wTGS exhibits large speed-ups over wTGS and that these speed-ups translate to improved statistical efficiency.
Indeed Subset wTGS with anchor set size $A=\tfrac{1}{2}S$ exhibits a relative statistical efficiency $\sim\!20$ larger than wTGS. 
Subset wTGS with $A=0$ exhibits somewhat marginal improvements above wTGS, highlighting the importance of the anchor
set $\AAA$ in Algorithm~\ref{algo}.

Next we demonstrate the feasibility of scaling Subset wTGS to $P \sim 10^6$, see Fig.~\ref{fig:big} for results. 
To extend the maize data to $P>98385$ we append random covariates drawn from a unit Normal distribution.
Thanks to the $O(S)$ iteration cost of Subset wTGS, GPU memory is the main bottleneck to accommodating large(r) datasets.
Indeed the time needed to obtain $50$k MCMC samples for $P=7.5 \times 10^5$ is $\sim 35$ minutes on a Tesla T4 GPU.
By contrast wTGS does not scale to this regime unless conditional PIPs are computed sequentially 
in batches.\footnote{We estimate a $\sim24$ hour runtime.}
We find high precision and sensitivity in identifying causal covariates
across the entire range of $P$ considered, highlighting the value of scalable Bayesian variable selection algorithms.

\subsection{PG-wTGS and correlated covariates}
\label{sec:corr}

\begin{figure*}[ht]
    \includegraphics[width=0.245\linewidth]{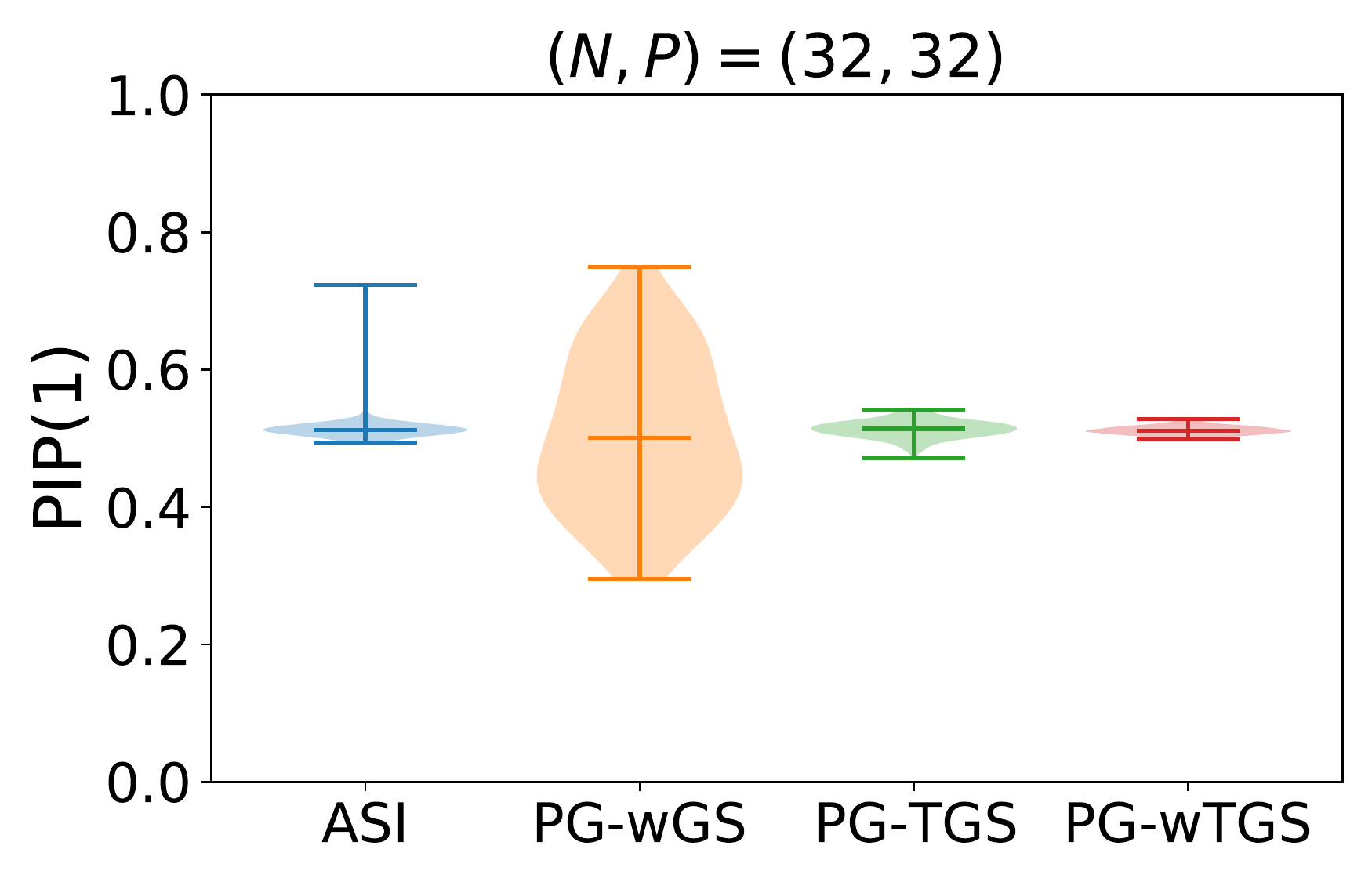}
    \includegraphics[width=0.245\linewidth]{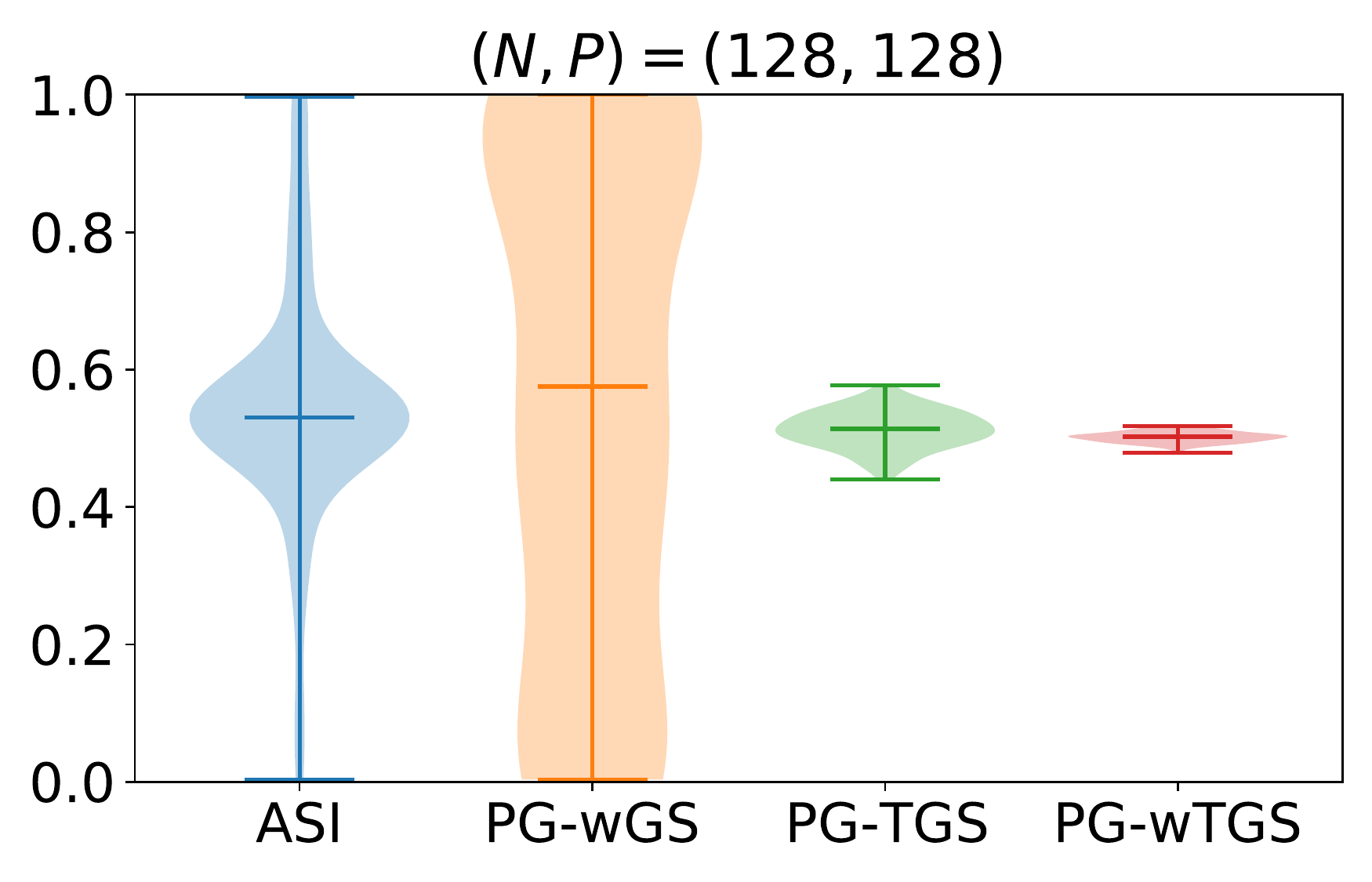}
    \includegraphics[width=0.245\linewidth]{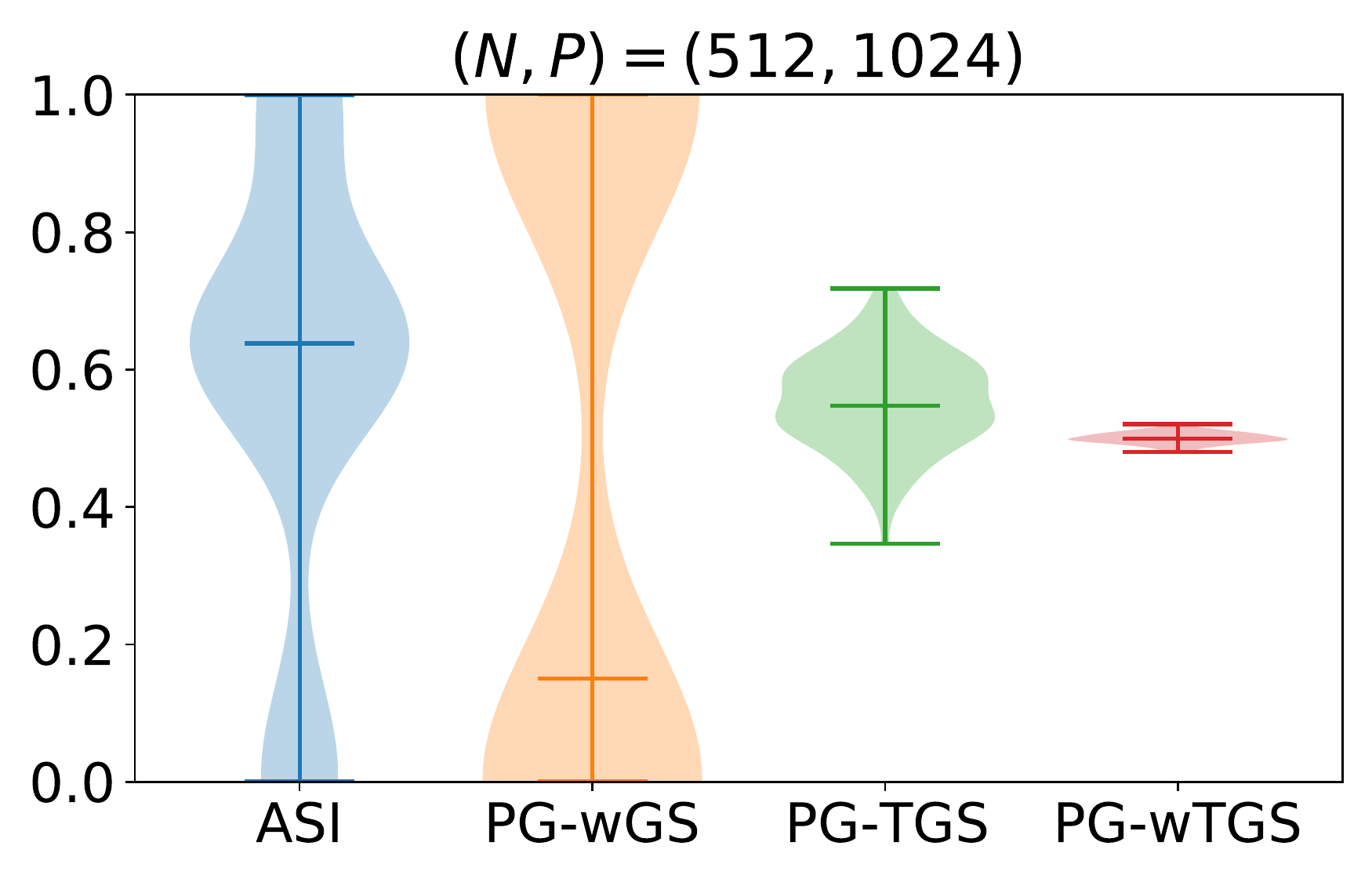}
    \includegraphics[width=0.245\linewidth]{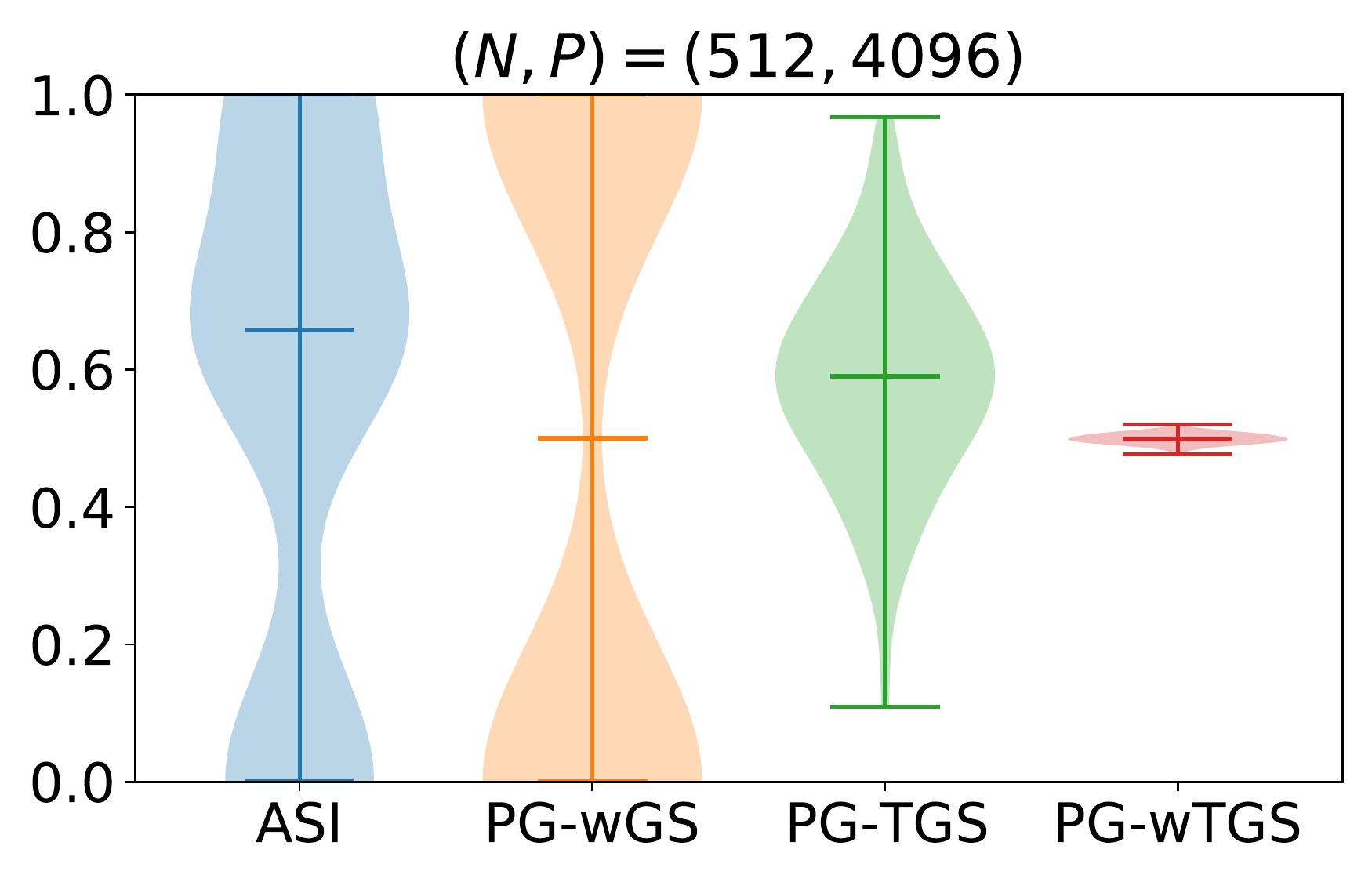}
    \caption{We depict violin plots for PIP (posterior inclusion probability) estimates for the first covariate obtained with $10^5$ MCMC samples
    for four different methods on four datasets with varying numbers of data points $N$
    and covariates $P$. Horizontal bars denote the minimum, median, and maximum PIP estimates
    obtained from 100 independent MCMC runs. See Sec.~\ref{sec:corr} for details and Fig.~\ref{fig:tracecorrapp} in the
    supplement for corresponding trace plots.
    }
\label{fig:corrviolin}
\end{figure*}

We consider simulated Binomial regression datasets in which two covariates ($i\in\{1,2\}$) are highly correlated
and each alone can explain the response. 
This can be a challenging regime, since it is easy to get stuck in one mode and fail to explore the other mode.
We consider four datasets with $32 \le N \le 512$, $32\le P \le4096$,
and $\TC_n=10$ for all data points.
See Fig.~\ref{fig:corrviolin} for the results. 

To better understand the performance of PG-wTGS, we consider two variants, PG-TGS and PG-wGS, which
do without weighting by $\eta(\gmj, \omega)$ and tempering, respectively. 
In addition we compare to ASI \citep{wan2021adaptive}, an adaptive MCMC scheme that also uses P\olya-Gamma augmentation.

We see that PG-wGS does poorly on all datasets, including the smallest one with $P=32$ covariates.
PG-TGS does well for $P=32$ and $P=128$ but exhibits large variance for $P\ge 1024$. 
By contrast PG-wTGS yields low-variance PIP estimates in all cases, demonstrating the benefits of $\eta$-weighting and tempering.
ASI estimates exhibit low variance for $P=32$ (apart from a single outlier) but are high variance for larger $P$. 
This outcome is easy to understand. Since ASI adapts its proposal distribution during warmup using a running estimate of 
each PIP, it is vulnerable to a rich-get-richer phenomenon in which covariates with
large initial PIP estimates tend to crowd out covariates with which they are highly correlated.
In the present case the result is that the ASI PIP estimates for the first two covariates 
are strongly anti-correlated. That this anti-correlation is ultimately due to suboptimal adaptation
is easily verified. For example for $P=1024$ ($P=4096$) the Pearson correlation coefficient
between the difference of final PIP estimates, i.e.~PIP(1) - PIP(2), and the difference of the
corresponding initial PIP estimates that define the proposal distribution is $0.904$ ($0.998$), respectively.

\subsection{PG-wTGS and cancer data}
\label{sec:cancer}

\begin{figure*}[ht]
    \centering
    \includegraphics[width=0.495\linewidth]{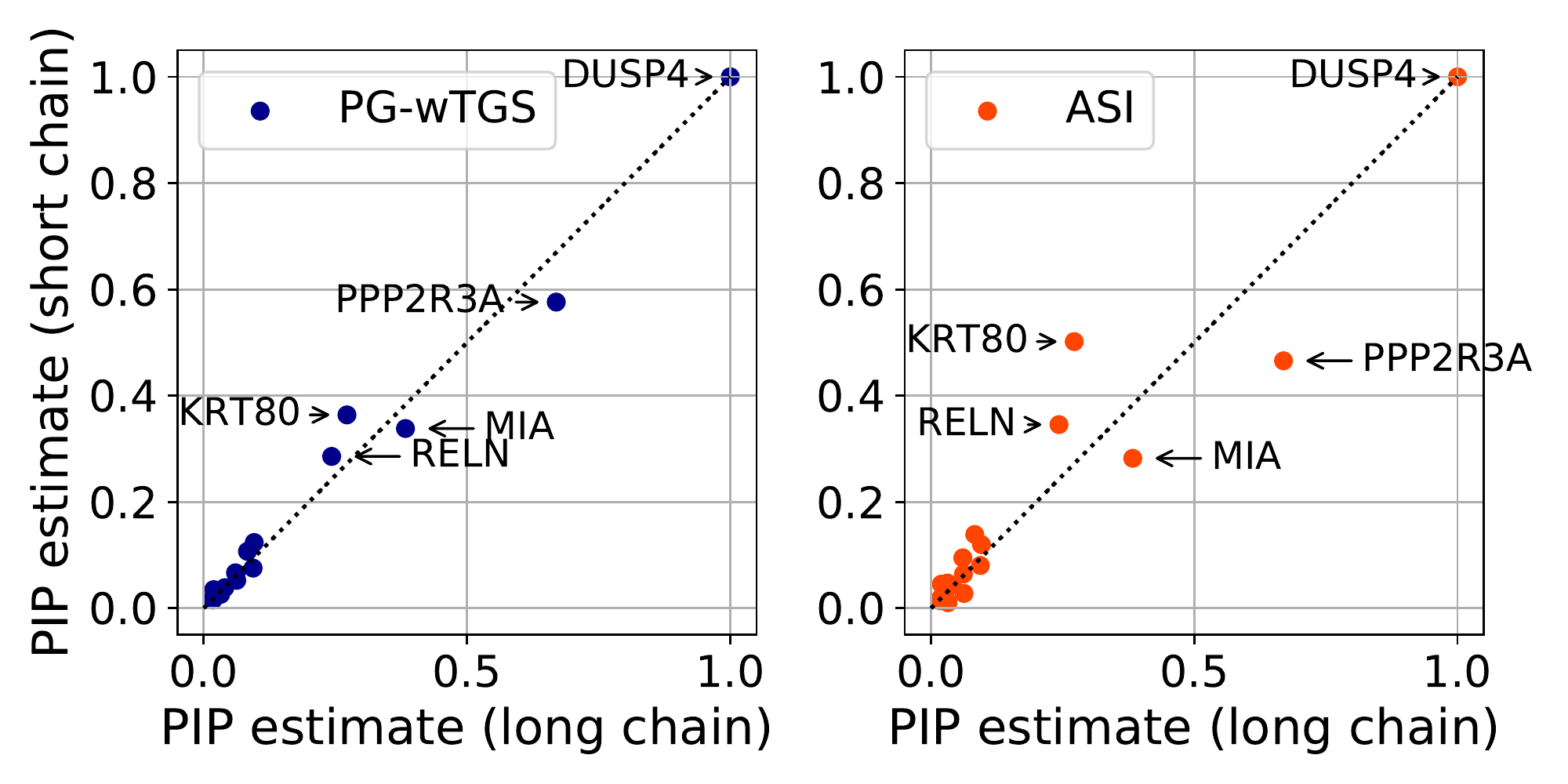}
    \includegraphics[width=0.495\linewidth]{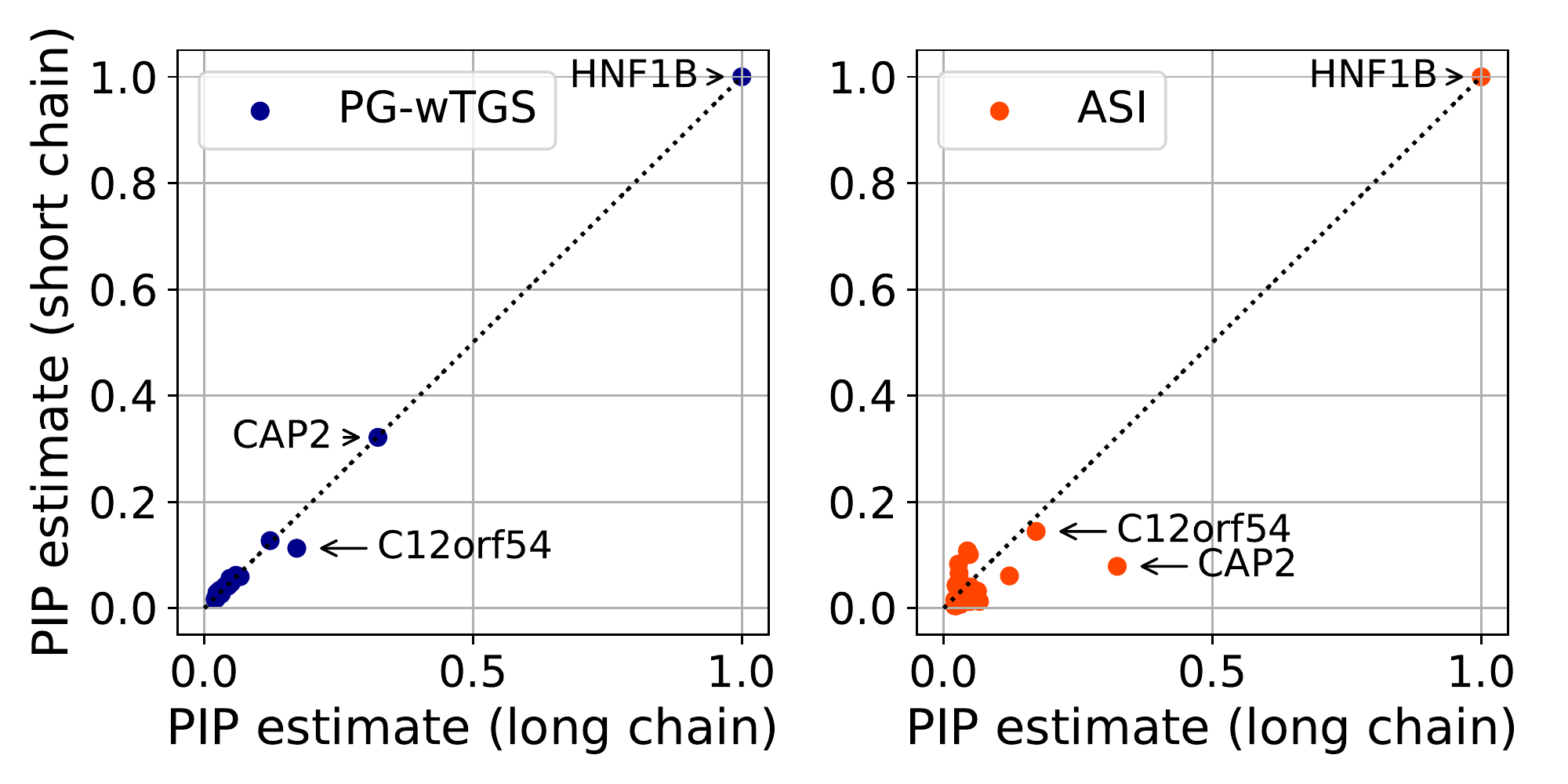}
    \caption{We compare PIP estimates obtained from PG-wTGS and ASI chains with $2.5 \times 10^5$ samples
    to a long PG-wTGS chain with $5 \times 10^6$ samples.
    For each method we depict the top 20 PIPs from the long chain paired with estimates from the short chain. 
    The PG-wTGS estimates are significantly more accurate than is the case for ASI.
    See Sec.~\ref{sec:cancer} for details and Fig.~\ref{fig:cancer}-\ref{fig:cancerbigfull} in the supplement for additional comparisons.
    }
\label{fig:cancerbigpartial}
\end{figure*}

We consider data collected from 900+ cancer cell lines in the Cancer Dependency Map project \citep{meyers2017computational,behan2019prioritization,pacini2021integrated}. 
Each cell line has been subjected to a loss-of-function genetic screen that uses CRISPR-Cas9
to identify genes essential for cancer proliferation and survival. 
Genes identified by such screens are thought to be promising candidates for genetic vulnerabilities
that can be used to guide the development of novel therapeutics.

In more detail, we consider a subset of the data that includes $N=907$ cell lines
and $P=17273$ covariates, with each covariate encoding the RNA expression level for a given gene.
We consider two gene knockouts: DUSP4 and HNF1B.\footnote{This choice serves as a sanity check, since      
 for both knockouts the RNA expression level of the corresponding gene is known to be
 highly predictive of cell viability.}
 For each knockout the dataset contains a real-valued response that encodes the effect of knocking out 
that particular gene. 
We binarize this response variable by using the 20\% quantile as a cutoff.

What makes this dataset particularly challenging is that the covariates 
are strongly correlated.
For example, DUSP4 RNA expression exhibits a correlation
greater than $0.40$ ($0.30)$ with $19$ ($203$) other covariates, respectively.
Similarly the HNF1B covariate has a correlation greater than $0.70$ ($0.50)$ with $2$ ($33$) other covariates, respectively.
In Fig.~\ref{fig:cancerbigpartial} we compare PIP estimates obtained with PG-wTGS and ASI,
in both cases comparing to estimates obtained with long PG-wTGS runs.
The much lower variance of PG-wTGS estimates as compared to ASI estimates is apparent.
Indeed the mean absolute PIP error in the top hits is about $\sim5x$ larger for ASI
(see Table~\ref{table:cancer} in the supplemental materials for details and a list of all the top hits).

\subsection{Inferring the inclusion probability $h$}
\label{sec:infh}

\begin{figure}[ht]
\centering
    \includegraphics[width=0.78\linewidth]{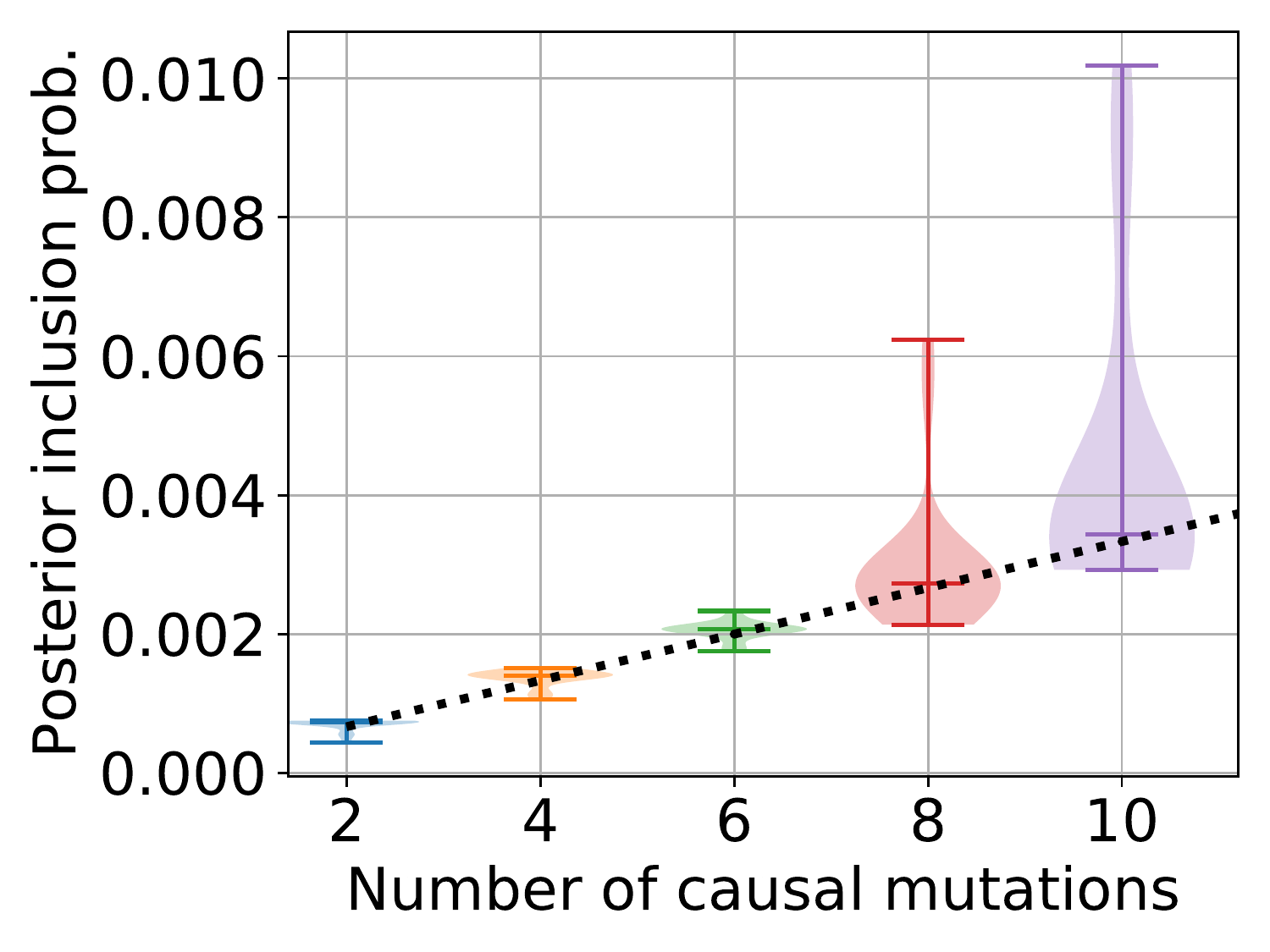}
\caption{We depict posterior estimates of the inclusion probability $h$ for the experiment
    in Sec.~\ref{sec:infh}. Middle horizontal bars depict median posterior estimates across $20$ simulations.
    The black dotted line indicates the proportion of mutations that are causal. }
\label{fig:infh}
\end{figure}

We consider an application of variable selection to viral transmission \citep{jankowiak2022inferring}.
Each covariate encodes the presence of a particular mutation in a virus like SARS-CoV-2,
only a small number of which are assumed to affect viral fitness. Modeling viral
transmission as a diffusion process results in a tractable Gaussian likelihood.
We consider a virus with $P=3000$ mutations and simulate a pandemic occurring in $30$ geographic regions. 
We place a Beta prior on $h$ and vary the number of causal mutations (i.e.~those with non-zero effects)
and investigate whether the inferred inclusion probability $h$ reflects the true number of
causal mutations. As we would expect, see Fig.~\ref{fig:infh}, this is indeed the case.
See Sec.~\ref{app:exp} for additional details.

\section{Discussion}
\label{sec:disc}

We have shown that Bayesian variable selection can be efficiently scaled
to $P\sim10^6$ and can accommodate count-based likelihoods. Given the extremely large $N$ and $P$ 
that can be found in some genomics datasets, an interesting direction for future work would be
to devise algorithms that can support $N$ and $P$ in the tens of millions. Doing so
would likely require new algorithmic ideas (e.g.~deterministic screening of covariates)
as well as linear algebra speed-ups (e.g.~incrementally caching computations as $\gamma$ space is explored).

\section*{Acknowledgments}
We thank Jim Griffin for clarifying some of the details of the methodology described in \citet{wan2021adaptive} and Zolisa Bleki for help with 
the \texttt{polyagamma} package. 
We also thank James McFarland, Joshua Dempster, and Ashir Borah for help with DepMap data. 
We thank Niloy Biswas for sharing the genomic data we used in our experiments in Sec.~\ref{sec:largepexp}.
Finally we kindly thank Giacomo Zanella for interesting discussion about Bayesian variable selection.

\bibliographystyle{plainnat}
\bibliography{ref}

\clearpage

\onecolumn

\appendix
\section{Appendix}

This appendix is organized as follows.
In Sec.~\ref{app:si} we discuss societal impact. 
In Sec.~\ref{app:inferh} we discuss how we infer the inclusion probability $h$. 
In Sec.~\ref{app:subsetpgwtgs} we discuss how we combine Subset wTGS and PG-wTGS. 
In Sec.~\ref{app:mll} we discuss conditional marginal log likelihood computations.
In Sec.~\ref{app:cc} we discuss computational complexity.
In Sec.~\ref{app:mg} we motivate the tempering scheme that underlies wTGS. 
In Sec.~\ref{app:wtgsdisc} we discuss the nature of the local moves made by wTGS. 
In Sec.~\ref{app:iw} we briefly discuss the role played by importance weighting in our MCMC methods. 
In Sec.~\ref{app:proof} we provide a proof of Proposition~\ref{prop:large}. 
In Sec.~\ref{app:rao} we discuss Rao-Blackwellized PIP estimators. 
In Sec.~\ref{app:omega} we discuss $\omega$-updates.
In Sec.~\ref{app:xi} we discuss $\xi$-adapation. 
In Sec.~\ref{app:anchor} we discuss how we adapt the anchor set $\AAA$. 
In Sec.~\ref{app:nb} we discuss the modifications of PG-wTGS that are needed to accommodate negative binomial likelihoods.
In Sec.~\ref{app:fig} we include additional figures and tables accompanying the experimental results in Sec.~\ref{sec:exp}.
In Sec.~\ref{app:addexp} we report additional experimental results. 
In Sec.~\ref{app:exp} we discuss experimental details.

\subsection{Societal impact}
\label{app:si}

We do not anticipate any negative societal impact from the methods described in this work, although we
note that they inherit the risks that are inherent to any algorithm that can be used for hypothesis testing and/or prediction. 
In more detail there is the possibility of the following risks.
First, predictive algorithms can be deployed in ways that disadvantage vulnerable groups in a population.
Even if these effects are unintended, they can still arise if deployed algorithms are poorly vetted with respect to their fairness implications. The same applies to any hypotheses investigated with a variable selection algorithm, especially if variables are correlated with
indicators that encode the identity of vulnerable groups.
Second, algorithms that offer uncertainty quantification may be misused by users who place unwarranted confidence in the
uncertainties produced by the algorithm. This can arise, for example, in the presence of undetected covariate shift.

\begin{algorithm*}[t!]
\DontPrintSemicolon 
\KwIn{Dataset $\DD = \{X, Y\}$ with $P$ covariates;
     prior inclusion probability $h$;
     prior precision $\tau$;
     total number of MCMC iterations $T$; 
     number of burn-in iterations $\Tburn$
     }
    \KwOut{Approximate weighted posterior samples $\{\rho^{(t)}, \gamma^{(t)} \}^T_{t=\Tburn +1}$}
    Let $\gamma^{(0)} = (0, ..., 0)$. \\
\For{$t =1, ..., T$} {
    Sample $i^{(t)} \sim f(\cdot | \gamma^{(t-1)})$ using Eqn.~\ref{eqn:iupdatelin} \\
    Let $\gamma^{(t)} = {\rm flip}(\gamma^{(t-1)} | i^{(t)})$ where ${\rm flip}( \gamma | i)$ 
        flips the $i^{\rm th}$ coordinate of $\gamma$: $\gamma_i \to 1 - \gamma_i$. \\
    Compute the unnormalized weight $\tilde{\rho}^{(t)} = \phi(\gamma^{(t)})^{-1}$ using Eqn.~\ref{eqn:phidefnlin}. \\
}
    Compute the normalized weights $\rho^{(t)} = \frac{ \tilde{\rho}^{(t)}  } { \sum_{s > \Tburn} \tilde{\rho}^{(s)} }$ for 
    $t=\Tburn + 1,...,T$. \\
    \Return{$\{\rho^{(t)}, \gamma^{(t)} \}_{t=\Tburn + 1}^T$}
    \caption{We outline the main steps in wTGS \citep{zanella2019scalable}. See Sec.~\ref{sec:wtgs} for discussion.
    Note that we use superscripts to indicate MCMC iterations.}
\label{linalgo}
\end{algorithm*}

\subsection{Inferring the inclusion probability $h$}
\label{app:inferh}

Consider the following (unnormalized) target distribution
\begin{align}
\label{eqn:fulltargeth}
f(\gamma, i, h)
    &\equiv p(\gamma | h, \DD) p(h | \alpha_h, \beta_h) \left \{ \delta_{i0} \xi +
    \tfrac{1}{P} \Sigma_{j=1}^P \frac{ \delta_{ij} \eta(\gmj, h) U(\gamma_j)}{ p(\gamma_j | \gmj, h, \DD)} \right\}
\end{align}
where we have introduced a hyperparameter $\xi > 0$ and $\alpha_h>0$ and $\beta_h>0$ parameterize the prior over $h$.
We define the inverse importance weight
\begin{align}
 \phi(\gamma, h) \equiv \xi +  \frac{1}{P} \sum_{i=1}^P  \frac{\tfrac{1}{2} \eta(\gmi, h)}{ p(\gamma_i | \gmi, h, \DD)}
\label{eqn:phidefnh}
\end{align}
We can do $i$ updates using the Gibbs distribution
\begin{align}
\label{eqn:iupdateh}
f(i | \gamma, h)  \propto \delta_{i0} \xi + \frac{1}{P} \sum_{j=1}^P \delta_{ij} \frac{\tfrac{1}{2}  \eta(\gmj, h)}{ p(\gamma_j | \gmj, h, \DD)}
\end{align}
When $i=0$ we can do $h$ updates using the Gibbs distribution
\begin{align}
\label{eqn:updateh}
    f(h | \gamma, i=0) = {\rm Beta}(\alpha=\alpha_h + |\gamma|, \beta = \beta_h + P - |\gamma|) 
\end{align}
where $|\gamma|$ is the number of covariates included in the model in the current iteration.
See Algorithm~\ref{inferhalgo} for a complete description of wTGS with inference over $h$.

The above discussion assumes the linear regression case, Eqn.~\ref{eqn:linmodeldefn}. To accommodate
count-based likelihoods we simply use the untempered $i=0$ state to make $h$ updates \emph{and} $\omega$ updates in succession (and in random order).

\begin{algorithm*}[t!]
\DontPrintSemicolon 
\KwIn{Dataset $\DD = \{X, Y\}$ with $P$ covariates;
     prior precision $\tau$;
      hyperparameters $(\alpha_h, \beta_h)$;
     total number of MCMC iterations $T$;
     number of burn-in iterations $\Tburn$;
     hyperparameter $\xi > 0$ (optional)
     }
    \KwOut{Approximate weighted posterior samples $\{\rho^{(t)}, \gamma^{(t)}, h^{(t)} \}^T_{t=\Tburn +1}$}
    Let $\gamma^{(0)} = (0, ..., 0)$ and $h^{(0)} \sim {\rm Beta}(\alpha_h, \beta_h)$. \\
\For{$t =1, ..., T$} {
    Sample $i^{(t)} \sim f(\cdot | \gamma^{(t-1)}, h^{(t-1)})$ using Eqn.~\ref{eqn:iupdateh}. \\
    If $i^{(t)} > 0$ let $\gamma^{(t)} = {\rm flip}(\gamma^{(t-1)} | i^{(t)})$ and $h^{(t)} = h^{(t-1)}$. \\
    Otherwise if $i^{(t)} = 0$ let $\gamma^{(t)} = \gamma^{(t-1)}$ and 
         $h^{(t)} \sim f(\cdot | \gamma^{(t)}, i=0)$ using Eqn.~\ref{eqn:updateh}. \\
    Compute the unnormalized weight $\tilde{\rho}^{(t)} = \phi(\gamma^{(t)}, h^{(t)})^{-1}$ using Eqn.~\ref{eqn:phidefnh}. \\
    If $\xi$ is not provided and $t \le \Tburn$ adapt $\xi$ using the scheme described in Sec.~\ref{app:xi}.\\
}
    Compute the normalized weights $\rho^{(t)} = \frac{ \tilde{\rho}^{(t)}  } { \sum_{s > \Tburn} \tilde{\rho}^{(s)} }$ for
    $t=\Tburn + 1,...,T$. \\
    \Return{$\{\rho^{(t)}, \gamma^{(t)}, h^{(t)} \}_{t=\Tburn + 1}^T$}
    \caption{We extend wTGS (see Algorithm~\ref{linalgo}) to allow inference over the inclusion probability $h$,
    with $h$ governed by a ${\rm Beta}(\alpha_h, \beta_h)$ prior. See Sec.~\ref{app:inferh} for additional discussion.
    Note that we use superscripts to indicate MCMC iterations.
    }
\label{inferhalgo}
\end{algorithm*}

\subsection{Subset PG-wTGS}
\label{app:subsetpgwtgs}

We show how to combine the algorithmic ideas from Sec.~\ref{sec:largep} and Sec.~\ref{sec:pginf},
i.e.~how to scale Bayesian variable selection with a count-based likelihood to large $P$.
The target distribution is
\begin{align}
\label{eqn:fulltargetSSpg}
f(\gamma, i, \omega, \SSS)
    &= p(\gamma, \omega | \DD)  \left \{ \delta_{i0} \xi
                   + \tfrac{1}{P} \Sigma_{j=1}^P \frac{\delta_{ij} \eta(\gmj, \omega) U(\gamma_j)}{ p(\gamma_j | \gmj, \omega, \DD)} \right\}
                   U(\SSS | i, \AAA)
\end{align}
where we assume $\SSS$ ranges over size $S$ subsets of $\{0, ..., P\}$ and that $0 \in \AAA$.
We define
\begin{align}
\label{eqn:iupdateSSpg}
    f(i | \gamma, \omega, \SSS)  \propto \left( \delta_{i0} \xi + \frac{1}{P} \sum_{j \in \SSS, j >0} \delta_{ij} \frac{\tfrac{1}{2}  \eta(\gmj, \omega)}{ p(\gamma_j | \gmj, \omega, \DD)} \right) U(\SSS | i, \AAA)
\end{align} 
and
\begin{align}
    \phi(\gamma, \omega, \SSS) \equiv \left( \xi \times U(\SSS | 0, \AAA) + \frac{1}{P} \sum_{i\in \SSS, i >0}  \frac{\tfrac{1}{2} \eta(\gmi, \omega)}{ p(\gamma_i | \gmi, \omega, \DD)} U(\SSS | i, \AAA) \right)
\label{eqn:phidefnSSpg}
\end{align}
The algorithm then follows the same logic as in Subset wTGS and PG-wTGS; see
Algorithm~\ref{pglargealgo} for a complete description.

We note an important implementation detail that is common to Algorithm~\ref{algo} and Algorithm~\ref{pglargealgo}.
Here we deal with the case of Algorithm~\ref{algo} for concreteness. 
Besides the value of zero, the probability $U(\SSS | i, \AAA)$ takes on two possible values:
\begin{align}
\label{eqn:ucomb}
    &U(\SSS | i, \AAA) = \frac{(S-A)!(P-S)!}{(P-A)!}  \qquad &&{\rm if} \qquad i \in \AAA \\ \nonumber
    &U(\SSS | i, \AAA) = \frac{(S-A-1)!(P-S)!}{(P-A-1)!}  \qquad &&{\rm if} \qquad i \notin \AAA  
\end{align}
Since, however, we always use normalized weights $\{ \rho^{(t)} \}$ when computing approximate posterior expectations,
any overall constant factor in $U(\SSS | i, \AAA)$ is irrelevant. Consequently we only need to keep track of the 
ratio of the two values in Eqn.~\ref{eqn:ucomb}, namely $\frac{S-A}{P-A}$. In particular there is no need to compute
factorials.

\begin{algorithm*}[t!]
\DontPrintSemicolon 
\KwIn{Dataset $\DD = \{X, Y, \TC\}$ with $P$ covariates;
     prior inclusion probability $h$;
     prior precision $\tau$;
     subset size $S$; anchor set size $A$;
     total number of MCMC iterations $T$; 
     number of burn-in iterations $\Tburn$;
     hyperparameter $\xi > 0$ (optional)
     }
    \KwOut{Approximate weighted posterior samples $\{\rho^{(t)}, \gamma^{(t)}, \omega^{(t)} \}^T_{t=\Tburn +1}$}
    Let $\gamma^{(0)} = (0, ..., 0)$ and $\omega^{(0)} \sim {\rm PG}(\TC, 0)$.  \\
    Choose $\AAA$ to include $\{0\}$ as well as the $A-1$ covariate indices exibiting the largest correlations with the response $Y$. \\
    Choose $i^{(0)}$ randomly from $\{1, ..., P\}$ and $\SSS^{(0)} \sim  U(\cdot | i^{(0)}, \AAA)$.\\
\For{$t =1, ..., T$} {
    Sample $i^{(t)} \sim f(\cdot | \gamma^{(t-1)}, \omega^{(t-1)}, \SSS^{(t-1)})$ using Eqn.~\ref{eqn:iupdateSSpg}. \\
    If $i^{(t)} >0$ let $\omega^{(t)}= \omega^{(t-1)}$ and $\gamma^{(t)} = {\rm flip}(\gamma^{(t-1)} | i^{(t)})$.\\
    Otherwise if $i^{(t)} =0$ let $\gamma^{(t)}= \gamma^{(t-1)}$ and 
               sample $\omega^{\prime (t)} \sim p(\cdot | \gamma^{(t-1)}, \bhat(\gamma^{(t-1)}, \omega^{(t-1)}), \DD)$.
    Set $\omega^{(t)} = \omega^{\prime(t)}$ with probability $\alpha(\omega^{(t)}  \! \to \! \omega^{\prime(t)} | \gamma^{(t)})$ given in Eqn.~\ref{eqn:alphaomega}.
    Otherwise set $\omega^{(t)} = \omega^{(t-1)}$. \\
    Compute the unnormalized weight $\tilde{\rho}^{(t)} = \phi(\gamma^{(t)}, \omega^{(t)}, \SSS^{(t)})^{-1}$ using Eqn.~\ref{eqn:phidefnSSpg}. \\
    If $\xi$ is not provided and $t \le \Tburn$ adapt $\xi$ using the scheme described in Sec.~\ref{app:xi}.
}
    Compute the normalized weights $\rho^{(t)} = \frac{ \tilde{\rho}^{(t)}  } { \sum_{s > \Tburn} \tilde{\rho}^{(s)} }$ for 
    $t=\Tburn + 1,...,T$. \\
    \Return{$\{\rho^{(t)}, \gamma^{(t)}, \omega^{(t)} \}_{t=\Tburn + 1}^T$}
    \caption{We outline the main steps in Subset PG-wTGS, which combines Algorithm \ref{algo} \& \ref{pgalgo}.
    See Sec~\ref{app:subsetpgwtgs} for additional discussion.
    Note that we use superscripts to indicate MCMC iterations.
    }
\label{pglargealgo}
\end{algorithm*}

\subsection{Efficient linear algebra for the (conditional) marginal log likelihood}
\label{app:mll}

Here we focus on computing the marginal log likelihood in the case of count-based likelihoods as required for Algorithm~\ref{pgalgo}.
The linear algebra required for
the linear regression case is essentially identical. See \citet{chipman2001practical,zanella2019scalable} for discussion
of the linear case.

\noindent The conditional marginal log likelihood $\log p(Y | X, \TC, \gamma, \omega)$ can be computed in closed form where, up to irrelevant constants, we have
\begin{align}
    \label{eqn:mll}
    \log p(Y | X, \TC, \gamma, \omega) &= \tfrac{1}{2} \ZZ_\gone \TT (X_\gone \TT \Omega X_\gone  + \tau \id_\gone)\inv \ZZ_\gone \\
    &-\tfrac{1}{2} \log \det(X_\gone \TT \Omega X_\gone   + \tau \id_\gone)    -\tfrac{1}{2} \log \det(\tau \inv \id_\gone)  \nonumber
\end{align}
where $\ZZ \in \RR^{P+1}$ with $\ZZ_j = \sum_{n=1}^N \kappa_n X_{n,j}$ 
for $j=1,...,P$ and the final component $\ZZ_{P+1} = \sum_{n=1}^N \kappa_n $ corresponds to the bias. Here and elsewhere $X$ is augmented with a column of all ones where necessary
and $\kappa_n \equiv Y_n - \tfrac{1}{2} \TC_n$, 
$\Omega$ is the $N \times N$ diagonal matrix formed from $\omega$,  and $\gone$ is used to refer to the active
indices in $\gamma$ as well as the bias, which is always included in the model by assumption.
Using a Cholesky decomposition the quantity in Eqn.~\ref{eqn:mll} can be computed in $\OO(\gabs^3 + \gabs^2 N)$ time. 
If done naively this becomes expensive in cases where Eqn.~\ref{eqn:mll} needs to be computed for many values of $\gamma$ (as is needed e.g.~to compute Rao-Blackwellized PIPs). 
Luckily, and as is done by \citep{zanella2019scalable} and others in the literature, the computational cost can be
reduced significantly since we can exploit the fact that in practice we always consider `neighboring' values of $\gamma$
and so we can leverage rank-1 update structure where appropriate.
In the following we provide the formulae necessary for doing so. We keep the derivation generic
and consider the case of adding arbitrarily many variables to $\gamma$ even though in practice we only make use of the rank-1 formulae.

In more detail we proceed as follows.
Let $\II$ be the active indices in $\gamma$ together with the bias index $P+1$ (i.e.~we conveniently
augment $X$ by an all-ones feature column in the following). Let $\kk$ be a non-empty set of indices not in $\II$
and let $\II_\kk = \II \cup \kk$.
We let $\XX = \Omega^{\tfrac{1}{2}} X$
 and rewrite 
$F_\ik \equiv (\XX_\ik\TT \XX_\ik  + \tau \id_\ik)\inv$ in terms of $F_\II \equiv (\XX_\II \TT \XX_\II + \tau \id_\II)\inv$ as follows:
\begin{equation}
F_\ik =
  \begin{pmatrix}
  \XX_\II \TT \XX_\II  + \tau \id_\II& \XX_\II \TT \XX_\kk \\
  \XX_\kk \TT \XX_\II & \XX_\kk \TT \XX_\kk + \tau \id_\kk
  \end{pmatrix}\inv = 
    \begin{pmatrix}
  F_\II + F_\II \XX_\II \TT \XX_\kk  G_\kk \XX_\kk \TT X_\II   F_\II        &        -F_\II \XX_\II \TT \XX_\kk    G_\kk   \\
  -G_\kk   \XX_\kk \TT \XX_\II    F_\II                                                 &               G_\kk
  \end{pmatrix}
\end{equation}
where $G_\kk \inv \equiv \XX_\kk \TT \XX_\kk + \tau \id_\kk - \XX_\kk \TT \XX_\II F_\II \XX_\II \TT \XX_\kk$.

To efficiently compute the quadratic term in Eqn.~\ref{eqn:mll} we need
 to compute $\ZZ_\ik \TT F_\ik \ZZ_\ik $ in terms of $\ZZ_\II \TT F_\II \ZZ_\II$.
Write $\ZZ_\ik = (\ZZ_\II, \ZZ_\kk)$ so we have 
\begin{align}
\ZZ_\ik \TT F_\ik \ZZ_\ik =&
   \begin{pmatrix} \ZZ_\II \\ \ZZ_\kk \end{pmatrix} \TT
    \begin{pmatrix}
  F_\II + F_\II \XX_\II \TT \XX_\kk  G_\kk \XX_\kk \TT \XX_\II   F_\II        &        -F_\II \XX_\II \TT \XX_\kk    G_\kk   \\
  -G_\kk   \XX_\kk \TT \XX_\II    F_\II                                                 &               G_\kk
  \end{pmatrix}
  \begin{pmatrix} \ZZ_\II \\ \ZZ_\kk \end{pmatrix} \\
  =\;& \ZZ_\II \TT F_\II \ZZ_\II + \ZZ_\II \TT F_\II \XX_\II \TT \XX_\kk  G_\kk \XX_\kk \TT \XX_\II   F_\II  \ZZ_\II  \\
   &- \ZZ_\II \TT F_\II \XX_\II \TT \XX_\kk    G_\kk   \ZZ_\kk - \ZZ_\kk \TT G_\kk   \XX_\kk \TT \XX_\II    F_\II  \XX_\II    +  \ZZ_\kk \TT G_\kk   \ZZ_\kk   \\
   =\;& \ZZt_\II \TT \ZZt_\II  + (\XXt_\kk \TT \XXt_\II \ZZt_\II )\TT (\XXt_\kk \TT \XXt_\II \ZZt_\II ) -2 (\XXt_\kk \TT \XXt_\II \ZZt_\II )\TT \ZZt_\kk +  \ZZt_\kk \TT \ZZt_\kk  \\
   =\;& \ZZt_\II \TT \ZZt_\II  + \left \vert \left\vert \XXt_\kk \TT \XXt_\II \ZZt_\II - \ZZt_\kk  \right\vert\right\vert^2. 
\end{align}
where $ \left \vert \left\vert \cdot  \right\vert\right\vert$ is the 2-norm in $\RR^{|\kk|}$ and we define
\begin{align}
\label{eqn:mllchols}
L_\II  L_\II \TT&= \XX_\II \TT \XX_\II + \tau \id_\II= F_\II \inv      \\ 
L_\kk  L_\kk \TT &= \XX_\kk \TT \XX_\kk + \tau \id_\kk - \XX_\kk \TT \XX_\II F_\II \XX_\II \TT \XX_\kk 
                             = \XX_\kk \TT \XX_\kk + \tau \id_\kk  - \XX_\kk \TT \XXt_\II \XXt_\II \TT \XX_\kk  = G_\kk \inv \nonumber \\
\ZZt_\II &\equiv L_\II \inv \ZZ_\II \qquad \qquad        \ZZt_\kk \equiv L_\kk \inv \ZZ_\kk \qquad       \XXt_\II \equiv  \XX_\II L_\II \invT   \qquad     \XXt_\kk \equiv  \XX_\kk L_\kk \invT \nonumber
\end{align}
Here $L_\II$ and $L_\kk$ are Cholesky factors.
This can be rewritten as
\begin{align}
\ZZ_\ik \TT F_\ik \ZZ_\ik =  \ZZt_\II \TT \ZZt_\II  + \left \vert \left\vert  W_\kk  \right\vert\right\vert^2
\qquad {\rm with} \qquad   W_\kk \equiv L_\kk \inv  \left(  \XX_\kk \TT  \XXt_\II \ZZt_\II - \ZZ_\kk  \right)
\end{align}
Together these formulae can be used to compute the quadratic term efficiently.

Next we turn to the log determinant in Eqn.~\ref{eqn:mll}.
We begin by noting that 
\begin{align}
 \log \det \left( \XX_\ik \TT \XX_\ik + \tau \id_\ik \right) + \log\det( \tau \inv \id_\ik) =  \log\det(\Omega) + \log \det \left(X_\ik X_\ik\TT/\tau +\Omega\inv \right)
\end{align}
and
\begin{align}
\log \det \left(X_\ik X_\ik\TT/\tau +\Omega\inv \right) &= \log \det \left(X_\II X_\II \TT/\tau +\Omega\inv \right) + 
\log \det \left( \id_\kk/\tau\right) \\ &+ \log \det \left(\tau\id_\kk + X_\kk \TT (X_\II X_\II \TT/\tau +\Omega\inv)\inv X_\kk \right)
\end{align}
which together imply
\begin{align}
\{  \log \det \left( \XX_\ik \TT \XX_\ik + \tau \id_\ik \right) + \log\det( \tau \inv \id_\ik)  \} - \nonumber \\ \nonumber
\{  \log \det \left( \XX_\II \TT \XX_\II + \tau \id_\II \right) + \log\det( \tau \inv \id_\II)  \} &= \nonumber
\log \det \left(X_\ik X_\ik\TT/\tau +\Omega\inv \right) - \log \det \left(X_\II X_\II \TT/\tau +\Omega\inv \right) \\
&=  \nonumber
 \log \det \left(\id_\kk +  \tau\inv X_\kk \TT (X_\II X_\II \TT/\tau +\Omega\inv)\inv X_\kk \right)
\end{align}
While these equations can be used to compute the log determinant reasonably efficiently, they 
exhibit cubic computational complexity w.r.t.~$N$. So instead we write
\begin{align}
    \label{eqn:detid}
\det( \XX_\ik \TT \XX_\ik  + \tau \id_\ik) &= \det
  \begin{pmatrix}
  \XX_\II \TT \XX_\II  + \tau \id_\II& \XX_\II \TT \XX_\kk \\
  \XX_\kk \TT \XX_\II & \XX_\kk \TT \XX_\kk + \tau \id_\kk
  \end{pmatrix} \\ 
  &= \det(\XX_\kk \TT \XX_\kk + \tau \id_\kk -  \XX_\kk \TT \XX_\II  ( \XX_\II \TT \XX_\II  + \tau \id_\II)\inv \XX_\II \TT \XX_\kk ) \times \det( \XX_\II \TT \XX_\II  + \tau \id_\II) \nonumber \\
  &= \det( G_\kk \inv ) \times \det( \XX_\II \TT \XX_\II  + \tau \id_\II)   \nonumber
\end{align}
This form is convenient because it relies on the term $G_\kk$ that we in any case need to compute the quadratic form. Similarly $ \det( \XX_\II \TT \XX_\II  + \tau \id_\II)$ is easily computed
from the Cholesky factor $L_\II$.

Above we considered the case of turning on covariates, i.e.~$\gamma_i=0 \rightarrow \gamma_i=1$.
Since we assume that $\gabs \ll P$ these computations tend to dominate the computational cost.
However, we must also consider the case of turning off covariates, i.e.~$\gamma_i=1 \rightarrow \gamma_i=0$.
To efficiently compute the required terms we make extensive use of the following identity.
Let $A$, $B$, $C$, and $D$ be appropriate $(M-1) \times (M-1)$, $(M-1) \times 1$, $1 \times (M-1)$, and $1 \times 1$ matrices,
respectively. Then the identity
\begin{align}
    \label{eqn:invid}
\begin{pmatrix}
  A & B \\ C & D 
\end{pmatrix}\inv = 
\begin{pmatrix}
  \tilde{A} & \tilde{B} \\ \tilde{C} & \tilde{D} 
\end{pmatrix} \qquad 
    \implies \qquad A \inv = \tilde{A} - \tilde{B} \tilde{D}\inv \tilde{C}
\end{align}
can be used to cheaply compute $A\inv$ if the inverse of the block matrix $((A,B),(C,D))$ is available.
In other words once we have computed $F_\II \equiv (\XX_\II \TT \XX_\II + \tau \id_\II)\inv$ using the
Cholesky factor $L_\II$ we can
use submatrices of $F_\II$ to cheaply compute the inverse of submatrices of $\XX_\II \TT \XX_\II + \tau \id_\II$,
which are precisely the quantities we need to compute Eqn.~\ref{eqn:mll} for downdates of $\gamma$.
In particular once the quadratic term has been computed, we can compute the log determinant by again appealing to Eqn.~\ref{eqn:detid},
using Eqn.~\ref{eqn:invid} to compute $F_\II$ in 
$G_\kk \inv = \XX_\kk \TT \XX_\kk + \tau \id_\kk - \XX_\kk \TT \XX_\II F_\II \XX_\II \TT \XX_\kk$
for redefinitions of $\II$ and $\KK$ appropriate to a downdate.

\subsection{Computational complexity}
\label{app:cc}

The primary computational cost in Subset wTGS, PG-wTGS, ASI, and the other MCMC algorithms considered in the main text
arises in computing conditional PIPs of the form $p(\gamma_j = 1 | \gmj, \DD)$ (linear regression case) 
or $p(\gamma_j = 1 | \gmj, \omega, \DD)$ (count-based likelihood case) for $j=1,...,P$,
the principal ingredient for which are conditional marginal log likelihoods as in Eqn.~\ref{eqn:mll}.
In the case of PG-wTGS, PG-TGS, PG-wGS, and ASI the next largest computational cost is usually sampling P\olya-Gamma variables, although
this is $\OO(N)$ and so the cost is moderate in most cases. For PG-wTGS, PG-TGS, PG-wGS, and ASI 
computing the MH acceptance probability (e.g.~Eqn.~\ref{eqn:alphaomega} for the case of PG-wTGS)
is another subdominant but non-negligible cost.

The precise computational cost of computing $p(\gamma_j = 1 | \gmj, \DD)$  
and $p(\gamma_j = 1 | \gmj, \omega, \DD)$ depends on the details
of how the formulae in Sec.~\ref{app:mll} are implemented. 
For example in the linear regression setting it can be
advantageous to pre-compute $X \TT X$ if the result can be stored in memory.
In our experiments we do so whenever this is feasible (for a mid-grade GPU this is typically possible for $P \lesssim 4 \times 10^4$).
In the case of PG-wTGS where $\omega$ changes every few MCMC iterations, pre-computing $\XX \TT \XX$ is not advantageous. 
Note that to avoid possible accumulation of numerical errors we do not compute $F_\II$ or other quantities using 
computations from the previous MCMC iteration, although doing so is possible in principle for the linear regression
 case (see e.g.~\citet{zanella2019scalable}).

\paragraph{Linear regression case (wTGS)}

Using the various rank-1 update/downdate formulae from Sec.~\ref{app:mll} the result is $\OO(\gabs N P + N \gabs^2 + \gabs^3)$ computational complexity per MCMC iteration if pre-computing $X \TT X$ is not possible. If pre-computing $X \TT X$ is possible
the computational complexity per MCMC iteration is instead $\OO(P \gabs^2 + \gabs^3)$ along with a one-time $\OO(NP^2)$ cost
to compute $X \TT X$.

\paragraph{Linear regression case (Subset wTGS)}

Using the various rank-1 update/downdate formulae from Sec.~\ref{app:mll} the result is $\OO(\gabs N S + N \gabs^2 + \gabs^3)$ computational complexity per MCMC iteration if pre-computing $X \TT X$ is not possible. If pre-computing $X \TT X$ is possible
the computational complexity per MCMC iteration is instead $\OO(S \gabs^2 + \gabs^3)$ along with a one-time $\OO(NP^2)$ cost
to compute $X \TT X$.

\paragraph{PG-wTGS for Binomial and Negative Binomial regression}

Using the various rank-1 update/downdate formulae from Sec.~\ref{app:mll} the result is $\OO(\gabs N P + N \gabs^2 + \gabs^3)$ computational complexity per MCMC iteration with $i>0$ and $\OO(N + N \gabs^2 + \gabs^3)$ per MCMC iteration with $i=0$.

We note that the asymptotic formulae reported above are somewhat misleading in practice, since
most of the necessary tensor ops are highly-parallelizable and very efficiently implemented on modern hardware. 
For this reason Fig.~\ref{fig:cancerlarge} and Fig.~\ref{fig:runtime} are particularly useful for understanding 
the runtime in practice, since
the various parts of the computation will be more or less expensive depending on the precise regime and the underlying
low-level implementation and hardware.

\subsection{TGS motivation: binary variables and Metropolized-Gibbs}
\label{app:mg}

To provide intuition for the Tempered Gibbs Sampling (TGS) strategy that underlies wTGS, we consider a single latent binary variable $x$
governed by the probability distribution $p(x) = {\rm Bernoulli}(q)$. A Gibbs sampler for this distribution simply samples $x \sim p$ in
each iteration of the Markov chain. An alternative strategy is to employ a so-called Metropolized-Gibbs move w.r.t.~$x$ \citep{liu1996peskun}.
For binary $x$ this results in a proposal distribution that is deterministic in the sense that it always proposes a flip: $0 \to 1$ or $1 \to 0$.
The corresponding Metropolis-Hastings (MH) acceptance probability for a move $x \to x^\prime$ is given by
\begin{align}
\alpha(x \! \to \! x^\prime) =
    \begin{cases}
      \min(1, \tfrac{q}{1-q}) & {\rm if} \; x = 0\\
     \min(1, \tfrac{1- q}{q}) & {\rm if} \; x = 1
    \end{cases}
    \label{eqn:accept}
\end{align}
As is well known, this update rule is more statistically efficient than the corresponding Gibbs move \citep{liu1996peskun}.
For our purposes, however, what is particularly interesting is the special case where $q=\tfrac{1}{2}$. In this
case the acceptance probability in Eqn.~\ref{eqn:accept} is identically equal to one. Consequently
the Metropolized-Gibbs chain is deterministic:
\begin{align}
... \to 0\to 1 \to 0 \to 1 \to 0 \to 1 \to ...
\end{align}
Indeed this Markov chain can be described as \emph{maximally non-sticky}.
This shows why building tempering into inference algorithms for binary latent variable models like that in Bayesian variable selection
might be an attractive strategy for avoiding the stickiness of a vanilla Gibbs sampler.

\subsection{The nature of local moves in wTGS} 
\label{app:wtgsdisc}

wTGS samples an auxiliary variable $i$ controlled by the Gibbs update in Eqn.~\ref{eqn:iupdatelin}.
To better understand how wTGS and its variants Subset wTGS and PG-wTGS are designed to efficiently explore regions of high posterior mass
it is important to take a closer look at the form of these $i$ updates. To do so we compute 
$ \frac{\eta(\gmi)}{p(\gamma_i | \gmi, \DD)} \propto f(i | \gamma)$ in four regimes, see Table~\ref{table:wtgs}. 
We see that if covariate $i$ is not included in the model ($\gamma=0$) and has a small PIP covariate $i$ will 
be chosen to be updated only infrequently and, furthermore, that the probability of $i$ being chosen depends on $\epsilon$; thus $\epsilon$ controls
the amount of exploration. By contrast if $i$ has a large PIP and is currently excluded from the model ($\gamma=0$) or
if $i$ has a small PIP and is currently included in the model ($\gamma=1$), then $f(i | \gamma) \sim \OO(1)$, with the consequence
that $i$ is likely to be flipped in the next move. This reflects the greedy nature of wTGS, which focuses much of its
computational budget on turning on likely covariates and/or turning off unlikely covariates (i.e.~un/likely under the posterior). 
Finally, if $i$ has a large PIP and is currently on ($\gamma=1$) it will occasionally be turned off (especially if no other covariates
satisfy the `greedy' condition described in the previous two sentences), which promotes exploration in and around posterior modes. 
In particular if covariate $i$ is turned off and covariate $i$ is highly correlated with $j$ then turning off $i$ allows
for the possibility that $j$ is turned on instead in the next MCMC iteration; indeed there will be a $\sim 50$\% chance of doing
so if $i$ and $j$ are the only covariates that satisfy the greedy condition.
Taken together the behavior of $f(i | \gamma)$ reflected in Table~\ref{table:wtgs} 
results in a satisfying balance between exploration and exploitation.

\begin{table}[]\centering
    \resizebox{0.59\columnwidth}{!}{%
\begin{tabular}{lllll}
\multicolumn{1}{l|}{} & \multicolumn{1}{l|}{$\gamma = 0$ } & $\gamma=1$  &  &  \\ \cline{1-3}
    \multicolumn{1}{l|}{$p(\gamma_i =1 | \DD) \approx 0$} & \multicolumn{1}{l|}{$\epsilon / P \ll 1$} & $\frac{\epsilon/P}{p(\gamma_i  =1 |\DD)}\gg 1$ &  &  \\ \cline{1-3}
    \multicolumn{1}{l|}{$p(\gamma_i  =1 | \DD) \approx 1$} & \multicolumn{1}{l|}{$\tfrac{1}{1 - p(\gamma_i  =1 | \DD)} \gg 1$} & $\approx 1$ &   &  \\
                      &                       &  &  &
\end{tabular} }
    \caption{We explore how the quantity $\frac{\eta(\gmi)}{p(\gamma_i | \gmi, \DD)} = \frac{p(\gamma_i  =1 | \DD) + \tfrac{\epsilon}{P}}{p(\gamma_i | \gmi, \DD)}$ 
    varies as a function of $\gamma$ and ${\rm PIP}(i) = p(\gamma_i = 1 | \DD)$
    under the approximation $p(\gamma_i | \gmi, \DD) \approx \gamma_i p(\gamma_i  =1 | \DD) + (1 - \gamma_i)( 1 - p(\gamma_i  =1 | \DD))$.
    We further assume that either $p(\gamma_i  =1 | \DD) \ll \epsilon/ P$ or $1-p(\gamma_i  =1 | \DD) \ll \epsilon/ P$. 
    Off-diagonal entries in the table correspond to `greedy' moves that are given large weight by wTGS.
    See Sec.~\ref{app:wtgsdisc} for discussion.
    }
    \label{table:wtgs}
\end{table}

\subsection{Importance weights}
\label{app:iw}

Importance weights $\rho \sim \phi^{-1}$ in wTGS and its variants (see e.g.~Eqn.~\ref{eqn:phidefnlin} and Algorithm \ref{algo})
are bounded from above. For example for wTGS in the linear regression case we have
\begin{align}
    \phi(\gamma) = \tfrac{1}{2} \sum_{i=1}^P  \frac{{\rm PIP}(i) + \epsilon /P }{ p(\gamma_i | \gmi, \DD)} \ge \frac{\epsilon}{2}
\label{eqn:phidefnlinapp}
\end{align}
with the consequence that (unnormalized) importance weights are bounded from above by $\frac{2}{\epsilon}$;
note that in experiments we typically use $\epsilon=5$.
We also note that the bound in Eqn.~\ref{eqn:phidefnlinapp} is somewhat loose.
In practice the variance of importance weights normalized so that 
$\sum_{t=1}^T \rho^{(t)} = T$ is $\OO(1)$; see the rightmost panels in Fig.~\ref{fig:applargepnoisy}-\ref{fig:applargep} for variances observed in practice.

\subsection{Proof of Proposition 1}
\label{app:proof}



In the main text we made use of an auxiliary variable representation in which the state $i$ is explicitly included 
in the state space. For the present purpose it is more convenient to think of Subset wTGS, Algorithm \ref{algo},
as acting on the space $\{0,1\}^P \times \PPP$, where $\PPP$ is the set of all subsets of $\{1, ..., P\}$
of size $S$ that contain the anchor set $\AAA$. The transition kernel can be written as 
\begin{align}
K((\gamma, \SSS) \rightarrow (\gamma^\prime, \SSS^\prime)) = 
    \sum_{i \in \SSS} f(i | \gamma, \SSS) \delta(\gamma^\prime - {\rm flip}(\gamma | i)) U(\SSS^\prime | i, \AAA)
    \label{eqn:k}
\end{align}
where $f(i | \gamma, \SSS)$ is the posterior conditional probability in Eqn.~\ref{eqn:iupdateSS} and $\delta(\cdot)$
is the Dirac delta function. We first show that $K$ is reversible w.r.t.~the auxiliary target 
$f(\gamma, \SSS) = p(\gamma | \DD) \phi(\gamma, \SSS)$, see Eqn.~\ref{eqn:fphiSS}. As is evident from Eqn.~\ref{eqn:k}, 
$K$ is zero unless $\gamma$ and $\gamma^\prime$ differ in exactly one coordinate---call it $i$---so that
we have $\gmi = \gmi^\prime$.
Thus for non-zero $K$ we have
\begin{align}
    K((\gamma, \SSS) \rightarrow (\gamma^\prime, \SSS^\prime)) &= 
    f(i | \gamma, \SSS) U(\SSS^\prime | i, \AAA) \\
    &= \phi(\gamma, \SSS)^{-1}\frac{\tfrac{1}{2}\eta(\gmi)}{ p(\gamma_i | \gmi, \DD)}  U(\SSS | i, \AAA) U(\SSS^\prime | i, \AAA)
\end{align}
which implies that
\begin{align}
    \frac{K((\gamma, \SSS) \rightarrow (\gamma^\prime, \SSS^\prime))}{K((\gamma^\prime, \SSS^\prime) \rightarrow (\gamma, \SSS))}
    &= \frac{\phi(\gamma^\prime, \SSS^\prime)p(\gamma_i^\prime | \gmi^\prime, \DD)}{\phi(\gamma, \SSS)p(\gamma_i | \gmi, \DD)} \\
    &= \frac{\phi(\gamma^\prime, \SSS^\prime)p(\gamma^\prime | \DD)}{\phi(\gamma, \SSS)p(\gamma | \DD)}  \\
    &= \frac{f(\gamma^\prime, \SSS^\prime)}{f(\gamma, \SSS)} 
\end{align}
where we used that $p(\gmi^\prime | \DD) = p(\gmi | \DD)$. 
Since reversibility is trivially satisfied if $K((\gamma, \SSS) \rightarrow (\gamma^\prime, \SSS^\prime))$ is zero,
we have thus shown that $K$ is reversible w.r.t.~$f(\gamma, \SSS)$ and therefore $f$-invariant. 
Since our state space is finite and $f(i | \gamma, \SSS)>0$ if $i \in \SSS$
it is also clear that our Markov chain is both irreducible and Harris recurrent. Thus our Markov chain
satisfies the conditions of Theorem $17.0.1$ in \citet{meyn2012markov} so that the Law of Large Numbers holds
for any test function $h(\gamma, \SSS) : \{0,1\}^P \times \PPP \rightarrow \RR$.
In particular for any test function $h(\gamma) : \{0,1\}^P \rightarrow \RR$ we can
apply the Law of Large Numbers twice, once to $h \phi^{-1}$ and once to $\phi^{-1}$ (note that $\phi$ is bounded away from
zero and bounded from above). 
If we let $Z_f$ be the partition function of $f(\gamma, \SSS)$, i.e.~$Z_f \equiv \sum_{\gamma, \SSS} f(\gamma, \SSS)$, then
\begin{align}
    \tfrac{1}{T}{\textstyle \sum}_{t=1}^T  h(\gamma^{(t)}) \phi^{-1}(\gamma^{(t)}, \SSS^{(t)})  
    \rightarrow \EE_{f(\gamma, \SSS) / Z_f} \left[ h(\gamma) \phi^{-1}(\gamma, \SSS)\right] 
    = \EE_{p(\gamma | \DD)} \left[ h(\gamma) \right]  / Z_f
\end{align}
and
\begin{align}
    \tfrac{1}{T}{\textstyle \sum}_{t=1}^T \phi^{-1}(\gamma^{(t)}, \SSS^{(t)})  
    \rightarrow \EE_{f(\gamma, \SSS) / Z_f} \left[ \phi^{-1}(\gamma, \SSS)\right] = Z_f^{-1}
\end{align}
It follows that
\begin{align}
    \frac{ \tfrac{1}{T}{\textstyle \sum}_{t=1}^T  h(\gamma^{(t)}) \phi^{-1}(\gamma^{(t)}, \SSS^{(t)})}{\tfrac{1}{T}{\textstyle \sum}_{t=1}^T \phi^{-1}(\gamma^{(t)}, \SSS^{(t)})} \rightarrow \EE_{p(\gamma | \DD)} \left[ h(\gamma) \right] 
\end{align}
or equivalently utilizing normalized weights $\{ \rho^{(t)} \}$ 
\begin{align}
    {\textstyle \sum}_{t=1}^T  \rho^{(t)} h(\gamma^{(t)})
    \rightarrow \EE_{p(\gamma | \DD)} \left[ h(\gamma) \right] \;\; {\rm as} \;\; T \rightarrow \infty
\end{align}
This finishes the proof of the central claim of Proposition~\ref{prop:large}.
For the specific claim about Rao-Blackwellized PIP estimators see the next section.

\subsection{Rao-Blackwellized PIP estimators}
\label{app:rao}

A naive estimator for ${\rm PIP}(i) = p(\gamma_i = 1 | \DD)$ directly uses weighted samples
$\{ (\rho^{(t)}, \gamma^{(t)} ) \}$ provided by Algorithm~\ref{linalgo}:
\begin{align}
    \label{eqn:raw}
    {\rm PIP}(i) \approx \sum_t \rho^{(t)} \gamma_i^{(t)}
\end{align}
However, since wTGS and its variants compute conditional PIPs as part of inference, it is preferable to use
a lower variance Rao-Blackwellized estimator instead:
\begin{align}
    \label{eqn:rao}
    {\rm PIP}(i) \approx \sum_t \rho^{(t)} p(\gamma_i = 1 | \gmi^{(t)}, \DD)
\end{align}
We use the appropriate version of Eqn.~\ref{eqn:rao} in all experiments.
In the case of Subset wTGS, Algorithm~\ref{algo}, only $S$ conditional PIPs are computed in each MCMC iteration.
Using the analog of Eqn.~\ref{eqn:rao} would inflate the computational cost from $\OO(S)$ to $\OO(P)$, entirely defeating
the purpose of Subset wTGS. Thus for Subset wTGS we use a partially Rao-Blackwellized estimator instead:
\begin{align}
    \label{eqn:raopartial}
    {\rm PIP}(i) \approx \sum_t \rho^{(t)} \left\{ \II(i \in \SSS^{(t)})  p(\gamma_i = 1 | \gmi^{(t)}, \DD)  +
    \II(i \notin \SSS^{(t)}) \gamma_i^{(t)} \right\}
\end{align}
where $\II(\cdot)$ is an indicator function.
In other words we use conditional PIPs if they are computed as part of inference (because $i \in \SSS$) and otherwise use raw $\gamma$
samples. It is easy to see that the estimator in Eqn.~\ref{eqn:raopartial} is unbiased, since the test statistic under consideration
factorizes between $\gamma$ and $\SSS$. 
Indeed if we let $q(\SSS)$ denote the uniform distribution on $\PPP$ and 
$\zeta = \EE_{q(\SSS)} \left[ \II(i \in \SSS)\right]$
then the proof in Sec.~\ref{app:proof} makes it clear that the partially Rao-Blackwellized estimator in Eqn.~\ref{eqn:raopartial}
converges to
\begin{align}
    &\EE_{p(\gamma | \DD)} \EE_{q(\SSS)} \left[ \II(i \in \SSS)  p(\gamma_i = 1 | \gmi, \DD) +(1 - \II(i \in \SSS)) \gamma_i\right] \\
    &= \zeta \EE_{p(\gamma | \DD)} \left[ p(\gamma_i = 1 | \gmi, \DD) \right] + (1 - \zeta) \EE_{p(\gamma | \DD)} \left[\gamma_i\right] \\
    &= \zeta  {\rm PIP}(i) + (1 - \zeta) {\rm PIP}(i) = {\rm PIP}(i) 
\end{align}
It is also evident that Eqn.~\ref{eqn:raopartial} is lower variance than the raw estimator Eqn.~\ref{eqn:raw}.

\subsection{$\omega$-update in PG-wTGS}
\label{app:omega}

The acceptance probability for the $\omega$-update in Sec.~\ref{sec:pgwtgs} is given by
\begin{align}
\label{eqn:appalphaomega}
\alpha(\omega  \! \to \! \omega^\prime | \gamma) &=  \min\left(1,
      \frac{p(Y | \gamma, \omega^\prime, X, \TC) p(\gamma) p(\omega^\prime|\TC) }{p(Y | \gamma, \omega, X, \TC) p(\gamma) p(\omega|\TC)}
      \frac{p(\omega | \gamma, \bhat(\gamma, \omega^\prime), \DD)}{p(\omega^\prime | \gamma, \bhat(\gamma, \omega), \DD)} \right) 
\end{align}
where the ratio of proposal densities is given by
\begin{align}
 \frac{p(\omega | \gamma, \bhat(\gamma, \omega^\prime), \DD)}{p(\omega^\prime | \gamma, \bhat(\gamma, \omega), \DD)} &=
          \frac{p(Y | \gamma, \omega,  \bhat(\gamma, \omega^\prime), X, \TC) p(\gamma) p(\omega|\TC) p(\bhat(\gamma, \omega^\prime))}
      {\int \! d \omhat \, p(Y | \gamma, \omhat,  \bhat(\gamma, \omega^\prime), X, \TC) p(\gamma) p(\omhat|\TC) p(\bhat(\gamma, \omega^\prime))} \times \nonumber \\
           & \left\{ \frac{p(Y | \gamma, \omega^\prime,  \bhat(\gamma, \omega), X, \TC) p(\gamma) p(\omega^\prime|\TC) p(\bhat(\gamma, \omega))}
      {\int \! d \omhat \, p(Y | \gamma, \omhat,  \bhat(\gamma, \omega), X, \TC) p(\gamma) p(\omhat|\TC) p(\bhat(\gamma, \omega))} \right \} \inv  
\end{align}
Simplifying we have that the ratio in $\alpha(\omega  \! \to \! \omega^\prime | \gamma)$ is given by
\begin{align}
    &\frac{p(Y | \gamma, \omega^\prime, X, \TC) }{p(Y | \gamma, \omega, X, \TC)}
          \frac{p(Y | \gamma, \omega,  \bhat(\gamma, \omega^\prime), X, \TC)}
      {\int \! d \omhat \, p(Y | \gamma, \omhat,  \bhat(\gamma, \omega^\prime), X, \TC) p(\omhat|\TC)}
            \frac{\int \! d \omhat \, p(Y | \gamma, \omhat,  \bhat(\gamma, \omega), X, \TC) p(\omhat|\TC)}{p(Y | \gamma, \omega^\prime,  \bhat(\gamma, \omega), X, \TC) }
\nonumber \\
&= \frac{p(Y | \gamma, \omega^\prime, X, \TC) }{p(Y | \gamma, \omega, X, \TC)}
          \frac{p(Y | \gamma, \omega,  \bhat(\gamma, \omega^\prime), X, \TC)}
      {p(Y | \gamma, \bhat(\gamma, \omega^\prime), X, \TC) }
            \frac{ p(Y | \gamma, \bhat(\gamma, \omega), X, \TC)}{p(Y | \gamma, \omega^\prime,  \bhat(\gamma, \omega), X, \TC) }
\nonumber \\
&= \frac{p(Y | \gamma, \omega^\prime, X, \TC) }{p(Y | \gamma, \omega, X, \TC)}
\frac{p(Y | \gamma, \omega, \bhat(\gamma, \omega^\prime), X, \TC) }{p(Y | \gamma, \omega^\prime, \bhat(\gamma, \omega), X, \TC)}
\frac{p(Y | \gamma,  \bhat(\gamma, \omega), X, \TC) }{p(Y | \gamma, \bhat(\gamma, \omega^\prime), X, \TC)} \nonumber
\end{align}
which is Eqn.~\ref{eqn:alphaomega} in the main text.
Here
\begin{align}
\label{eqn:alphaomega2}
\frac{p(Y | \gamma, \omega, \bhat(\gamma, \omega^\prime), X, \TC) }{p(Y | \gamma, \omega^\prime, \bhat(\gamma, \omega), X, \TC)} &=
\frac{\exp(\kappa \cdot \psihat(\gamma, \omega^\prime) -\tfrac{1}{2} \omega \cdot \psihat(\gamma, \omega^\prime)^2) }{\exp(\kappa \cdot \psihat(\gamma, \omega) -\tfrac{1}{2} \omega^\prime \cdot \psihat(\gamma, \omega)^2)}
\end{align}
and
\begin{align}
\frac{p(Y | \gamma,  \bhat(\gamma, \omega), X, \TC) }{p(Y | \gamma, \bhat(\gamma, \omega^\prime), X, \TC)} &=
\frac{\prod_n \exp(\psihat(\gamma, \omega)_n))^{Y_n}}{\prod_n (1 + \exp(\psihat(\gamma, \omega)_n))^{\TC_n}}
\frac{\prod_n (1 + \exp(\psihat(\gamma, \omega^\prime)_n))^{\TC_n}}{\prod_n \exp(\psihat(\gamma, \omega^\prime)_n))^{Y_n}}
\end{align}
where
\begin{align}
(\psihat(\gamma, \omega))_n \equiv \bhat(\gamma, \omega)_0 + \bhat(\gamma, \omega)_\gamma  \cdot X_{n\gamma}
\end{align}
and
\begin{align}
    \bhat(\gamma, \omega) = (X_\gone \TT \Omega X_\gone + \tau \mathbb{1}_\gabsone)\inv X_\gone \TT \kappa
    \in \RR^{\gabsone}
    \label{eqn:bhatdefn}
\end{align}
where as in Sec.~\ref{app:mll} $X$ is here augmented with a column of all ones.
As detailed in \citep{polson2013bayesian} the (approximate) Gibbs proposal distribution
that results from conditioning on $\bhat$ is given by a P\olya-Gamma distribution determined by $\TC$ and $\psihat$:
\begin{align}
    p(\omega^\prime | \gamma, \bhat(\gamma, \omega), \DD) = {\rm PG}(\omega^\prime | \TC, \psihat(\gamma, \omega))
\end{align}
In practice we do without the MH rejection step for $\omega$ in the early stages of burn-in to allow the 
MCMC chain to more quickly reach probable states.

\subsection{$\xi$-adaptation in PG-wTGS and other wTGS variants}
\label{app:xi}

Here we discuss how $\xi>0$ in Eqn.~\ref{eqn:fulltargetpg} can be adapted during burn-in. 
The same adaptation scheme (mutatis mutandis) can also be used for
Algorithm~\ref{inferhalgo}, where the $i=0$ state is introduced to allow for $h$-updates.

The magnitude of $\xi$ controls the frequency of $\omega$ updates. Ideally
$\xi$ is such that an $\OO(1)$ fraction of MCMC iterations result in a $\omega$ update,
with the remainder of the computational budget being spent on $\gamma$ updates.
Typically this can be achieved by choosing $\xi$ in the range $\xi \sim 1-5$.
Here we describe a simple scheme for choosing $\xi$ adaptively 
during burn-in to achieve the desired behavior.

We introduce a hyperparameter $f_\omega \in (0, 1)$ that controls the desired $\omega$
update frequency. Here $f_\omega$ is normalized such that $f_\omega=1$ corresponds to a situation
in which all updates are $\omega$ updates, i.e.~all states in the MCMC chain are in the $i=0$ state
(something that would be achieved by taking $\xi \to \infty$). Since $\omega$ updates are of somewhat
less importance for obtaining accurate PIP estimates than $\gamma$ updates, we recommend a somewhat
moderate value of $f_\omega$, e.g. $f_\omega \sim 0.1-0.4$. 
For all experiments in this paper we use $f_\omega = 0.25$.

Our adaptation scheme proceeds as follows.
We initialize $\xi^{(0)} = 5$. At iteration $t$ during the burn-in a.k.a.~warm-up phase
we update $\xi^{(t)}$ as follows: 
\begin{align}
    \xi^{(t+1)} = \xi^{(t)} +  \frac{f_\omega - \frac{\xi^{(t)}}{\phi(\gamma^{(t)}, \omega^{(t)})} }{\sqrt{t + 1}}
\end{align}
By construction this update aims to achieve that a fraction $f_\omega$ of MCMC states
satisfy $i=0$, since the quantity
\begin{align}
\phi(\gamma, \omega) = \xi +  \frac{1}{P} \sum_{i=1}^P  \frac{\tfrac{1}{2} \eta(\gmi, \omega)}{ p(\gamma_i | \gmi, \omega, \DD)}
\end{align}
encodes the total probability mass assigned to states $i=0$ and $i>0$.

\subsection{Anchor set $\AAA$ adaptation in Subset wTGS}
\label{app:anchor}

We adopt a simple adaptation scheme for the anchor set $\AAA$. During burn-in we keep a running PIP estimate for each covariate using
the partially Rao-Blackwellized estimator described in Sec.~\ref{app:rao}. Periodically---in our experiments every 100 iterations---we update $\AAA$ to be the $A$ covariates exhibiting the largest PIPs according to the current running PIP estimate. At the end of the burn-in period the anchor set is updated one last time and remains fixed thereafter.

\subsection{PG-wTGS for Negative Binomial regression}
\label{app:nb}

We specify in more detail how we can accommodate the negative binomial likelihood using
P\olya-Gamma augmentation. Using the identity
\begin{align}
    \frac{(e^\psi)^a}{(1 + e^\psi)^b} = \tfrac{1}{2^b} e^{(a-\tfrac{1}{2}b)\psi} \EE_{{\rm PG}(\omega |  b, 0)} \left[ \exp(-\tfrac{1}{2} \omega \psi^2) \right]
\end{align}
we write
\begin{align}
    {\rm NegBin}(Y_n | \psi_n, \nu) &= \frac{\Gamma(Y_n + \nu)}{\Gamma(Y_n + 1)\Gamma(\nu)}
    \left(\frac{\exp(\psi_n + \psi_0 - \log \nu)}{1 + \exp(\psi_n + \psi_0 - \log \nu)} \right)^{Y_n}
    \left(\frac{1}{1 + \exp(\psi_n + \psi_0- \log \nu)} \right)^{\nu} \\
    &\propto \frac{1}{2^{\nu}} e^{\tfrac{1}{2}(Y_n - \nu)(\psi_n + \psi_0- \log \nu)} 
    \EE_{p(\omega_n |Y_n + \nu, 0)} \left[ \exp(-\tfrac{1}{2} \omega_n (\psi_n + \psi_0 - \log \nu)^2) \right] \nonumber
\end{align}
where as before $\psi_n = \beta_0 + \beta_\gamma  \cdot X_{n\gamma}$
and $\psi_0$ is a user-specified offset. Here $\nu >0$ controls the overdispersion of the negative binomial likelihood.
We note that by construction the mean of ${\rm NegBin}(Y_n | \psi_n, \nu)$ is given by $\exp(\psi_n + \psi_0)$.
Thus $\psi_0$ (which can potentially depend on $n$) can be used to specify a prior mean for $Y$.
This is equivalent to adjusting the prior mean of the bias $\beta_0$ in the case of constant $\psi_0$.

Comparing to Sec.~\ref{sec:pg} we see that $\kappa_n$ is now given by $\kappa_n = \tfrac{1}{2}(Y_n - \nu)$.
When computing $\log p(Y | X, \gamma, \omega, \nu, \psi_0)$ the quantity $\ZZ$ now becomes 
$\ZZ_j = \sum_n X_{n,j} \left(\kappa_n - \omega_n(\psi_0 - \log \nu) \right)$, see Sec.~\ref{app:mll}. 
One also picks up an additional factor of 
\begin{align}
    \exp(\kappa \cdot (\psi_0 - \log \nu) - \tfrac{1}{2} \omega \cdot (\psi_0 - \log \nu)^2) 
\end{align}
In particular we have the formula
\begin{align}
    \label{eqn:mllnu}
    \log p(Y | X, \gamma, \omega, \nu, \psi_0) &= \tfrac{1}{2} \ZZ_\gone(\omega, \nu) \TT (X_\gone \TT \Omega X_\gone  + \tau \id_\gone)\inv \ZZ_\gone(\omega, \nu) \\
    &-\tfrac{1}{2} \log \det(X_\gone \TT \Omega X_\gone   + \tau \id_\gone)    -\tfrac{1}{2} \log \det(\tau \inv \id_\gone)  \nonumber \\
    &+\textstyle{\sum}_n  \left( \log \Gamma(Y_n + \nu) - \log \Gamma(\nu) - \log 2^{\nu} \right) \nonumber \\
    &+ \textstyle{\sum}_n  \kappa_n(\nu) \left( \psi_0 - \log \nu \right)   
    -\tfrac{1}{2}  \textstyle{\sum}_n \omega_n   \left( \psi_0 - \log \nu \right)^2     \nonumber
\end{align}

In our experiments we infer $\nu$, which we assume to be unknown. For simplicity we put a flat (i.e.~improper)
prior on $\log \nu$, although other choices are easily accommodated. To do so we modify the $\omega$ update
described in Sec.~\ref{app:omega} to a joint $(\omega, \log \nu)$ update. In more detail we use a simple
gaussian random walk proposal for $\log \nu$ with a user-specified scale (we use $0.03$ in our experiments).
Conditioned on a proposal $\log \nu^\prime$ we then sample a proposal $\omega^\prime$. Similar
to the binomial likelihood case, we do this by computing
\begin{align}
    \bhat(\gamma, \omega, \nu) \equiv \EE_{p(\beta | \gamma, \omega, \nu, \DD)} \left[ \beta \right]
\end{align}
and use a proposal distribution $\omega^\prime \sim p(\cdot | \gamma, \bhat, \nu^\prime, \DD)$.
In the negative binomial case the formula for $\bhat$ in Eqn.~\ref{eqn:bhatdefn} becomes
\begin{align}
    \bhat(\gamma, \omega, \nu) = (X_\gone \TT \Omega X_\gone + \tau \mathbb{1}_\gabsone)\inv X_\gone \TT
    \left(\kappa - \omega(\psi_0 - \log \nu) \right)
    \in \RR^{\gabsone}
\end{align}
Additionally the proposal distribution is given by 
\begin{align}
    p(\omega^\prime | \gamma, \bhat(\gamma, \omega, \nu), \nu^\prime, \DD) = 
    {\rm PG}(\omega^\prime | Y + \nu^\prime, \psihat(\gamma, \omega, \nu))
\end{align}
The acceptance probability can then be computed as in Sec.~\ref{app:omega}, although in this 
case the resulting formulae are somewhat more complicated because of the need to 
keep track of $\nu$ and $\nu^\prime$ as well as the fact that there is less scope for cancellations
so that we need to compute quantities like $\Gamma(\nu)$. Happily, just like in the binomial regression case,
the acceptance probability can be computed without recourse to the P\olya-Gamma density.
In more detail the acceptance probability can be computed with help of the following expressions.
\begin{align}
\label{eqn:appalphaomeganu}
\alpha(\omega, \nu  \! \to \! \omega^\prime, \nu^\prime | \gamma) &=  \min\left(1,
     \tilde{\alpha}(\omega, \nu  \! \to \! \omega^\prime, \nu^\prime | \gamma) \right) 
\end{align}
where $\tilde{\alpha}(\omega, \nu  \! \to \! \omega^\prime, \nu^\prime | \gamma)$ is given by 
\begin{align}
\tilde{\alpha}(\omega, \nu  \! \to \! \omega^\prime, \nu^\prime | \gamma) = \tilde{\alpha}_1 \times \tilde{\alpha}_2 \times \tilde{\alpha}_3
\end{align}
with
\begin{align}
\tilde{\alpha}_1 &= 2^{N(\nu^\prime - \nu)}  \frac{p(Y | X, \gamma, \omega^\prime, \nu^\prime, \psi_0)}{p(Y | X, \gamma, \omega, \nu, \psi_0)} \\
\tilde{\alpha}_2 &= \frac{\textstyle{\prod_n} \left( e^{\psihat_n(\gamma, \omega, \nu) + \psi_0 - \log \nu^\prime} \right)^{Y_n}}{\textstyle{\prod_n} \left( e^{\psihat_n(\gamma, \omega^\prime, \nu^\prime) + \psi_0 - \log \nu} \right)^{Y_n}} 
\frac{\textstyle{\prod_n} \left(1 +  e^{\psihat_n(\gamma, \omega^\prime, \nu^\prime) + \psi_0 - \log \nu} \right)^{Y_n + \nu}}{\textstyle{\prod_n} \left(1 +  e^{\psihat_n(\gamma, \omega, \nu) + \psi_0 - \log \nu^\prime} \right)^{Y_n + \nu^\prime}} \\
\tilde{\alpha}_3 &= \frac{\textstyle{\prod_n}e^{\kappa_n(\nu)(\psihat_n(\gamma, \omega^\prime, \nu^\prime) + \psi_0 - \log \nu)}}{\textstyle{\prod_n}e^{\kappa_n(\nu^\prime)(\psihat_n(\gamma, \omega, \nu) + \psi_0 - \log \nu^\prime)}} 
\frac{\textstyle{\prod_n}e^{-\tfrac{1}{2}\omega_n(\psihat_n(\gamma, \omega^\prime, \nu^\prime) + \psi_0 - \log \nu)^2}}{\textstyle{\prod_n}e^{-\tfrac{1}{2}\omega^\prime_n(\psihat_n(\gamma, \omega, \nu) + \psi_0 - \log \nu^\prime)^2}} 
\end{align}
The correctness of these formulae can be checked numerically by comparing to the P\olya-Gamma density in regimes where the density can be easily and reliably computed. This is equally true for the binomial likelihood case.

\subsection{Additional figures and tables}
\label{app:fig}

\begin{figure*}[ht]
    \includegraphics[width=0.33\linewidth]{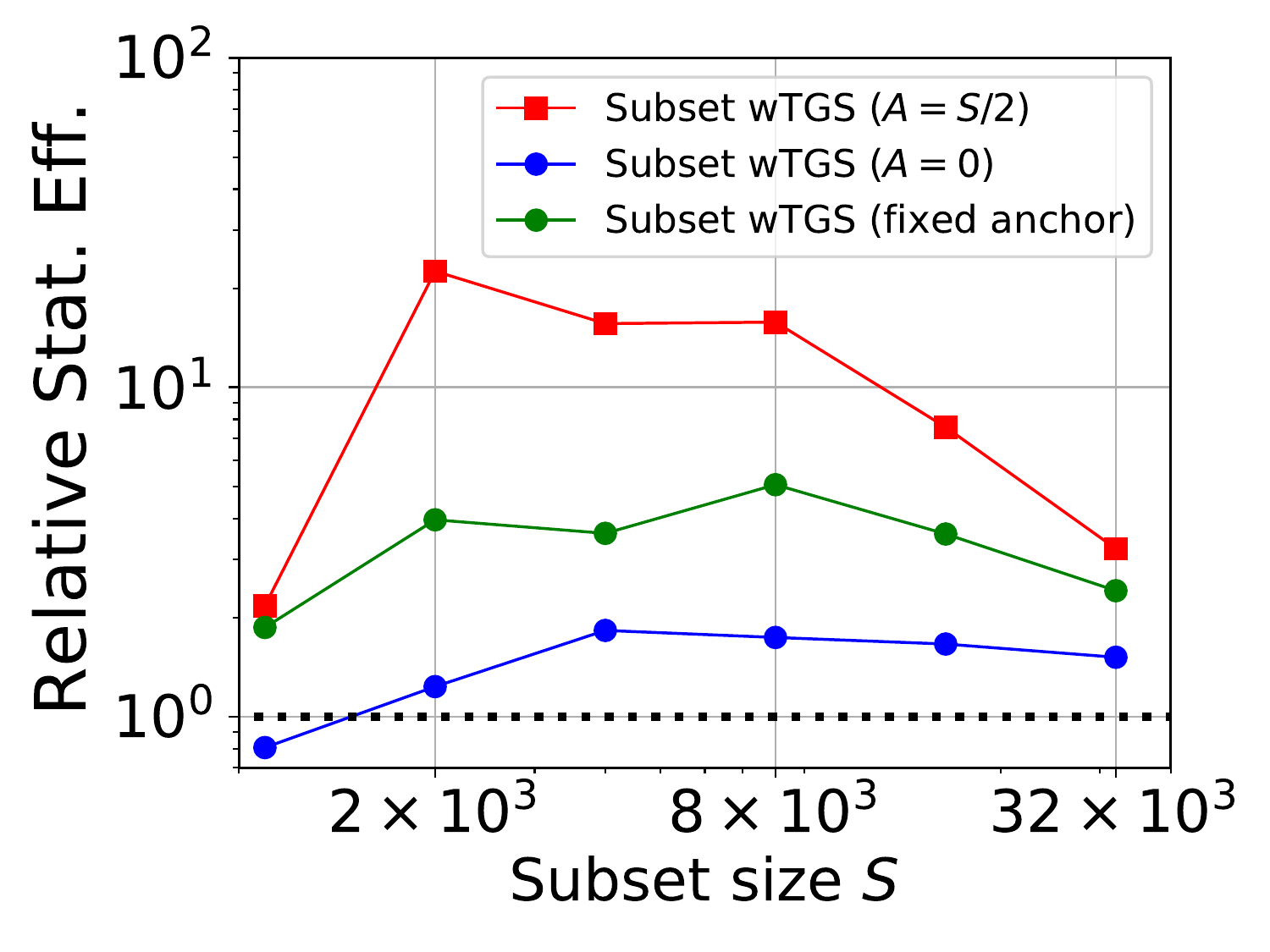}
    \includegraphics[width=0.33\linewidth]{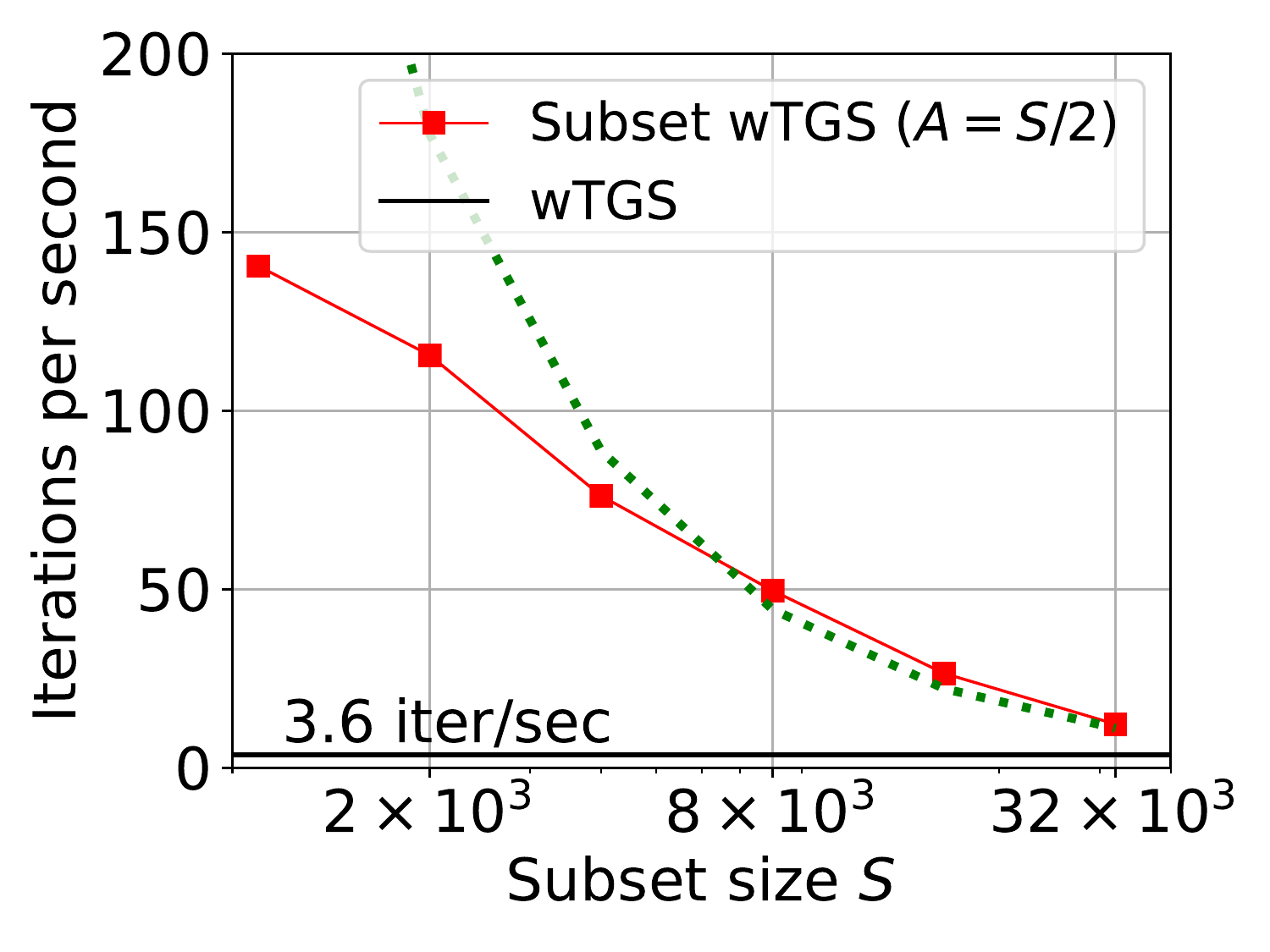}
    \includegraphics[width=0.33\linewidth]{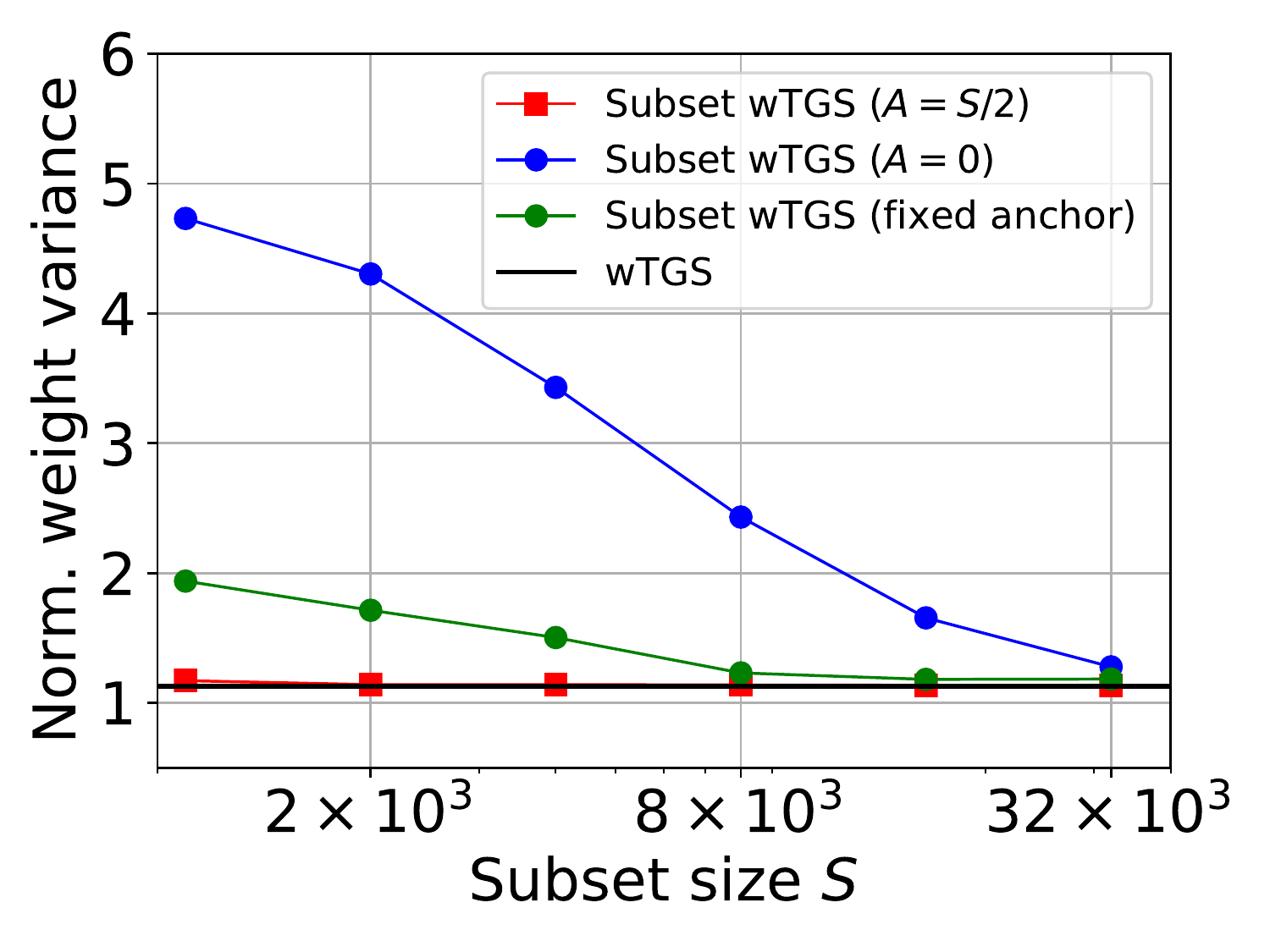}
    \caption{We report additional results for the experiment in Sec.~\ref{sec:largepexp} with $P=98385$.
    These results are directly analogous to the results reported in the main text except we use synthetic
    data with a lower signal to noise ratio.
    In addition to Subset wTGS with our default setting of $A=S/2$, we also report results with $A=0$ (i.e.~no anchor set)
    and a fixed anchor set with $A=S/2$ (i.e.~the anchor set is not updated during burn-in; instead $\AAA$ is chosen
    with the initialization heuristic in Algorithm \ref{algo}).
    See Sec.~\ref{app:exp} for additional details.
    {\bf (Left)}  We depict the relative statistical efficiency of Subset wTGS with subset size $S$ compared to wTGS.
    As in the main text, we restrict our focus to covariates with PIPs above a threshold of $0.001$.
    Subset wTGS with $A=S/2$ exhibits large gains in statistical efficiency relative to wTGS.
    The benefits of adapting $\AAA$ are also apparent: we expect $\AAA$ adapation to be particularly important for noisy datasets.
    {\bf (Middle)} We depict the number of iterations per second (IPS) for Subset wTGS as a function of $S$. The green dotted line
    depicts the IPS that would be expected if the latter scaled like $S^{-1}$.
    {\bf (Right)} We depict the variance of weights $\{ \rho^{(t)} \}$ obtained by running Subset wTGS.
    For the purposes of this figure, the weights are normalized so that the mean weight is equal to unity.
    We see that Subset wTGS with $A=0$ has substantially higher variance. Moreover, 
    in this regime (which is characterized by larger observation noise) the variance of Subset wTGS weights for $A=S/2$ is nearly
    identical to the variance of wTGS weights; importantly, both these variances are moderate.
    }
\label{fig:applargepnoisy}
\end{figure*}

\begin{figure*}[ht]
    \includegraphics[width=0.33\linewidth]{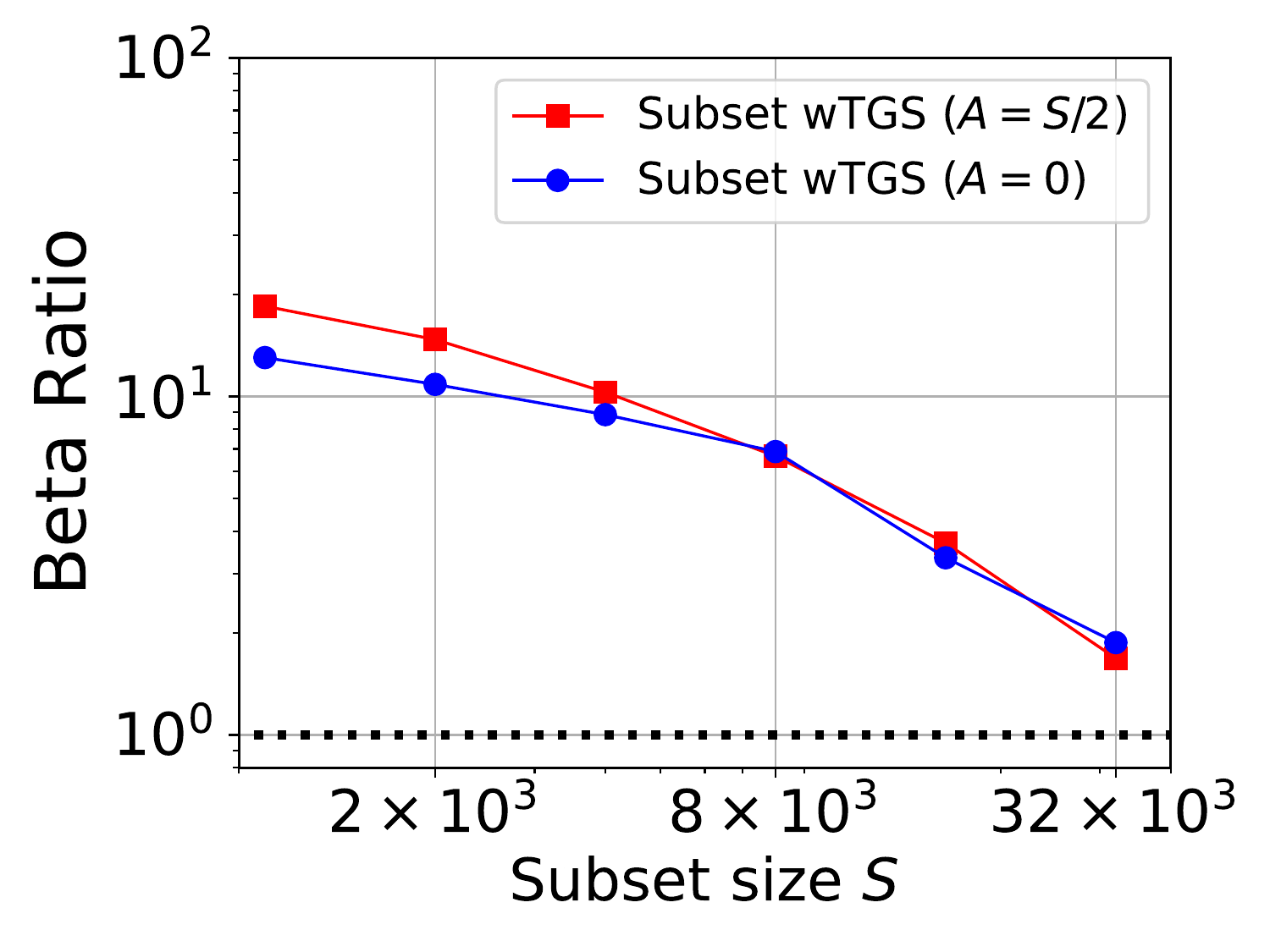}
    \includegraphics[width=0.33\linewidth]{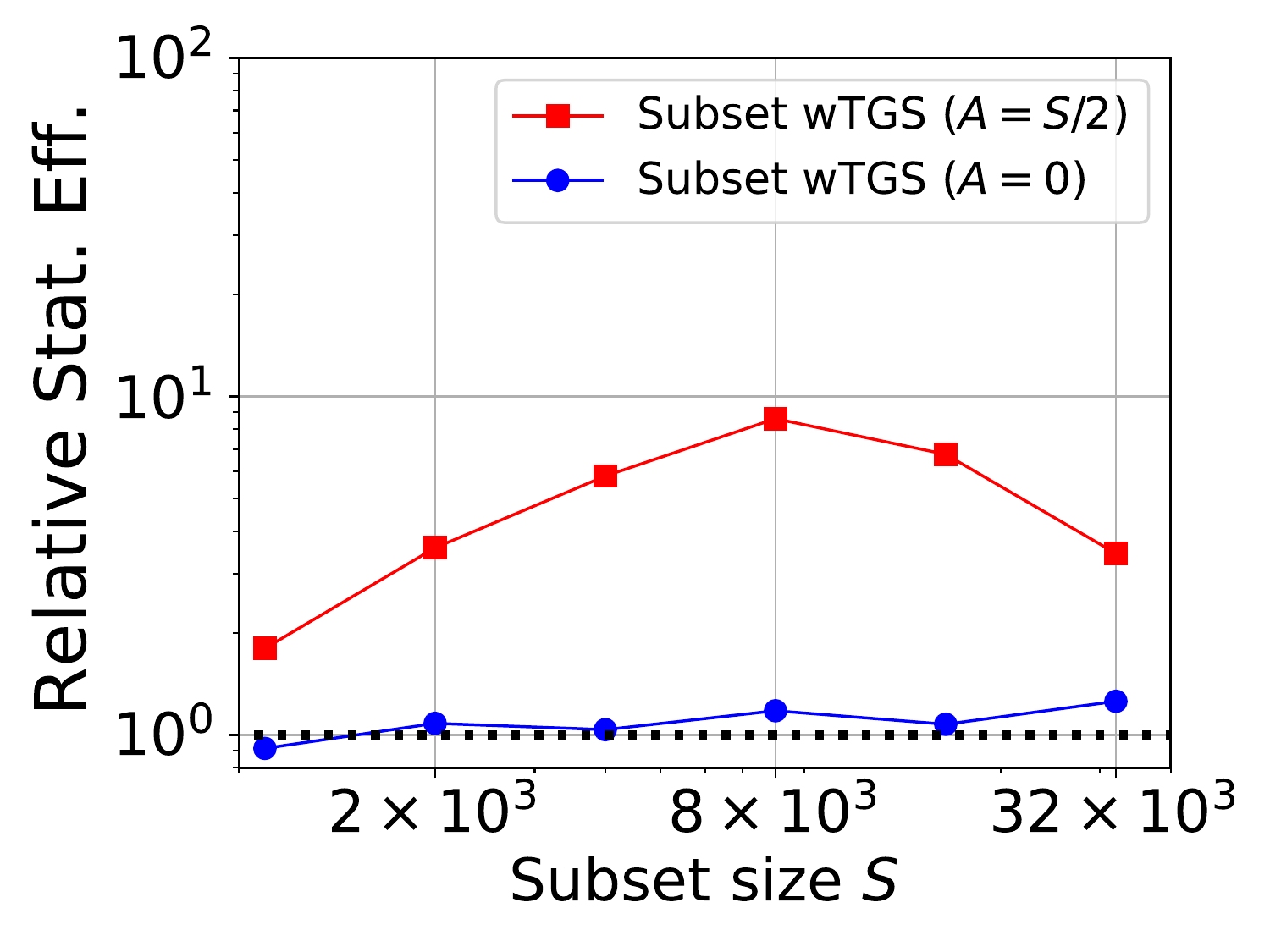}
    \includegraphics[width=0.33\linewidth]{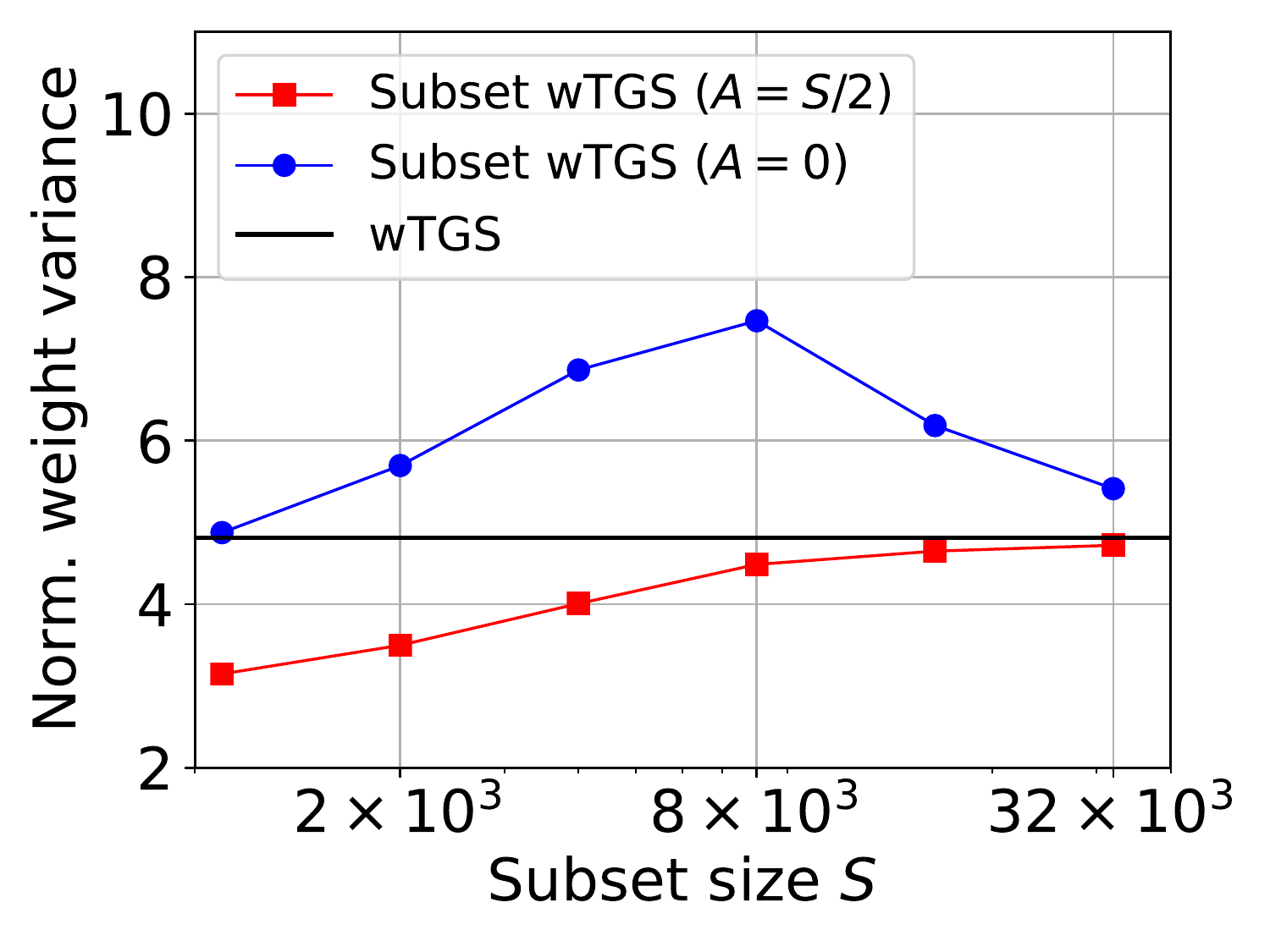}
    \caption{We report additional results for the experiment in Sec.~\ref{sec:largepexp} with $P=98385$.
    See Sec.~\ref{sec:exp} and Sec.~\ref{app:exp} for additional details.
    {\bf (Left)}   We report a statistical efficiency ratio (relative to wTGS) for the inferred coefficients $\beta$.
    {\bf (Middle)} This figure is identical to the top panel in Fig.~\ref{fig:largep}, except the reported
    statistical efficiency is w.r.t.~\emph{all} covariates, instead of those with PIP above $0.001$. Since the
    resulting (relative) statistical efficiency is somewhat less than that in Fig.~\ref{fig:largep}, we see that some of the improved
    statistical efficiency of Subset wTGS is due to a variance trade-off between large PIP and small PIP covariates. 
    We hypothesize that this is largely driven by the partial Rao-Blackwellization of the PIP estimator used in Subset wTGS:
     since low PIP covariates $i$ are in $\SSS$ only infrequently, the corresponding PIP estimate does not benefit much from 
     Rao-Blackwellization and is consequently higher variance.
    Since there is generally no compelling need to obtain precise PIP estimates of low PIP covariates (e.g.~distinguishing a PIP of $1.1 \times 10^{-5}$ from $1.2 \times 10^{-5}$), this trade-off is well worth the resulting speed-ups.
    {\bf (Right)}  We depict the variance of weights $\{ \rho^{(t)} \}$ obtained by running Subset wTGS.
    For the purposes of this figure, the weights are normalized so that the mean weight is equal to unity.
    We see that Subset wTGS with $A=0$ has substantially higher variance, which explains its suboptimal performance.
    Importantly the variance of Subset wTGS with $A=\tfrac{1}{2}S$ is moderate and is indeed less than that of wTGS.
    We also note that a variance of $4$ corresponds to a situation in which roughly half the weights are equal to $2$ and the 
    remainder are very small; this is indeed approximately what we observe in practice. A non-negligible number of 
    low weight samples is the price we pay for the exploration enabled by tempering.
    }
\label{fig:applargep}
\end{figure*}

\begin{figure}[ht]
\centering
\begin{minipage}[b]{0.38\linewidth}
    \includegraphics[width=\linewidth]{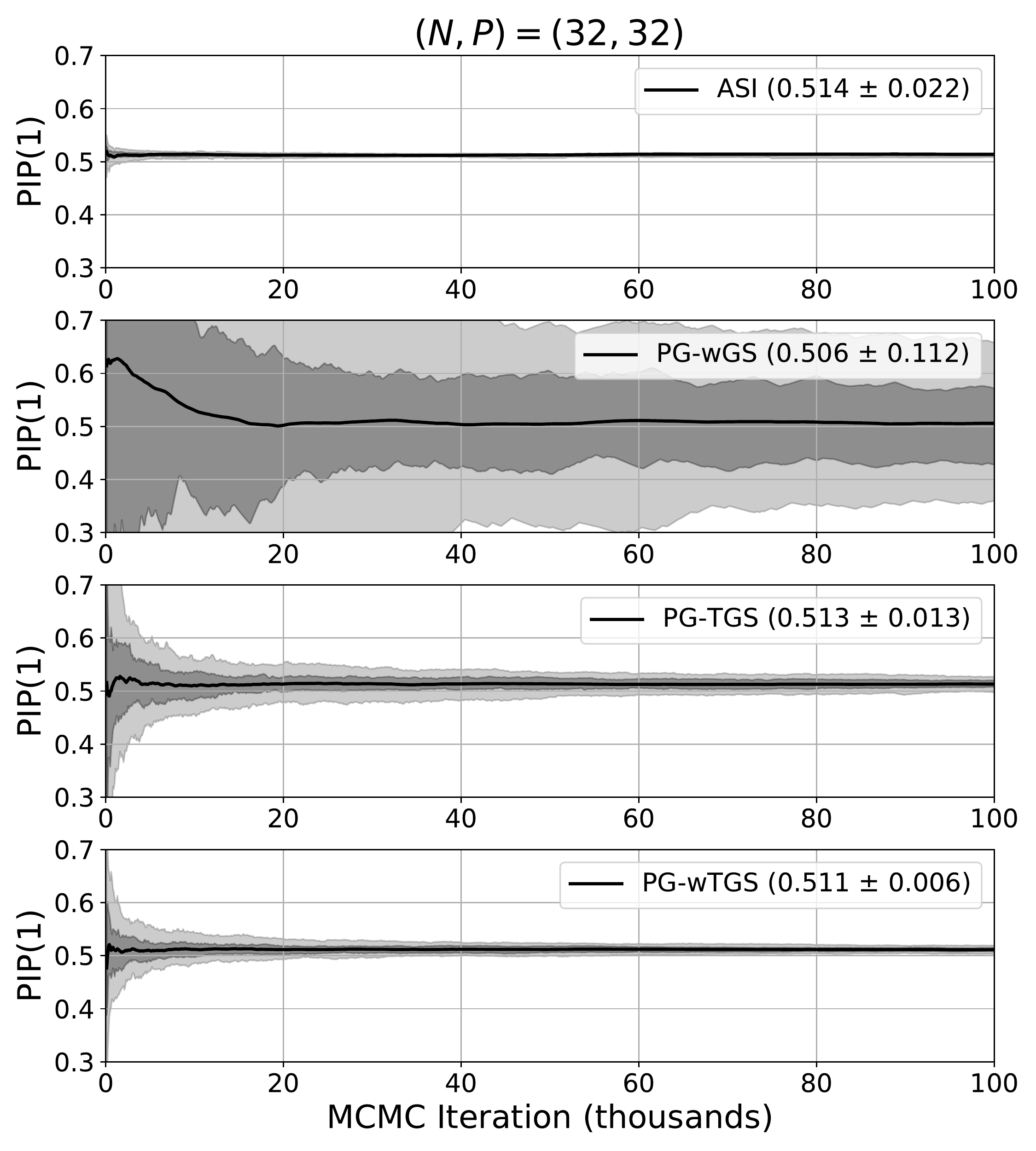}
\label{fig:corr32}
\end{minipage}
\begin{minipage}[b]{0.38\linewidth}
    \includegraphics[width=\linewidth]{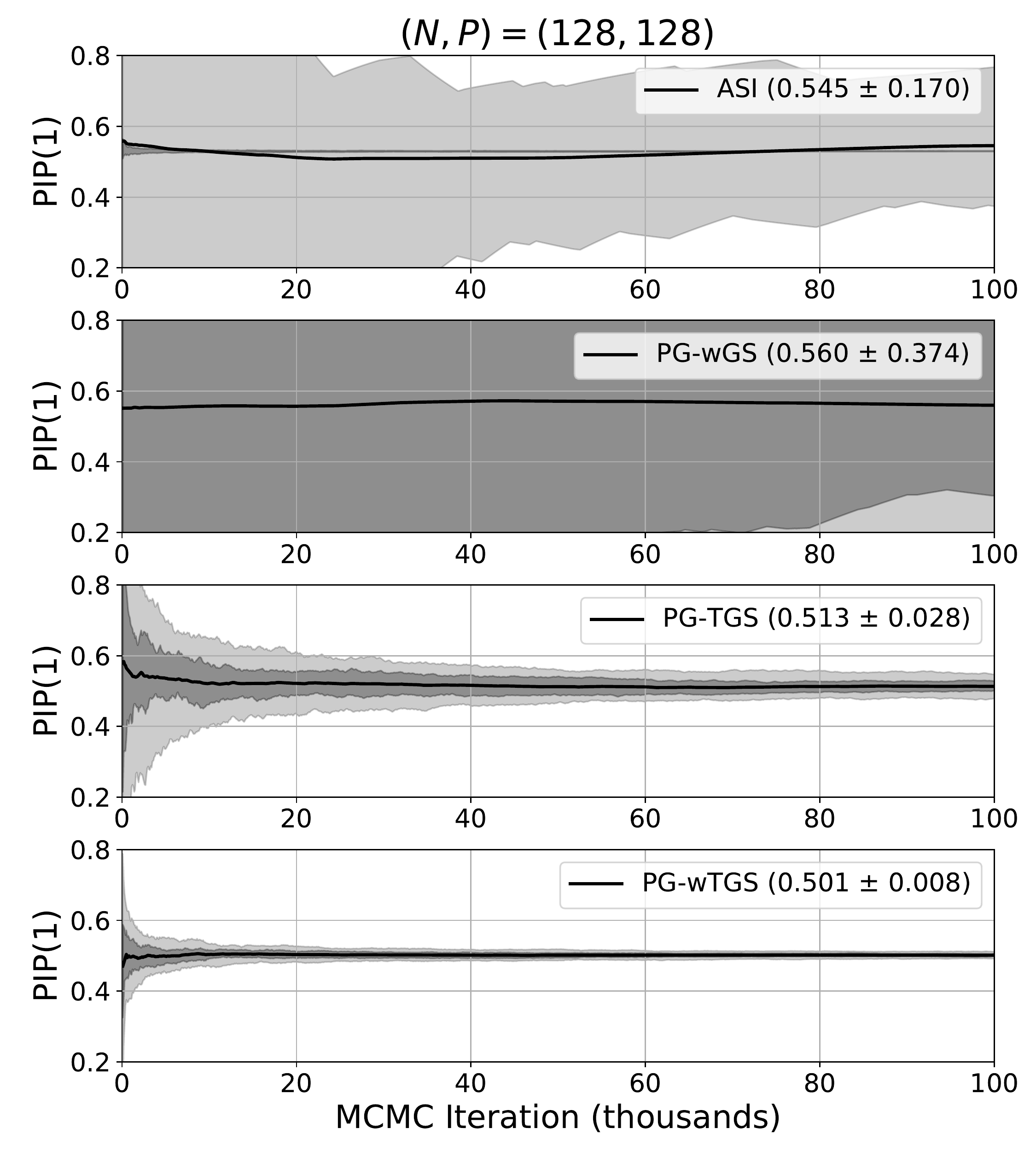}
\label{fig:corr128}
\end{minipage}
\quad
\begin{minipage}[b]{0.38\linewidth}
    \includegraphics[width=\linewidth]{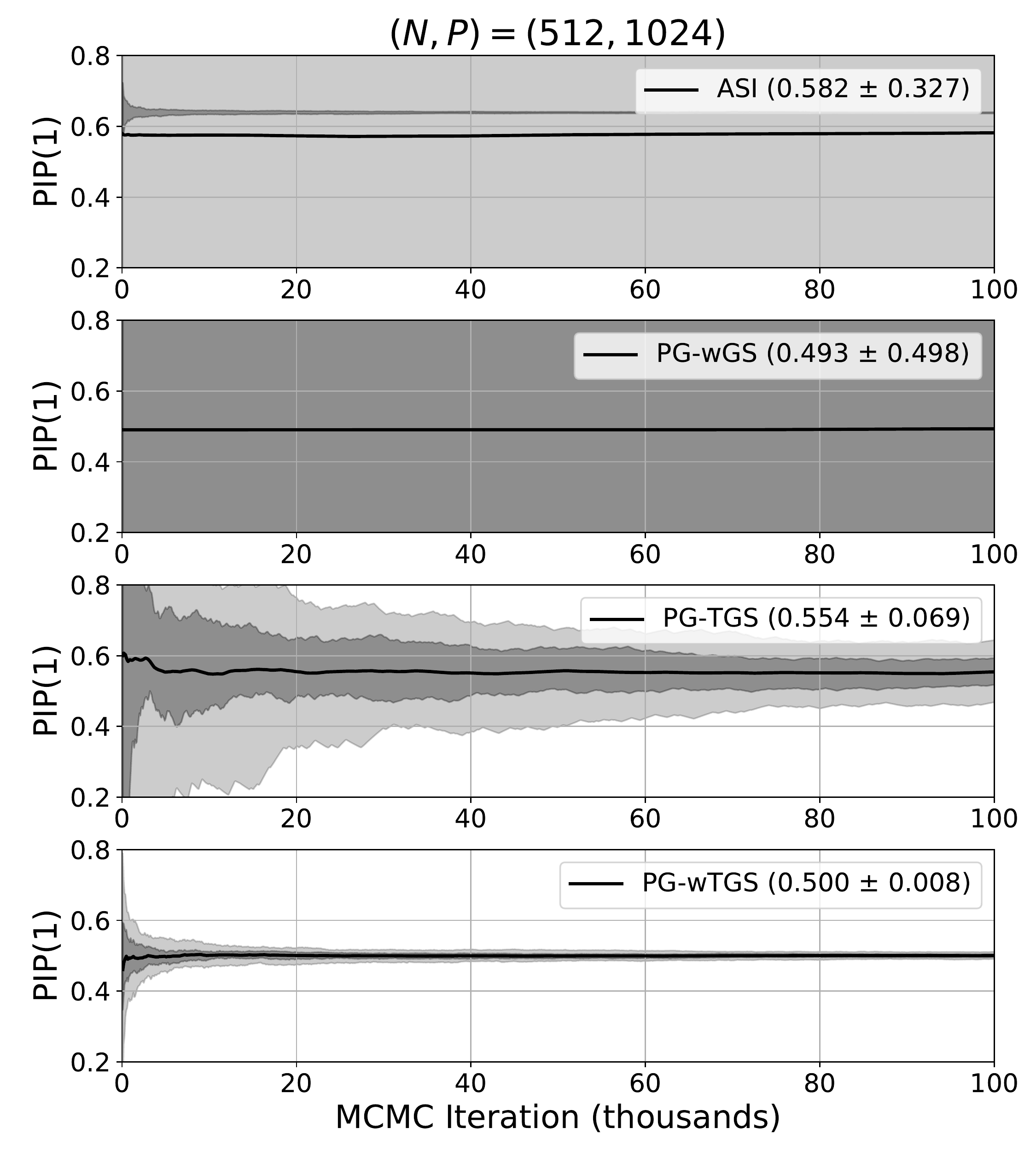}
\label{fig:corr512a}
\end{minipage}
\quad
\begin{minipage}[b]{0.38\linewidth}
    \includegraphics[width=\linewidth]{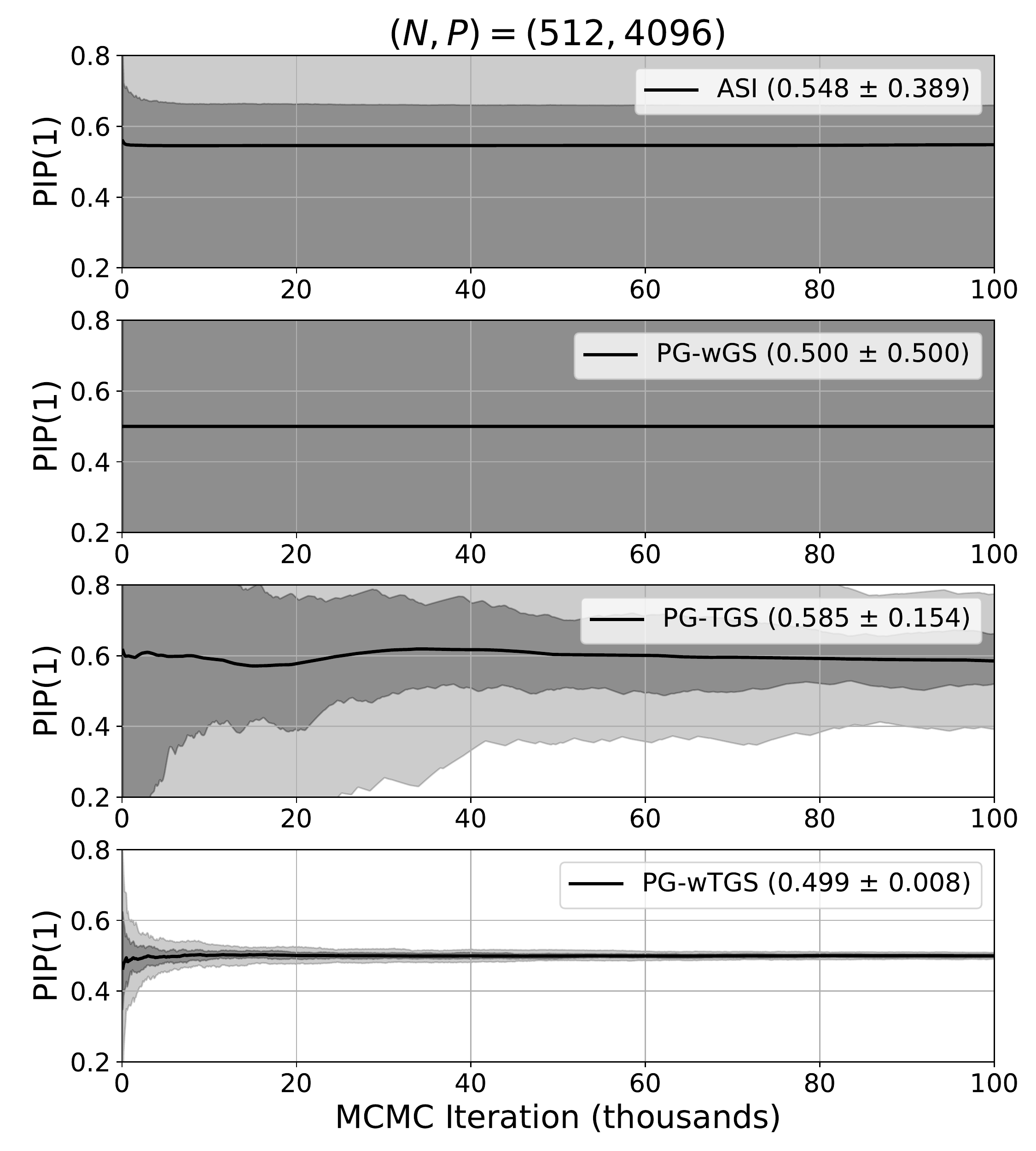}
\label{fig:corr512b}
\end{minipage}
    \caption{This is a companion figure to Fig.~\ref{fig:corrviolin} in the main text. 
    We depict posterior inclusion probability (PIP) estimates for the first covariate in the scenario
    described in Sec.~\ref{sec:corr} for four different MCMC methods and four different values of $(N, P)$.
    At each iteration $t$ the PIP is computed using all samples obtained through iteration $t$.
    The mean PIP is depicted with a solid black line and light and dark grey confidence intervals
    denote $10\%\!-\!90\%$ and $30\%\!-\!70\%$ quantiles, respectively.
    The true PIP is almost exactly $\tfrac{1}{2}$.
    In each case we run 100 independent chains.
    For each method we also report the final PIP estimate (mean and standard deviation) in parentheses.
    }
\label{fig:tracecorrapp}
\end{figure}

\begin{figure*}[ht]
    \centering
    \includegraphics[width=0.45\linewidth]{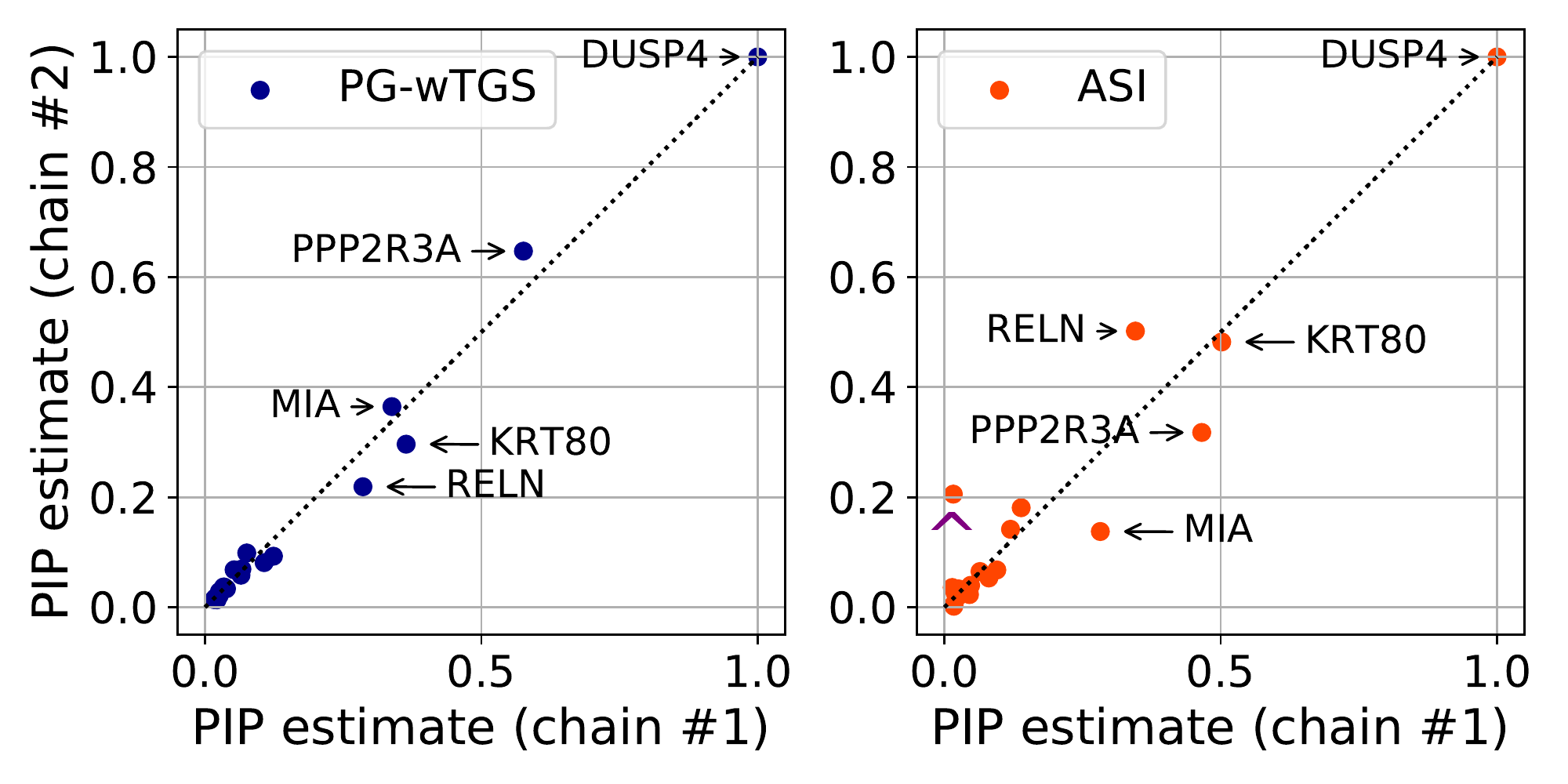}
    \includegraphics[width=0.45\linewidth]{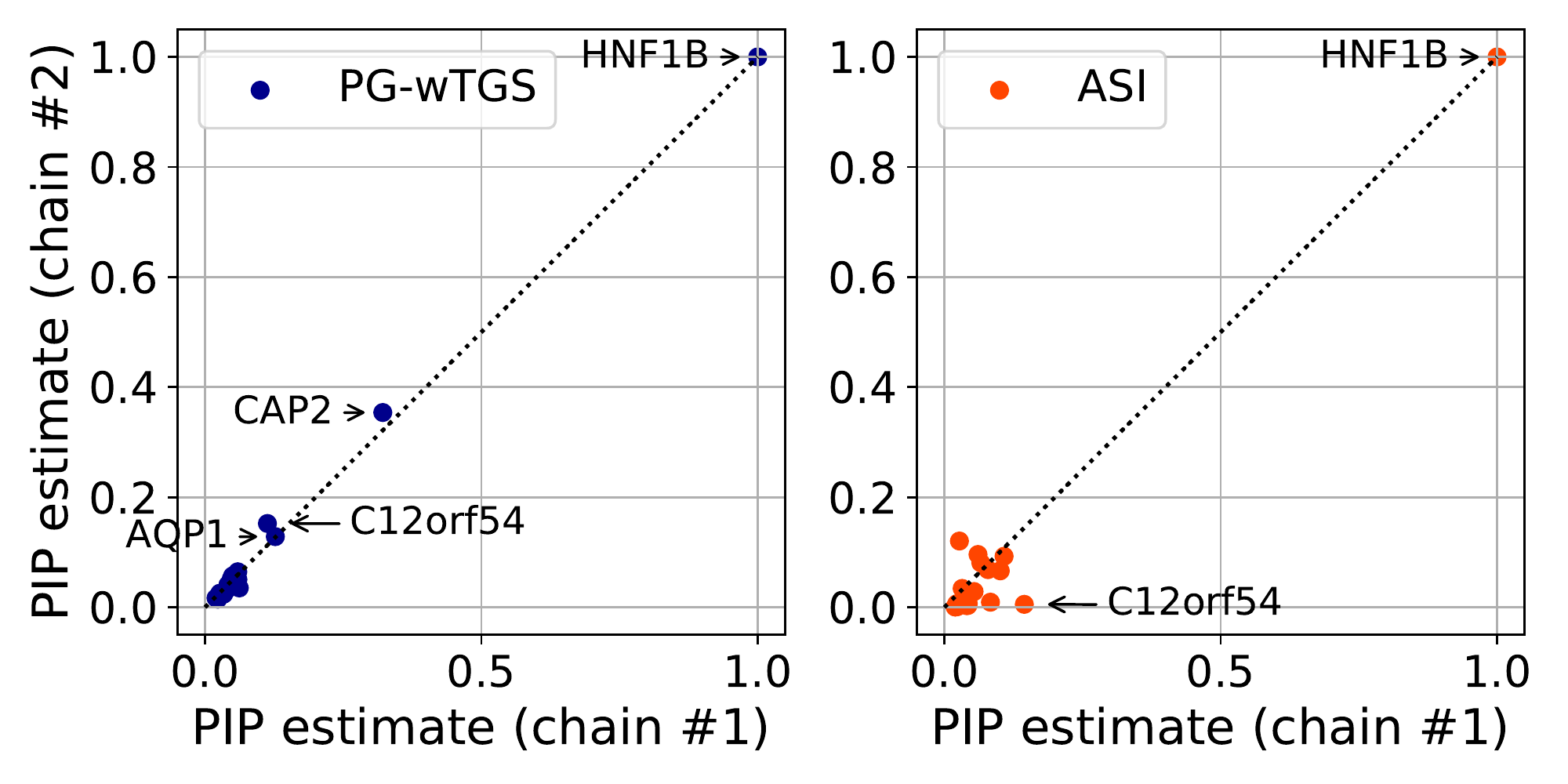}
    \caption{We depict PIP estimates for two independent MCMC chains
    for two cancer datasets (left: DUSP4; right: HNF1B) using two MCMC methods. For each method we depict
    the top 20 PIPs from chain \#1 paired with the estimate from chain \#2.
    The PG-wTGS estimates show much better inter-chain concordance.
    For example, the PIPs obtained with ASI for KRT7 on the DUSP4 dataset
    (marked with a purple caret \^{}) differ by a factor of $12.8$ between the two chains,
    while the two PG-wTGS estimates are $0.025$ and $0.021$.
    Similarly the PIPs obtained with ASI for C12orf54 on the HNF1B dataset
    differ by a factor of 26.9, while the two PG-wTGS estimates are $0.113$ and $0.152$.
    See Sec.~\ref{sec:cancer} for details.
    }
\label{fig:cancer}
\end{figure*}

\begin{figure}[ht]
    \includegraphics[width=0.49\linewidth]{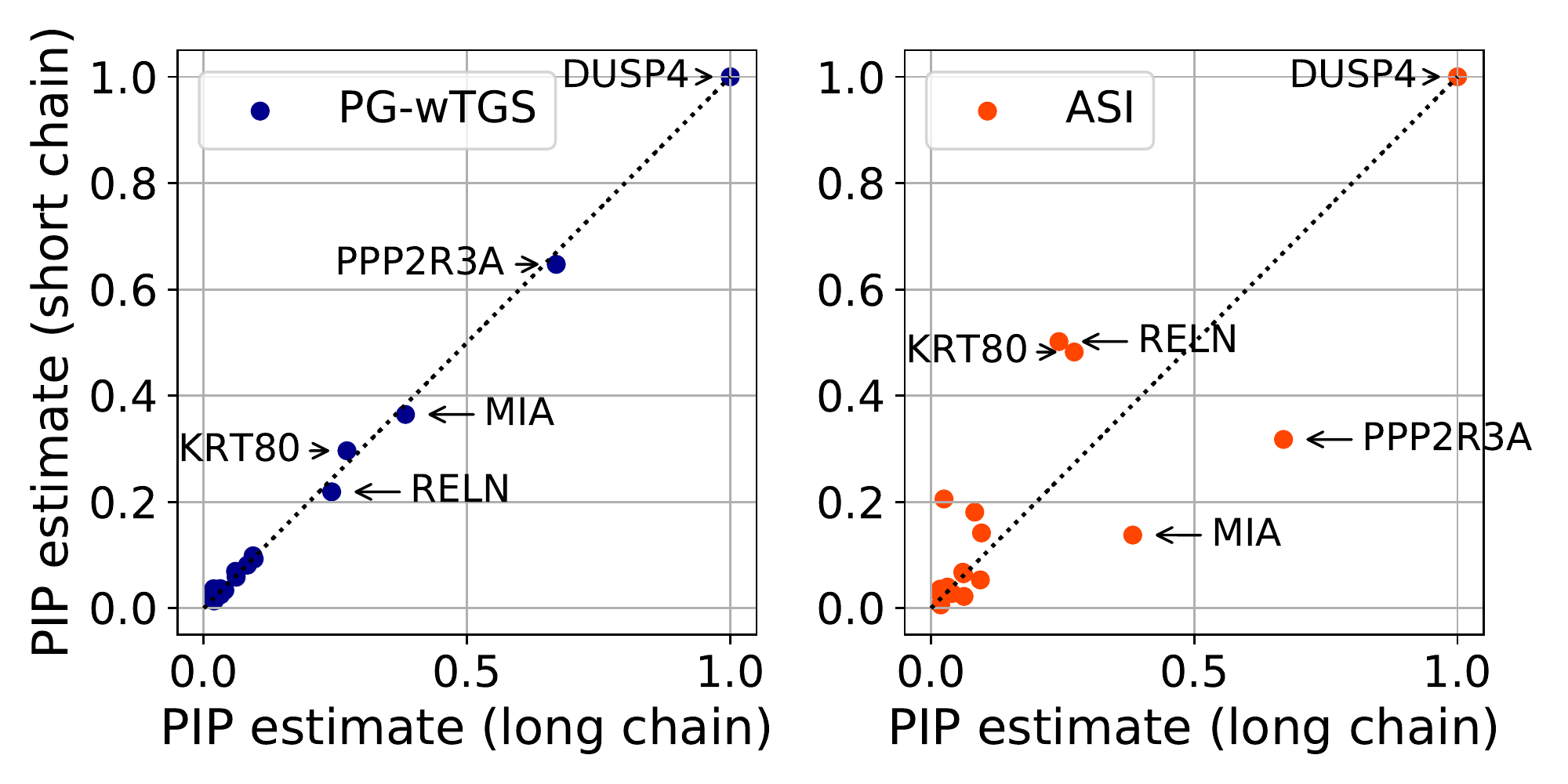}
    \includegraphics[width=0.49\linewidth]{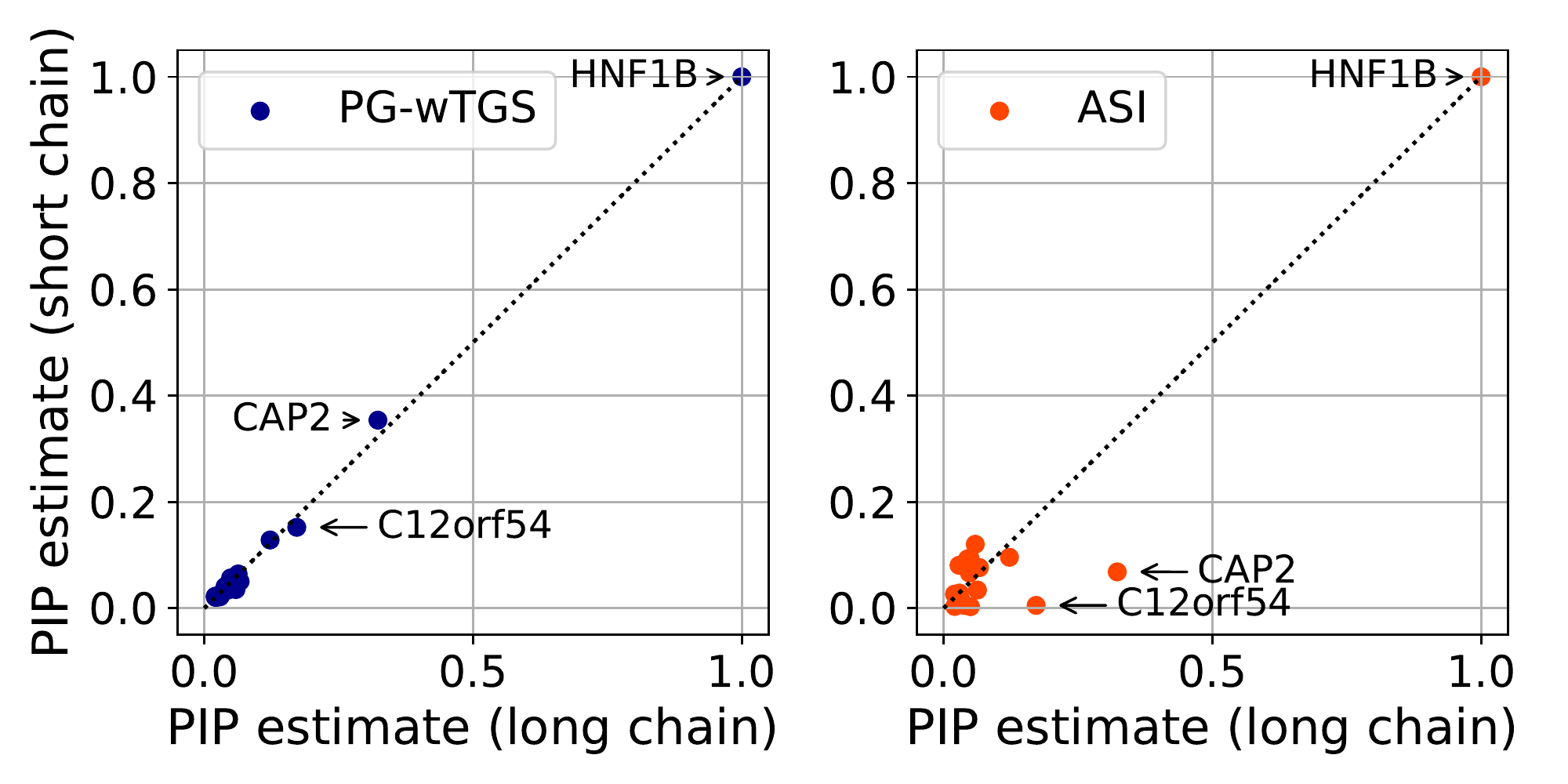}
    \caption{In this companion figure to Fig.~\ref{fig:cancerbigpartial} we compare PIP estimates 
    obtained from short PG-wTGS and ASI chains with $2.5 \times 10^5$ samples
    to a long PG-wTGS chain with $5 \times 10^6$ samples.
    For each method we depict the top 20 PIPs from the long chain paired with estimates from
    the short chains. Note that this figure is identical to Fig.~\ref{fig:cancerbigpartial} except the two short chains
    are independent of the two short chains in Fig.~\ref{fig:cancerbigpartial}. 
    The PG-wTGS estimates obtained with the short chains are significantly more accurate than is the case for ASI. 
    See Sec.~\ref{sec:cancer} for details. 
    }
\label{fig:cancerbigfull}
\end{figure}

Additional figures for the first experiment in Sec.~\ref{sec:largepexp} are depicted in Fig.~\ref{fig:applargepnoisy}-\ref{fig:applargep}.
Additional trace plots for the experiment in Sec.~\ref{sec:corr} are depicted in Fig.~\ref{fig:tracecorrapp}.
In Table \ref{table:cancer} we report PIP estimates for top hits in the cancer experiment
in Sec.~\ref{sec:cancer}; 
we also include Fig.~\ref{fig:cancer} and Fig.~\ref{fig:cancerbigfull}, where the latter is a companion of Fig.~\ref{fig:cancerbigpartial}.

\definecolor{Gray}{gray}{0.80}

\begin{table}[]
 \resizebox{.485\textwidth}{!}{
\begin{tabular}{l|l|l|l}
\rowcolor{Gray} Gene & PG-wTGS-5M & PG-wTGS-250k & ASI-250k \\ \hline
DUSP4 & 1.000 & 1.000 / 1.000 & 1.000 / 1.000 \\ \hline
PPP2R3A & 0.669 & 0.576 / 0.647 & 0.466 / 0.318 \\ \hline
MIA & 0.383 & 0.338 / 0.365 & 0.282 / 0.138 \\ \hline
KRT80 & 0.272 & 0.364 / 0.296 & 0.502 / 0.482 \\ \hline
RELN & 0.243 & 0.286 / 0.219 & 0.346 / 0.502 \\ \hline
ZNF132 & 0.096 & 0.124 / 0.093 & 0.120 / 0.142 \\ \hline
TRIM51 & 0.094 & 0.075 / 0.099 & 0.080 / 0.053 \\ \hline
ZNF471 & 0.083 & 0.107 / 0.081 & 0.139 / 0.181 \\ \hline
S100B & 0.063 & 0.053 / 0.068 & 0.028 / 0.022 \\ \hline
ZNF571 & 0.062 & 0.065 / 0.058 & 0.064 / 0.065 \\ \hline
ZNF304 & 0.060 & 0.067 / 0.069 & 0.095 / 0.068 \\ \hline
ZNF772 & 0.040 & 0.039 / 0.034 & 0.044 / 0.028 \\ \hline
RXRG & 0.032 & 0.026 / 0.025 & 0.010 / 0.028 \\ \hline
ZNF17 & 0.031 & 0.033 / 0.037 & 0.047 / 0.040 \\ \hline
ZNF134 & 0.026 & 0.026 / 0.029 & 0.026 / 0.033 \\ \hline
KRT7 & 0.025 & 0.025 / 0.021 & 0.016 / 0.206 \\ \hline
ZNF71 & 0.020 & 0.022 / 0.014 & 0.045 / 0.023 \\ \hline
CCIN & 0.019 & 0.035 / 0.037 & 0.019 / 0.026 \\ \hline
ZNF419 & 0.018 & 0.019 / 0.017 & 0.018 / 0.006 \\ \hline
ZMYM3 & 0.017 & 0.016 / 0.027 & 0.014 / 0.036 \\ \hline
\end{tabular} 
}
\resizebox{.505\textwidth}{!}{
\begin{tabular}{l|l|l|l}
\rowcolor{Gray} Gene & PG-wTGS-5M & PG-wTGS-250k & ASI-250k \\ \hline
HNF1B & 1.000 & 1.000 / 1.000 & 1.000 / 1.000 \\ \hline
CAP2 & 0.323 & 0.322 / 0.354 & 0.079 / 0.068 \\ \hline
C12orf54 & 0.172 & 0.113 / 0.152 & 0.145 / 0.005 \\ \hline
AQP1 & 0.122 & 0.127 / 0.128 & 0.061 / 0.096 \\ \hline
FAM43B & 0.067 & 0.059 / 0.050 & 0.013 / 0.077 \\ \hline
KLRF1 & 0.063 & 0.059 / 0.065 & 0.032 / 0.034 \\ \hline
ARMC4 & 0.059 & 0.062 / 0.035 & 0.027 / 0.120 \\ \hline
SERPINE1 & 0.050 & 0.047 / 0.052 & 0.040 / 0.003 \\ \hline
CLIC6 & 0.049 & 0.050 / 0.057 & 0.013 / 0.094 \\ \hline
GSDME & 0.048 & 0.056 / 0.050 & 0.101 / 0.066 \\ \hline
UGCG & 0.044 & 0.042 / 0.034 & 0.108 / 0.093 \\ \hline
NEK6 & 0.039 & 0.041 / 0.041 & 0.019 / 0.005 \\ \hline
SERPINA10 & 0.032 & 0.027 / 0.026 & 0.007 / 0.017 \\ \hline
ECH1 & 0.029 & 0.028 / 0.022 & 0.054 / 0.029 \\ \hline
KIF1C & 0.029 & 0.034 / 0.024 & 0.066 / 0.081 \\ \hline
S100A4 & 0.028 & 0.029 / 0.025 & 0.083 / 0.009 \\ \hline
MSANTD3 & 0.023 & 0.029 / 0.023 & 0.005 / 0.006 \\ \hline
PLIN3 & 0.023 & 0.019 / 0.020 & 0.043 / 0.007 \\ \hline
IL4R & 0.021 & 0.018 / 0.021 & 0.016 / 0.003 \\ \hline
SHBG & 0.020 & 0.016 / 0.022 & 0.005 / 0.027 \\ \hline
\end{tabular}
}
    \vspace{2mm}
\caption{These tables are companions to Fig.~\ref{fig:cancerbigpartial}, Fig.~\ref{fig:cancer}, and Fig.~\ref{fig:cancerbigfull}. 
    We depict PIP estimates for DUSP4 (left) and HNF1B (right). In each case we
    include the result from a PG-wTGS run with five million samples as well as two shorter
    runs from PG-wTGS and ASI (with the two results separated by a slash). We depict the top 20 genes as determined by the long PG-wTGS run.
    The much lower variance and higher accuracy of PG-wTGS are apparent. Indeed if we take
    the top 20 PIP estimates from the long run as truth then we can compute the mean absolute error (MAE)
    of the short run estimates. The resulting MAEs are $0.007$ ($0.014$) for PG-wTGS and $0.043$ ($0.061$) for ASI for
    HNF1B (DUSP4), respectively. In other words the ASI MAE is about five times larger than the PG-wTGS MAE.}
\label{table:cancer}
\end{table}

\subsection{Additional experiments}
\label{app:addexp}

\subsubsection{Subset wTGS and cancer data}
\label{app:subcan}

\begin{figure}[ht]
\centering
    \includegraphics[width=0.30\linewidth]{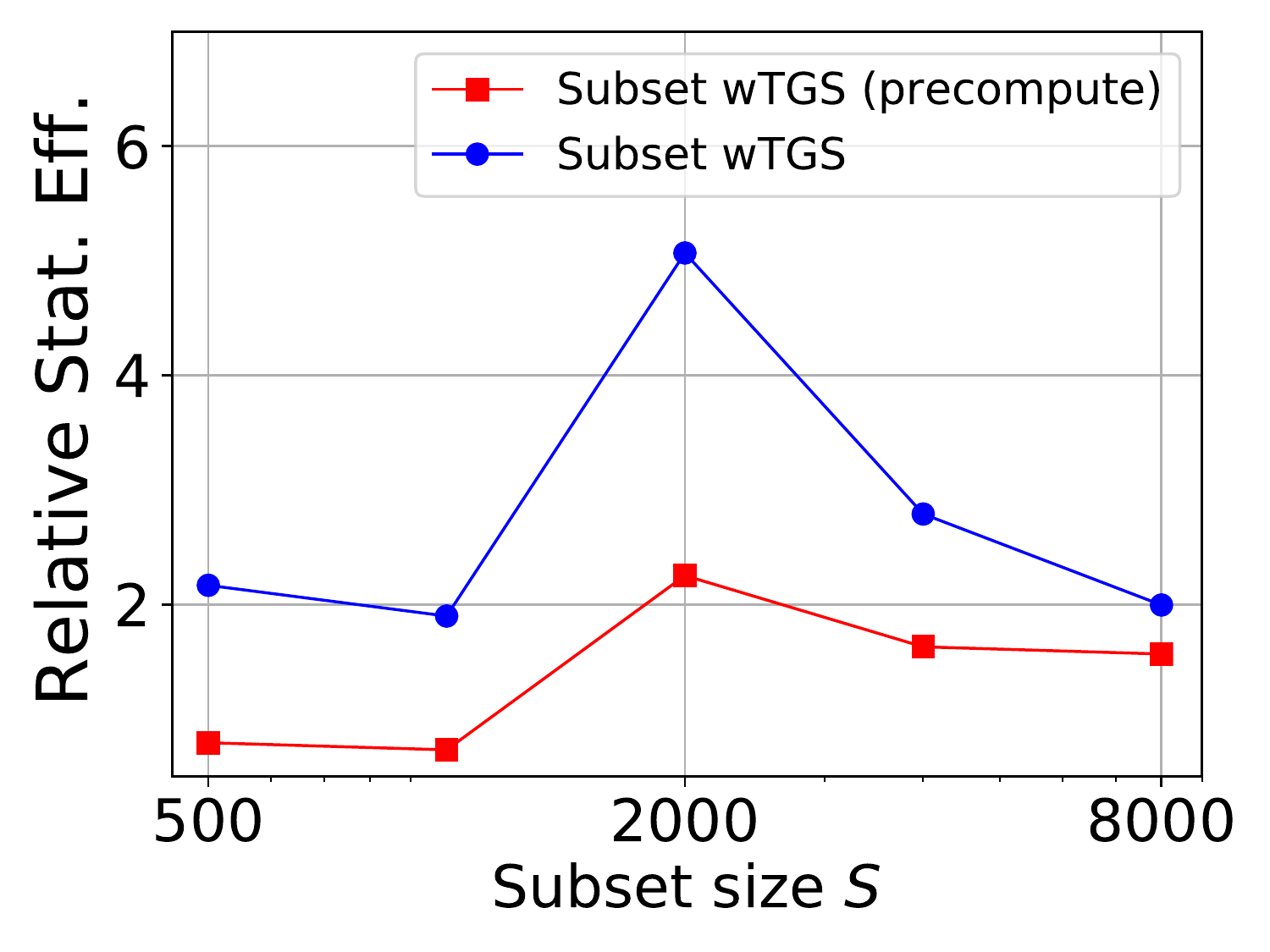}
    \includegraphics[width=0.30\linewidth]{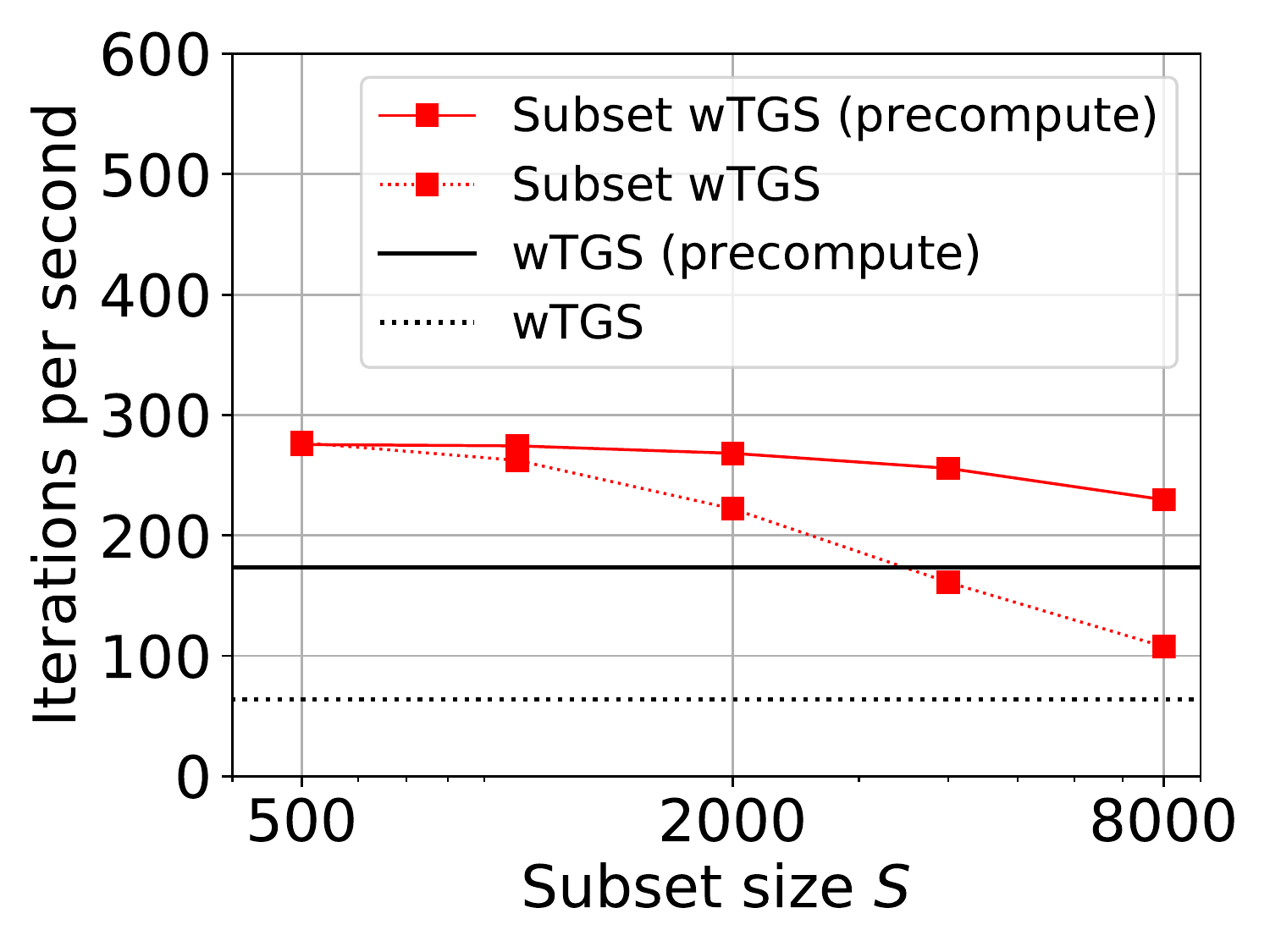}
\caption{
    We explore the moderately large $P$ regime using cancer data from a ZEB2 gene knockout experiment. 
    Since $P=17273$ is somewhat moderate, and since we are doing linear regression and not generalized linear regression
    with a non-linear link function, the covariate covariance matrix $X^{\rm T}X$ can be pre-computed and stored in memory
    on a mid-grade GPU; this leads to iteration speed-ups of a factor of $\sim\!2\!-\!3$.
    See Sec.~\ref{app:subcan} for additional details on the experiment.
    {\bf (Left)}  We depict the relative statistical efficiency of Subset wTGS with subset size $S$ compared to wTGS.
    The gain in statistical efficiency is largest when $X^{\rm T}X$ cannot be pre-computed (blue), which would be
    the case for Negative Binomial and Binomial Regression. Note that the moderate gain in statistical efficiency
    in ths regime is not surprising; the runtime results in the rightmost panel make it clear that GPU utilization is
    only moderate and so the speed-ups that result in switching from wTGS to Subset wTGS are limited.
    {\bf (Right)} We depict the number of iterations per second (IPS) for Subset wTGS as a function of $S$ as well as wTGS. 
    Results are obtained with a NVIDIA Tesla T4 GPU.
}
\label{fig:cancerlarge}
\end{figure}

We run an additional experiment to complement the experimental results presented in Sec.~\ref{sec:largepexp}.
We use the same cancer dataset as in Sec.~\ref{sec:cancer} (so $N=907$ and $P=17273$) except we look at the gene ZEB2.
We also use the continuous response as provided in the dataset (i.e.~without quantization).
See Fig.~\ref{fig:cancerlarge} for results.

\subsubsection{PG-wTGS runtime}
\label{app:runtime}

\begin{figure}[ht]
\centering
    \includegraphics[width=0.55\linewidth]{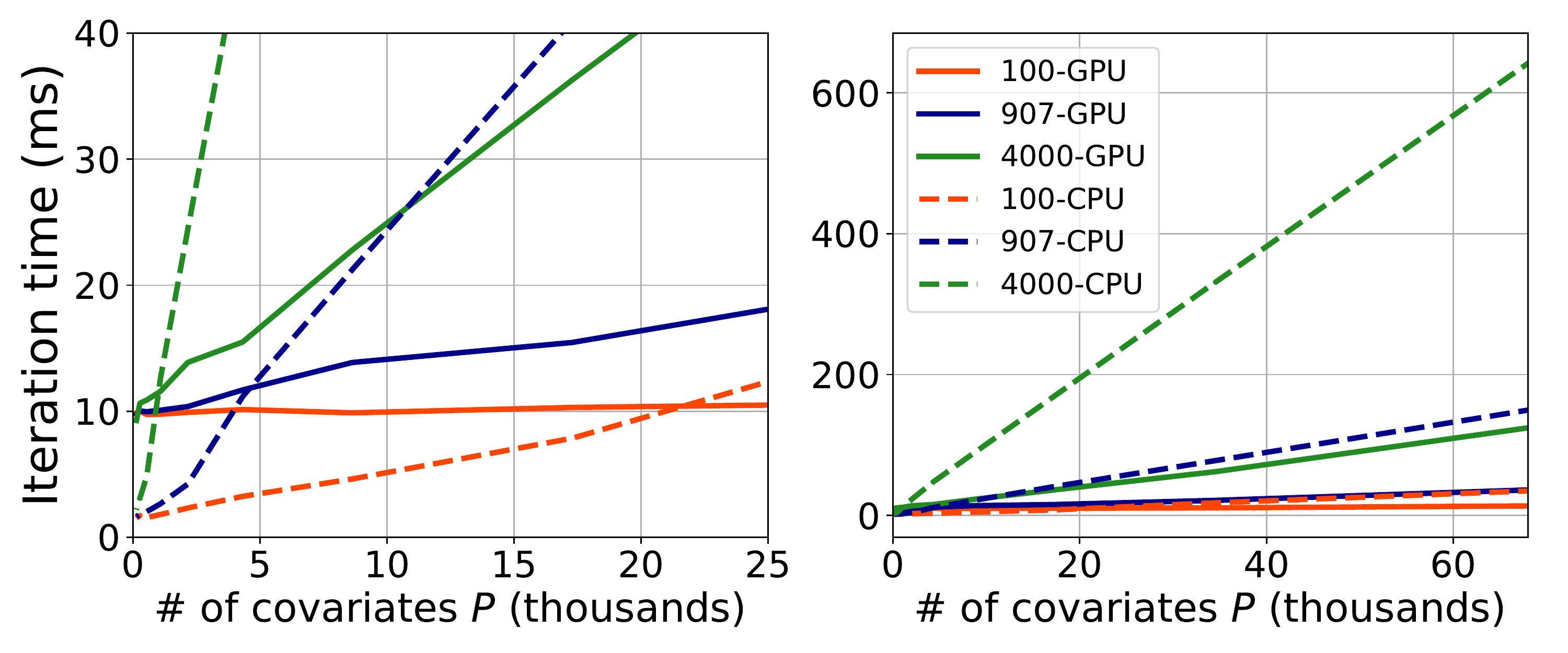}
\caption{We depict MCMC iteration times in milliseconds for PG-wTGS on CPU and GPU as the number of
    covariates $P$ is varied. We also vary the number of data points $N \in \{ 100, 907, 4000 \}$.
    See Sec.~\ref{app:runtime} for details.
    Note that the figure on the left is a magnified version of the figure on the right.
    The CPU has 24 cores (Intel Xeon Gold 5220R 2.2GHz) and the GPU is a NVIDIA Tesla K80 GPU.}
\label{fig:runtime}
\end{figure}

In Fig.~\ref{fig:runtime} we depict MCMC iteration times for PG-wTGS for various values
of $N$ and $P$. To make the benchmark realistic we use semi-synthetic data derived from
the DUSP4 cancer dataset ($N=907$, $P=17273$) used in Sec.~\ref{sec:cancer}. In particular
for $N \ne 907$ and $P \ne 17273$ we subsample and/or add noisy data point replicates and/or
add random covariates as needed. As discussed in more detail in Sec.~\ref{app:cc}, PG-wTGS, PG-wGS, PG-TGS, and ASI 
all have similar runtimes, since each is dominated by the $\OO(P)$ cost
of computing $p(\gamma_j = 1 | \gmj, \omega, \DD)$ for $j=1,...,P$. As can be seen in Fig.~\ref{fig:runtime}
for any given $N$ the iteration time is lower on CPU for small $P$, but GPU parallelization
is advantageous for sufficiently large $P$.\footnote{All P\olya-Gamma sampling is done on CPU.} We 
also note that the computational complexity of PG-wTGS is no worse than linear
in $N$, with the consequence that PG-wTGS can be applied to datasets with large $N$ \emph{and} large $P$ in practice, at least if the sparsity asumption holds
(i.e.~most variables are excluded in the posterior: $\gabs \ll P$).

\subsubsection{Hospital visit data and negative binomial regression}
\label{sec:hospital}

We consider a hospital visit dataset with $N=1798$ considered in \citep{hilbe2011negative} and gathered
from Arizona Medicare data. The response variable is length of hospital stay for patients
undergoing a particular class of heart procedure and ranges between 1 and 53 days.
We expect the hospital stay to exhibit significant dispersion and so we use a negative binomial likelihood.
There are three binary covariates: sex (female/male),
admission type (elective/urgent), and age (over/under 75).
To make the analysis more challenging we add 97 superfluous covariates
drawn i.i.d.~from a unit Normal distribution so that $P=100$.

\begin{figure}[ht]
    \centering
    \includegraphics[width=0.30\linewidth]{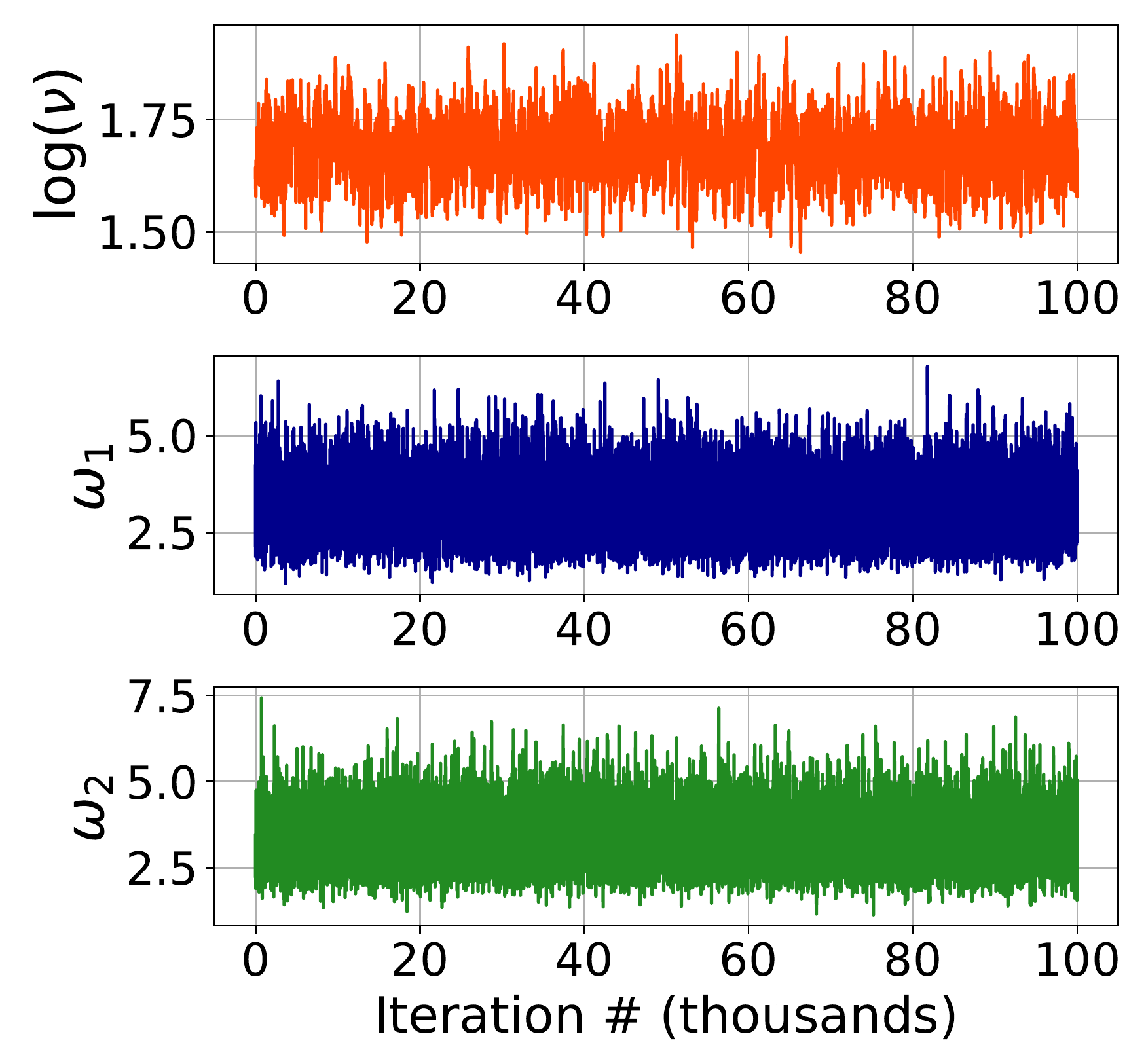}
    \includegraphics[width=0.28\linewidth]{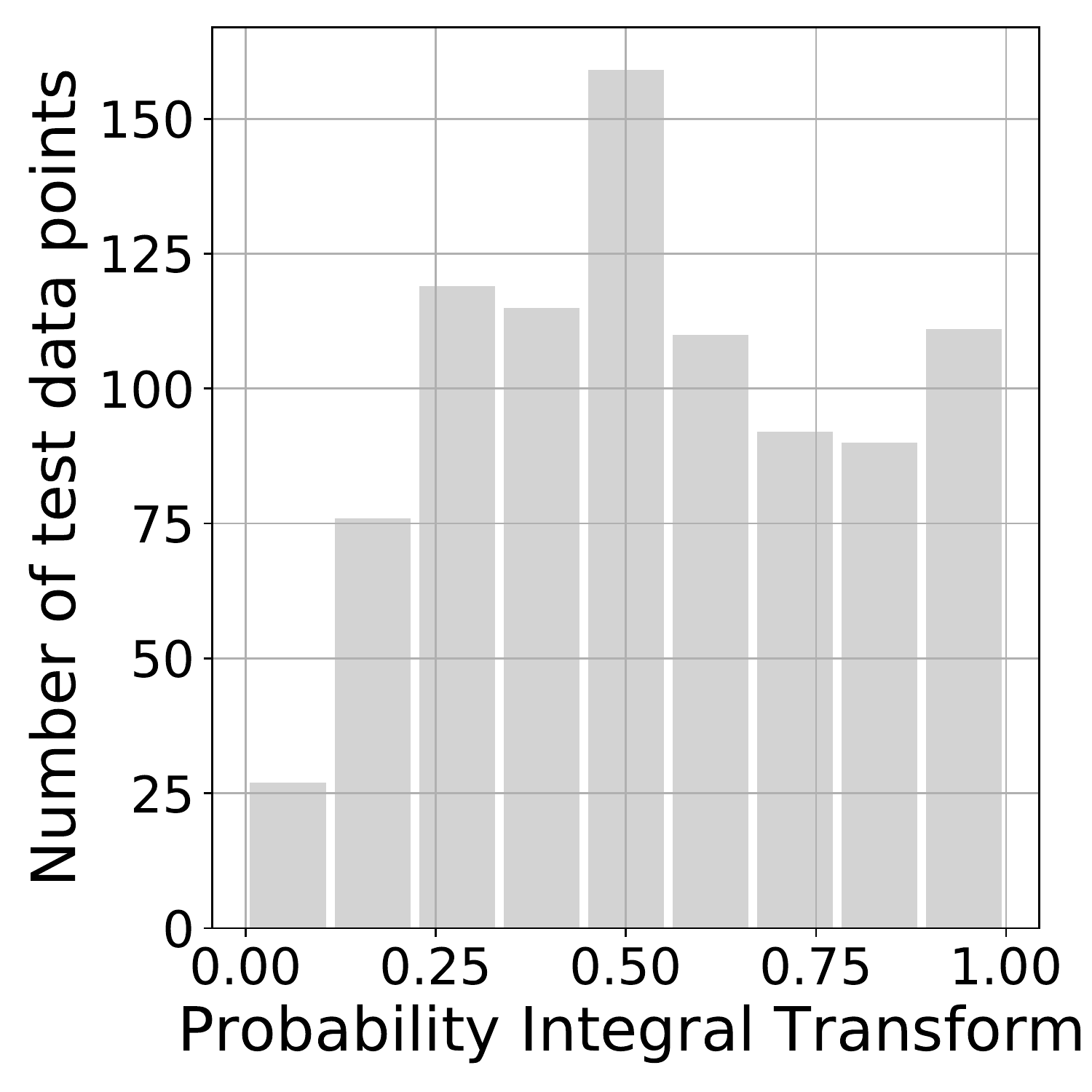}
    \caption{{\bf Left:} We depict trace plots for $\log(\nu)$ and two randomly chosen
    P\olya-Gamma variates $\omega_1$ and $\omega_2$ for a PG-wTGS run on the data
    in Sec.~\ref{sec:hospital}. The data is quite dispersed, with the posterior mean
    of the dispersion parameter $\nu$ being about $5.4$. {\bf Right:} We depict the Probability Integral Transform histogram
    for $899$ held-out test points for the hospital data in Sec.~\ref{sec:hospital}.
    }
\label{fig:nutraceplot}
\end{figure}

\begin{figure}[ht]
    \centering
    \includegraphics[width=0.7\linewidth]{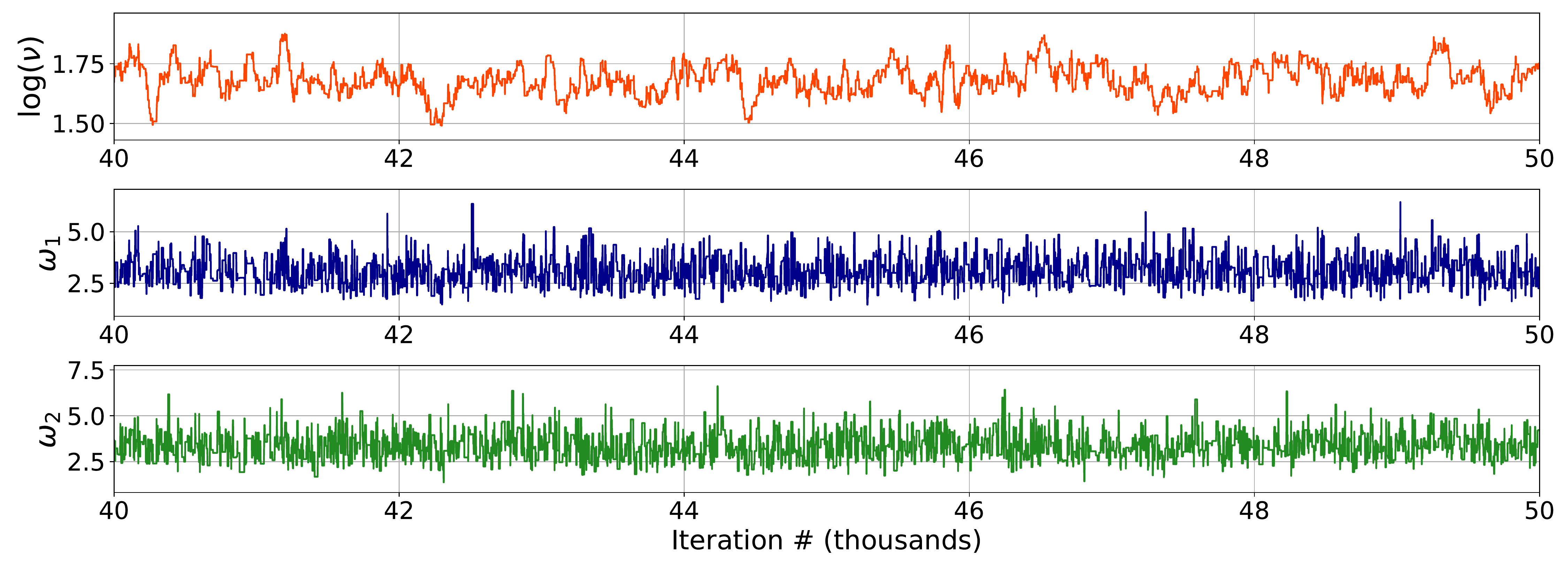}
    \caption{In this zoomed-in companion figure to Fig.~\ref{fig:nutraceplot} we depict trace plots for $\log(\nu)$ and two randomly chosen
    P\olya-Gamma variates $\omega_1$ and $\omega_2$ for a PG-wTGS run on the hospital data
    in Sec.~\ref{sec:hospital}.
    }
\label{app:longtrace}
\end{figure}

Running PG-wTGS on the full dataset we find strong evidence for inclusion of two of the covariates:
sex (PIP $\approx 0.95$) and admission type (PIP $\approx 1.0$).
The corresponding coefficients are negative ($-0.15 \pm 0.02$) and positive ($0.63 \pm 0.03$), respectively.\footnote{
    Each estimate is conditioned upon inclusion of the corresponding covariate in the model.}
This corresponds to shorter hospital stays for males and longer hospital stays for urgent admissions.
In Fig.~\ref{fig:nutraceplot} (left) we depict trace plots for a few latent variables, each of which
is consistent with good mixing; see also Fig.~\ref{app:longtrace} for a zoomed-in view.

Next we hold-out half of the dataset in order to assess the quality of the model-averaged predictive distribution.
We use the mean predicted hospital stay to rank the held-out patients and then partition them into two groups of equal
size. Comparing this predicted partition to the observed partition of patients into short- and long-stay patients,
we find a classification accuracy of 66.6\%.
In Fig.~\ref{fig:nutraceplot} (right) we depict a more fine-grained predictive diagnostic, namely Dawid's
Probability Integral Transform (PIT) \citep{dawid1984present}. Since the PIT values are approximately uniformly
distributed, we conclude that the predictive distribution is reasonably well-calibrated, although
probably somewhat overdispersed.

\subsubsection{Health survey data and negative binomial regression}
\label{sec:health}

We consider the German health survey with $N=1127$ considered in \citep{hilbe2007count}.
The response variable is the annual number of visits to the doctor
and ranges from $0$ to $40$ with a mean of $2.35$.
As in Sec.~\ref{sec:hospital}, we expect significant dispersion and thus use a negative binomial likelihood.
There are two covariates: i) a binary covariate for self-reported health status (not bad/bad); and
ii) an age covariate, which ranges from 20 to 60. We normalize the age covariate
so that it has mean zero and standard deviation one.
To make the analysis more challenging we add 198 superfluous covariates
drawn i.i.d.~from a unit Normal distribution so that $P=200$.

Running PG-wTGS on the full dataset we find strong evidence for inclusion of the health status covariate
(PIP $\approx 1.0$).
The health status coefficient is positive ($1.15 \pm 0.10$), suggesting
that patients whose health is self-reported as bad have $e^{1.15} \approx 3.17$ times as many visits to the doctor
as compared to those who report otherwise. This is consistent with the raw empirical ratio, which is about $3.16$.
We find that the data are very overdispersed and infer the dispersion parameter to be $\nu = 0.99 \pm 0.07$.
See Sec.~\ref{app:nb} for additional details on PG-wTGS for negative binomial regression.

\subsection{Experimental details}
\label{app:exp}

\paragraph{Large $P$ experiments}

For both experiments we create semi-synthetic datasets as follows. We first shuffle the covariate indices.
Next we divide the covariates into $20$ approximately equally sized blocks. Within each block we compute the
correlation between each pair of covariates and randomly select a pair with absolute correlation between $0.5$
and $0.9$; we then randomly choose one of the two indices. In this way we select $20$ covariates, each of which exhibits
non-trivial correlations with at least one other covariate. We then draw $20$ coefficients from
the uniform distribution on $[-1.0, -0.1] \cup [0.1, 1.0]$. We then use our synthetic coefficient
vector $\beta^*$ with $20$ non-zero coefficients to generate
a response $Y_n$ as $Y_n = \beta^* \cdot X_n + \epsilon_n$ for $n=1,...,N$ and where $\epsilon_n \sim \NN(0, \sigma_0^2)$
is i.i.d.~gaussian noise. We generate two datasets: one with $\sigma_0=0.5$ (these results are reported in
Fig.~\ref{fig:largep} in the main text) and one with $\sigma_0=2.5$ (these results are reported in Fig.~\ref{fig:applargepnoisy}).

For the first experiment with $P=98385$ we use all $N=2267$ datapoints and a single fixed dataset.
For the second experiment with $P>98385$ we also use a single fixed dataset, but run experiments for $5$ train/test splits,
where half the data is held-out for testing. As described in the main text, to obtain a dataset with $P>98385$ covariates
we augment the maize data with covariates drawn i.i.d.~from a unit Normal distribution.
We set the prior inclusion probability $h$ to $h=10/P$, the prior precision to $\tau = 10^{-4}$, and $\epsilon=5$.

The relative statistical efficiency reported in Fig.~\ref{fig:big} is defined as a ratio of effective samples sizes per unit time,
which is equivalent to a ratio of time-normalized variances.
It is computed as follows:
\begin{align}
    \frac{{\rm StatEff}({\rm Subset \; wTGS})}{{\rm StatEff}({\rm wTGS})} =  
    \frac{ \sigma_{\rm wTGS}^2 T_{\rm wTGS} } {  \sigma_{\rm Subset \; wTGS}^2 T_{\rm Subset \; wTGS} }
    \label{eqn:stateff}
\end{align}
where e.g.~$ T_{\rm wTGS}$ is the runtime of wTGS and $\sigma_{\rm wTGS}^2$ is the corresponding variance for the
estimator of interest. In Fig.~\ref{fig:big} the estimator of interest is the sum of PIPs over all covariates with
a PIP that exceeds a threshold of $0.001$, of which there are $53$. To determine these ``relevant'' covariates and compute
reference PIPs to compute the required variance in Eqn.~\ref{eqn:stateff}, 
we run $10$ independent wTGS chains with $50$k samples each and compute a mean PIP
across the $10$ chains (this requires about $40$ hours of GPU compute). 
Note that these long chains are independent of the shorter chains used to assess the
statistical efficiency of each method. For each short chain we collect $20$k post-adaptation samples, except
for wTGS where we collect $10$k. In all cases there are $5$k burn-in iterations. For each method we run $10$ independent
chains; the resulting variability determines the variance in Eqn.~\ref{eqn:stateff}. Together
with the runtime, this allows us to compute the (relative) statistical efficiency.

Runtime results are obtained with a NVIDIA Tesla T4 GPU. The predictive and coefficient RMSEs reported
in Fig.~\ref{fig:big} are normalized by the standard deviation of $Y$ and the 
euclidean norm of $\beta^*$, respectively, for interpretability: with this normalization
a RMSE less than unity is a strict improvement over guessing zero.

\paragraph{PG-wGS/PG-TGS/PG-wTGS/ASI}

For experiments with count-based likelihoods (unless specified otherwise) we set the prior precision $\tau = 0.01$ and $h = 5 /P$.
We choose the exploration parameter $\epsilon$ that enters $\eta(\cdot)$ to be $\epsilon=5$.
We use the $\xi$-adaptation scheme described in Sec.~\ref{app:xi}.

We note that PG-TGS uses $\eta(\cdot) = 1$ but still utilizes Metropolized-Gibbs moves to update $\gamma_i$; these
moves result in deterministic flips because of tempering.
By contrast PG-wGS uses the same weighting function $\eta(\cdot)$ as in PG-wTGS but there is no tempering,
with the consequence that $\gamma_i$ still undergoes Metropolized-Gibbs moves but the acceptance probabability is no longer
identically equal to one. See Eqn.~\ref{eqn:accept} for the resulting acceptance probability. 

ASI has several hyperparameters which we set as follows.
We set the exponent $\lambda_{\rm ASI}$ that controls adaptation to $\lambda_{\rm ASI} = 0.75$.
We set $\epsilon_{\rm ASI}=0.1/P$ as suggested by the authors. We target an acceptance probability of $\tau_{\rm ASI}=0.25$.

\paragraph{Correlated covariates scenario}

The covariates for $p=3,4,...,P$ are generated independently from a standard Normal distribution:
$X_{n,p} \sim \NN(0, 1)$ for all $n$. We then generate $z \in \RR^N$ with $z_n\sim \NN(0, 1)$ and
set $X_{n, p=1} \sim \NN(z_n, 10^{-4})$ and  $X_{n, p=2} \sim \NN(z_n, 10^{-4})$. That is the first
two covariates are almost identical apart from a small amount of noise. 
We then generate the responses $Y_n$ using success logits given by $\psi_n = z_n$. 
The total count $\TC_n$ for each data point is set to 10.
Consequently the true posterior concentrates on two modes with $\gamma = (1, 0, 0, ...)$ and  $\gamma = (0, 1, 0, ...)$. 
We set $h=1/P$ and run each algorithm for 10 thousand burn-in/warmup iterations and use the subsequent 100 thousand samples for analysis.

\paragraph{Cancer data}

All chains are run for 25 thousand burn-in/warmup iterations.

\paragraph{Inferring $h$}

We follow the discrete time branching process simulator setup described in the supplement of \citet{jankowiak2022inferring}. We use
identical hyperparameters to those used in the reference except we vary the number of causal effects in each simulation.
In addition for each simulation we choose effect sizes from the uniform distribution on $[-0.10, -0.02] \cup [0.02, 0.10]$.
We choose $\alpha_h = 0.25$ and $\beta_h = 250$ to define the Beta prior over $h$; this choice
corresponds to a relatively broad prior with prior mean $h = 0.001$ (which corresponds to $3$ causal mutations expected a priori).

We provide some intuition for the behavior observed in Fig.~\ref{fig:infh}.
Note that the diffusion-based likelihood that underlies \citet{jankowiak2022inferring} is an approximation
of the underlying discrete time branching process dynamics. Consequently the model is not perfectly well specified.
For this reason---and because of the inherent noisiness of the data---as $h$ increases, there
may be a tendency to push $h$ up further, since doing so allows the model to achieve a better fit of the observed pandemic,
even if some of the identified mutations may be spurious. This explains the larger tails observed for simulations with $10$
causal mutations. This is a general reminder that one needs to proceed with
caution when placing a prior on $h$; in some cases it may be more sensible to assume fixed values of $h$ and do a
sensitivity analysis to assess sensitivity to prior assumptions.

\paragraph{Subset wTGS and cancer experiment}

The experimental details closely follow the experiment in Sec.~\ref{sec:largepexp}, except in contrast the data we use here
is not semi-synthetic. We run $10$ independent chains with wTGS for $500$k iterations each to compute reference PIPs.
We then run $20$ additional independent chains for each method (i.e.~vanilla wTGS and Subset wTGS for various values of $S$)
for $50k$ iterations; the results
of these chains are then used to compute the relative statistical efficiency. To do so we use the PIP over all covariates
as the estimator of interest. In all cases we allow for $10$k burn-in iterations.

\paragraph{Runtime experiment}
For each value of $N$ and $P$ we run each MCMC chain for 2000 burn-in iterations and report iteration times averaged
over a subsequent $10^4$ iterations; we report results in Fig.~\ref{fig:runtime}.

\paragraph{Hospital data}

We run PG-wTGS for 10 thousand burn-in iterations and use the subsequent 100 thousand samples for analysis.
The $899$ held-out patients are chosen at random.
We use a random walk proposal scale for $\log \nu$ of $0.03$.
We set $\psi_0$ to be the logarithm of the mean of the observed $Y$ (this is equivalent to shifting the prior
mean of the bias $\beta_0$; see Sec~\ref{app:nb}).

\paragraph{Health survey data}

We run PG-wTGS for 10 thousand burn-in iterations and use the subsequent 100 thousand samples for analysis.
We use a random walk proposal scale for $\log \nu$ of $0.03$.
We set $\psi_0$ to be the logarithm of the mean of the observed $Y$ (this is equivalent to shifting the prior
mean of the bias $\beta_0$; see Sec~\ref{app:nb}).

\end{document}